\documentclass[a4paper,11pt]{article}
\pdfoutput=1 

\usepackage{jheppub} 
\usepackage{graphicx}
\usepackage{bbm}
\usepackage{amsmath,amssymb}
\allowdisplaybreaks
\usepackage{bbm}
\usepackage{slashed} 
\usepackage{pdflscape} 

\usepackage[T1]{fontenc} 

\usepackage{bm}         
\usepackage{array}
\usepackage{multirow}   
\usepackage{booktabs}   
\usepackage{float}      

\renewcommand{\d}{\mathrm d}

\renewcommand{\(}{\left(}
\renewcommand{\)}{\right)}
\renewcommand{\[}{\left[}
\renewcommand{\]}{\right]}

\newcommand {\beq} {\begin{equation}}
\newcommand {\eeq} {\end{equation}}
\newcommand {\bea} {\begin{eqnarray}}
\newcommand {\eea} {\end{eqnarray}}

\newcommand{\MET}{\slashed{E}_T}

\title{LHC Dark Matter Signals  from  Vector Resonances and  Top Partners}
    
\preprint{FERMILAB-PUB-17-226-PPD-T, CTPU-17-22, KIAS-P17048}

\author[a,b]{Alexander S. Belyaev}
\author[c]{Thomas Flacke} 
\author[d,e]{Bithika Jain}
\author[a]{Patrick B. Schaefers}

\affiliation[a]{School of Physics \& Astronomy, University of Southampton, Southampton SO17 1BJ, UK}
\affiliation[b]{Particle Physics Department, Rutherford Appleton Laboratory, Chilton, Didcot, Oxon OX11 0QX, UK}
\affiliation[c]{Center for Theoretical Physics of the Universe, Institute for Basic Science (IBS), Daejeon, 34051, Korea}
\affiliation[d]{ICTP South American Institute for Fundamental
	Research \& Instituto de F\'isica Te\'orica \\ 
	Universidade Estadual  Paulista, S\~ao Paulo, Brazil}
\affiliation[e]{School of Physics, Korea Institute for Advanced Study, Seoul 02455, Korea}

\emailAdd{A.Belyaev@soton.ac.uk}
\emailAdd{flacke@ibs.re.kr}
\emailAdd{bjain@kias.re.kr}
\emailAdd{P.Schaefers@soton.ac.uk}

\abstract{
Extensions of the Standard Model which address the hierarchy problem and dark matter (DM) often contain top partners and additional resonances at the TeV scale.
We explore the phenomenology of a simplified effective model with a vector resonance $Z'$, a fermionic vector-like coloured partner of the top quark $T'$ as well as a scalar DM candidate $\phi$ and provide publicly available implementations in  \texttt{CalcHEP} and \texttt{MadGraph}.
We 
study the  $pp \to Z' \to T'\overline{T'} \to t\bar{t}\,\phi\phi$ process
at the LHC and find that it plays an important role in addition to the 
$T'\overline{T'}$ production via strong interactions.
It turns out that the presence of the $Z'$ can  provide  a dominant contribution to the $t\bar{t}+\MET$ signature without conflicting with existing bounds from $Z'$ searches in di-jet and di-lepton final states.
We 
find that through this process, the
LHC is already probing DM masses up to about 900~GeV and top partner masses up to about 1.5~TeV, 
thus exceeding the current bounds from QCD production alone almost by a factor of two for both particles.}


\begin{document} 
\maketitle
\newpage
 
\section{Introduction}
The Higgs boson discovery in July 2012~\cite{Aad:2012tfa,Chatrchyan:2012xdj}
was a remarkable celebration of the  unprecedented 
success of the Standard Model (SM), which was missing this last particle. At the same time, this announcement has opened a new chapter in the exploration of physics beyond the Standard Model (BSM). BSM physics is necessary to solve the principal problems of the SM
among which are:  a) the naturalness/fine-tuning  problem on theory side and the related
hierarchy between the Higgs mass and the Planck scale; b) the Dark Matter (DM) problem on the experimental
side at the cosmological scale --- the SM does not provide any viable DM candidate, while the existence of DM
has been established beyond any reasonable doubt.
There are several appealing classes of theories which have the potential to
solve these problems, and in these theories, 
the properties of the the Higgs boson (either as a composite state or a fundamental particle)
are compatible with those of the 125 GeV scalar discovered at the LHC.

Among these theories is Supersymmetry (SUSY)~\cite{Golfand:1971iw,Ramond:1971gb,Neveu:1971iv,Volkov:1973ix}, which 
solves the hierarchy problem in the fundamental Higgs sector via fermion-boson symmetry and provides dark matter candidates. A very attractive alternative to SUSY is  Technicolor (TC)~\cite{Weinberg:1975gm,Susskind:1978ms}, in which the electroweak (EW) symmetry is broken by strong dynamics in analogy to QCD.
In these models, the  Higgs boson is the bound state of new fundamental particles involved in these new strong dynamics, however in spite of the qualitatively  different nature, the Higgs  properties can be similar to those of the SM Higgs
and consistent with the LHC data~\cite{Belyaev:2013ida}.
Another set of promising BSM theories are  Composite Higgs (CH) scenarios \cite{Kaplan:1983sm,Georgi:1984ef,Dugan:1984hq} (see also recent developments starting from \cite{Agashe:2004rs}), in which the new gauge dynamics do not break the electroweak symmetry, but spontaneously break a global symmetry of the high energy model.\footnote{It should be noted that CH models require UV completion with some new strong dynamics and therefore, TC and CH models can appear as different limits of strongly coupled theories~\cite{Cacciapaglia:2014uja}.}
Further alternatives include 
Randall-Sundrum models \cite{Randall:1999ee, Gherghetta:2000qt},    Little Higgs models \cite{ArkaniHamed:2002qx,ArkaniHamed:2002qy,Schmaltz:2005ky},
as well as Twin-Higgs models \cite{Chacko:2005pe} also known as neutral naturalness \cite{Craig:2014aea}.

Many of the non-supersymmetric models mentioned above include a top partner sector which often plays an important role in keeping the models technically natural. Furthermore, the models typically contain further BSM resonances at the TeV scale, and --- if they (or their extensions) also address DM  --- a parity (or a larger symmetry group) keeping dark matter stable as well as a parity-odd dark matter sector.

In this article, we explore the phenomenology of a simplified model, which incorporates some of these ingredients 
at the level of an effective theory with a vector resonance $Z'$, a fermionic vector-like coloured partner of the top quark $T'$ (which we take as part of the parity-odd dark sector) as well as a scalar dark matter candidate $\phi$.
In particular, we study the process $pp \to Z' \to T'\overline{T'} \to t\bar{t} \phi \phi$, in which  the $T'\overline{T'}$ pair coming from the $Z'$ decay gives rise to a $t\bar{t}$ + missing transverse momentum signature, $t\bar{t} +\MET$. 
The $t\bar{t} + \MET$ signature  from the $Z'$ resonance has not been studied previously and, as we show, its new topology has different kinematical distributions in comparison to the $t\bar{t} + \MET$ signature coming from QCD production of $T'\overline{T'}$, which has been studied in~\cite{Han:2008gy,Kraml:2016eti,Baek:2016lnv}.
Besides introducing kinematics different from QCD  $T'\overline{T'}$-pair production, the presence of the $Z'$ can also provide  an additional, potentially even dominant contribution to the pair production rate of top partners without conflicting with existing bounds from $Z'$ searches in di-jet and di-lepton final states. As we will show, the $Z'$,  $T'$ and DM masses can  be probed with the $t\bar{t} +\MET$ signature well beyond the reach coming only from QCD production.

This article is organised as follows: In section~\ref{sec:model}, we present the model and  discuss its parameter space. In section~\ref{sec:results}, we describe the analysis setup, present a parton-level analysis, study gluon-$Z'$ interference effects,
explore the model constraints from di-jet and di-lepton LHC searches and eventually present the LHC potential to explore the model parameter space including the masses $M_{Z'}$,  $M_{T'}$ and $m_{\text{DM}} \equiv m_\phi$ with the $t\bar{t} +\MET$ signature. Lastly, we draw our conclusions in section~\ref{sec:conclusions}.

\section{A Simplified Model with Vector Resonances, Top Partners and Scalar DM}
\label{sec:model}
\subsection{The Model}
As outlined in the introduction, many models like composite Higgs models or models  with extra-dimensions contain top partners as well as vector resonances as part of their TeV scale particle spectrum, and a dark matter candidate (whose stability is protected by a discrete symmetry) is desirable in order to explain the observed dark matter relic density.

For our study, we use a simplified effective model which contains these ingredients in order to study the implications of their interplay for LHC searches. We consider a $Z'$ model where the $Z'$ vector resonance couples to SM quarks and leptons and to top partners $T'_s$ (SU(2) singlet) or $Q'=(T'_d, B'_d)$ (SU(2) doublet). We also include a neutral scalar $\phi$. The top partners and $\phi$ are assumed to carry negative DM parity with $m_{T'} > m_\phi$, while SM particles and the $Z'$ carry positive DM parity. This makes $\phi$ a stable DM candidate, which couples to both top quarks and top partners. The only other renormalisable DM couplings comprise a Higgs-portal coupling term $HH^\dagger \phi^2$ 
and DM self-interactions $\phi^4$. The detailed Langrangian for the model, which we abbreviate as ZP-TP-DM model reads:
\begin{subequations}
\label{eq:full-Lagrangian}
\bea
\mathcal{L} &=& \mathcal{L}_{SM} + \mathcal{L}_{kin} + \mathcal{L}_{Z'q}  + \mathcal{L}_{Z'\ell}  +  \mathcal{L}_{Z'Q'} +  \mathcal{L}_{\phi Q'} - V_\phi \label{eq:Lag}\\
\mathcal{L}_{kin} &=& - \frac{1}{4} \left(\partial_\mu Z'_\nu -\partial_\nu Z'_\mu\right) \left(\partial^\mu Z'^\nu -\partial_\nu Z'^\mu\right) + \frac{M^2_{Z'}}{2} Z'_\mu Z'^\mu\nonumber\\
&& + \frac{1}{2}\partial_\mu\phi \, \partial^\mu \phi - \frac{m^2_{\phi}}{2} \phi^2\nonumber\\
&& + \overline{T'_s} \left(i\slashed{D} - M_{T'_s}\right) T'_s +  \overline{Q'_d} \left(i\slashed{D} - M_{T'_d}\right) Q'_d \, ,\label{eq:fermkin}\\
\mathcal{L}_{Z'q}& = &\lambda_{Z'q\bar{q},L/R} \, Z'_\mu \left(\bar{q}_{L/R} \, \gamma^\mu \, q_{L/R}\right) \, ,\\
\mathcal{L}_{Z'\ell}& = &\lambda_{Z'\ell^+\ell^-,L/R} \, Z'_\mu \left(\bar{\ell}_{L/R} \, \gamma^\mu \, \ell_{L/R}\right) \, ,\\
\mathcal{L}_{Z'Q'}& = &\lambda_{Z'T'_s\overline{T'_s},L/R} \, Z'_\mu \left(\overline{T'_s}_{,L/R} \, \gamma^\mu \, q_{L/R}\right)\nonumber\\
&& +  \lambda_{Z'T'_d\overline{T'_d},L/R} \, Z'_\mu \left(\overline{T'_d}_{,L/R} \, \gamma^\mu \, T'_{d,L/R}\right) \nonumber\\
&& +  \lambda_{Z'T'_d\overline{T'_d},L/R} \, Z'_\mu \left(\overline{B'_d}_{,L/R} \, \gamma^\mu \, B'_{d,L/R}\right)\, ,\\
\mathcal{L}_{\phi Q'}& = & \left(\lambda_{\phi T'_s t} \, \phi \, \bar{t}_R \, T'_{s,R} + \lambda_{\phi T'_d t} \, \phi \, \bar{t}_L \, T'_{d,L} + \lambda_{\phi T'_d t} \, \phi \, \bar{b}_L \, B'_{d,L}\right) + \mbox{h.c.}\, ,\\
V_\phi &=& \frac{\lambda_\phi}{4!} \phi^4 + \frac{\lambda_{\phi H}}{2} \phi^2 \left(|H|^2-\frac{v^2}{2}\right) .
\eea
\end{subequations}

In general,
the $Z'$ is not necessarily  associated to a gauge symmetry and the couplings are ``current-couplings'' (and thus not restricted by gauge-invariance). We therefore leave the couplings of the $Z'$ to SM quarks, leptons, and top partners as free parameters. We write the DM interaction with the Higgs doublet  such that the electroweak contribution to the mass of $\phi$ is absorbed, and $m_\phi$ is the physical mass of $\phi$. One should note that DM-Higgs interactions do not affect the LHC $t\bar{t}+\MET$ signature under study.
However, these interactions are important for the constraints on the model parameter space from  relic density, DM direct (DD)  and indirect (ID) detection experiments, as well as  for $h\rightarrow \phi\phi$ invisible Higgs boson decay limits at the LHC.

In spite of the many parameters appearing even in the simplified model given by equation~(\ref{eq:full-Lagrangian}),
the number of parameters which are relevant to the $t\bar{t}+\MET$ signature under study at the LHC, DM searches in DD and ID experiments as well as Higgs physics at the LHC is much more reduced as we discuss in the following section.

\subsection{The Model Parameter Space and Analysis Setup}

We are studying the $t\bar{t}+\MET$ final state in this article, which receives contributions from $T'$-pair production either through QCD interactions or through resonant $Z'$ production with $Z'\rightarrow T'\overline{T'}$. For the $t\bar{t}+\MET$ signature at the LHC coming from the QCD $T'\overline{T'}$-pair production, the only relevant parameters are  $M_{T'}$ and $m_\phi$. In this work, we perform a detailed study of singlet $T'$ ($T'_s$) pair production. Doublet $T'$ ($T'_d$) pair production is expected to have very similar phenomenology.\footnote{The production of $T'_d$-pairs through QCD and $Z'$ (with couplings $\lambda_{Z'T'_d\overline{T'}_d}$ instead of $\lambda_{Z'T'_s\overline{T'_s}}$) and their kinematical distributions of $T'_d$ are basically identical to those of $T'_s$ (see also the remark below about the small difference between various chirality combinations). As main difference, $T'_d$-pair production obtains an additional contribution from $Z^*\rightarrow T'_d\overline{T'_d}$ which is, however, highly suppressed. One should note, though,  that  in case of $T'$ doublet, the $T'_d$  is accompanied by the charge $-\frac13$ state $B'_d$, which by itself can be pair-produced through QCD or $Z'$ and yields a $b\bar{b}+\MET$ signature, similar to the general $jj+\MET$ signature studied in the context of the QCD production of vector-like quarks followed by their decay to light quarks and DM~\cite{Giacchino:2015hvk}.
}

The $Z'$ contribution to the signature under study adds $M_{Z'}$ and the $Z'$ couplings to SM quarks and leptons as well as $Z'$ couplings to $T'_s$ to the parameter space.
In section~\ref{sec:results}, we demonstrate that the differences between the four possible chiral coupling combinations for $\lambda_{Z'q\bar{q},L/R}$ and $\lambda_{Z'T'_s\overline{T'_s},L/R}$ (i.e. LL, RR, LR, RL) are negligible when studying the $t\bar{t}+\MET$ signature. Therefore, it is sufficient to consider just one coupling combination and we choose it to be LL, i.e. the case where $\lambda_{Z'q\bar{q},L}$ and $\lambda_{Z'T'_s\overline{T'_s},L}$ are non-zero and all right-handed couplings vanish. 
We also consider the case of a non-vanishing $Z'$ coupling to SM leptons and choose $\lambda_{Z'\ell^+\ell^-,L}=\lambda_{Z'q\bar{q},L}$, from which results for other coupling ratios can be inferred.

The complete set of model parameters relevant to our study of the $t\bar{t}+\MET$ signature at the LHC comprises five parameters:

\begin{equation}
M_{T'_s}, \ \ m_\phi, \ \  M_{Z'},\ \ \lambda_{Z'q\bar{q},L},\ \  \lambda_{Z'T'_s\overline{T'_s},L\, .}
\end{equation}

The DM phenomenology --- in particular the DM relic density as well as DM direct and indirect detection --- depends on two more parameters,
\begin{equation}
\lambda_{\phi H}\ \  {\rm and} \ \  \lambda_{\phi T'_s t} \, , 
\end{equation}
whose effects we illustrate below. Before doing so, we describe our analysis setup to gather the results on the DM relic density as well as direct and indirect detection limits in the remainder of this section. The analysis setup used to perform the collider analysis is described in the next section.

We have implemented the model described by the Lagrangian in eq.~(\ref{eq:full-Lagrangian}) using the \texttt{LanHEP}~\cite{Semenov:1998eb,Semenov:2008jy,Semenov:2010qt} and \texttt{FeynRules}~\cite{Christensen:2008py,Alloul:2013bka} packages for \texttt{CalcHEP}~\cite{Belyaev:2012qa} 
and \texttt{MadGraph5\_aMC@NLO} \cite{Alwall:2014hca}, respectively.
The implementations have been cross-checked against each other for scattering and decay processes and are available at {\sc HEPMDB}~\cite{hepmdb}
under hepmdb:0717.0253~\cite{dmtp-calchep} (\texttt{CalcHEP}) and hepmdb:0717.0253~\cite{dmtp-mg} (\texttt{Madgraph}).
For the  parton level studies and simulations, we use \texttt{MadGraph5\_aMC\@NLO 2.3.3}  and \texttt{CalcHEP 3.6.27} with the \texttt{NNPDF2.3QED} PDF \cite{Ball:2013hta}. 
For both QCD renormalisation and PDF factorisation scales, we used $Q=M_{Z'}$. 
In our study, we do not apply NLO k-factors to the signal, so our results on the exclusion of the parameter space are conservative.

Hadronisation and parton showering were performed via \texttt{Pythia\,v8.219}~\cite{Sjostrand:2007gs} with subsequent
fast detector simulation performed using  \texttt{Delphes 3}~\cite{deFavereau:2013fsa} and \texttt{FastJet v.3.1.3} \cite{Cacciari:2005hq,Cacciari:2011ma} with a cone radius $\Delta R = 0.4$ for the jet reconstruction. The detector level analysis was performed using \texttt{CheckMATE 2.0.0} \cite{Drees:2013wra,Dercks:2016npn}
to probe  the $t\bar{t} + \slashed{E}_T$ signature against the current $\sqrt{s} = 13$ TeV ATLAS and CMS constraints \cite{Aad:2016tuk,Aaboud:2016uro,Aaboud:2016tnv,Aaboud:2016zdn,Aad:2016qqk,Aad:2016eki,Aaboud:2016lwz,TheATLAScollaboration:2015nxu,TheATLAScollaboration:2016gxs,ATLAS:2016xcm,ATLAS:2016ljb,CMS:2015bsf}.

For illustration purposes and in order to stress  the complementarity of collider and non-collider searches,
we have evaluated the  DM relic density, $\Omega_{\rm DM} h^2$, with the latest version of the 
\texttt{micrOMEGAs v4.3.5} package~\cite{Belanger:2013oya,Belanger:2006is, Belanger:2010gh},
which directly reads the model files in \texttt{CalcHEP} format. We have also checked the model parameter space for consistency with the limits from  DM direct detection (DD) experiments. To do so, we have evaluated the spin-independent cross section of DM scattering off the proton, $\sigma_{SI}$, using the \texttt{micrOMEGAs} package and compared it to the latest and so far strongest DM DD limit from the Xenon 1 Ton experiment~\cite{Aprile:2017iyp}. Since digital data was not provided in the above paper, we have digitised the limit and uploaded it to the \textsc{PhenoData} database~\cite{phenodata-xenon1t}. One should also note that the latest version of \texttt{micrOMEGAs} mentioned above correctly evaluates the one-loop-induced DM scattering rates on nuclei.

\begin{figure}[htbp]
\centering
\includegraphics[width=0.8\textwidth]{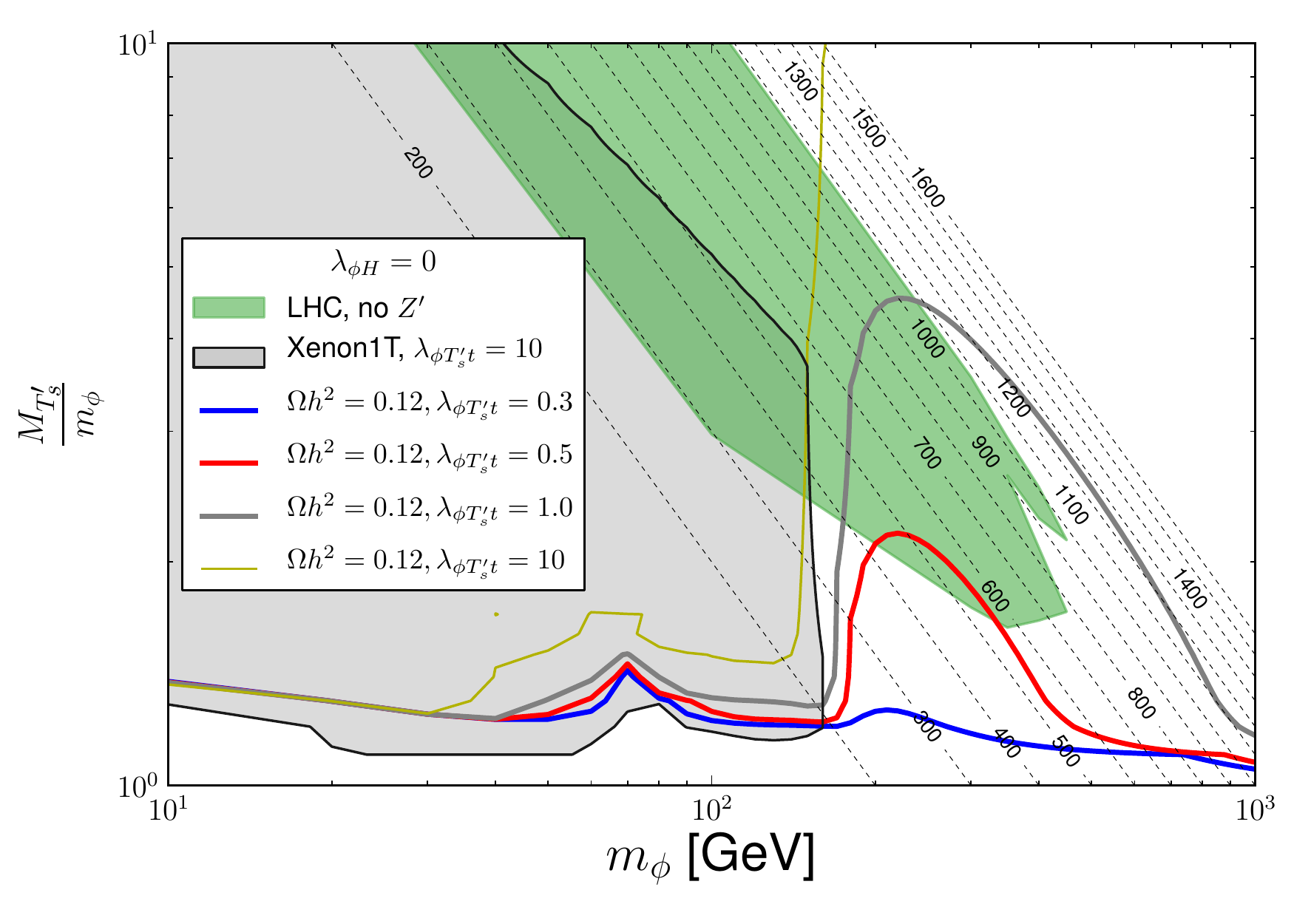}
\vskip -0.5cm
  \caption{LHC, DM Direct Detection and relic density constraints on the parameter space of the ZP-TP-DM  model in the $\(\frac{M_{T'_s}}{m_\phi}\text{ , }m_\phi\)$ plane for $\lambda_{\phi H}=0$:
  a) the green-shaded area indicates the current LHC exclusion region for the $t\bar{t}+\MET$ signature coming from the process $pp \to T'\overline{T'} \to t\bar{t}\phi\bar{\phi}$ mediated only by gluon exchange (no $Z'$ exchange);
  b) the grey-shaded area indicates the exclusion region from DM direct detection from the latest Xenon 1 Ton data~\cite{Aprile:2017iyp} for $\lambda_{\phi T'_s t}=10$;
  c) the parameter space above the blue, red, grey and yellow contours is excluded by the relic density constraints for $\lambda_{\phi T'_s t}=0.3$, 0.5, 1 and 10 respectively, with each contour corresponding to the $\Omega h^2=0.12$ iso-level. The thin dashed lines with the respective labels indicate the iso-levels of $M_{T'_s}$ in GeV.}
  \label{fig:DMTP-parspace1}
\end{figure}

In figure~\ref{fig:DMTP-parspace1}, we present LHC, DM Direct Detection (DD) and relic density constraints on the parameter space of the  ZP-TP-DM  model in the $\(\frac{M_{T'_s}}{m_\phi}\text{ , }m_\phi\)$ plane for $\lambda_{\phi H}=0$, i.e. for the case in which the relic density is fully determined by co-annihilation of $\phi$ with the $T'$, without any contribution from $\phi$ interactions with the Higgs.
The green-shaded area indicates the current LHC exclusion region for the $t\bar{t}+\MET$ signature coming from $pp \to T'\overline{T'} \to t\bar{t}\,\phi\phi$, mediated by gluon exchange only (no $Z'$ exchange). These bounds will be discussed in detail in section \ref{sec:QCD-Tp}. The thin dashed lines with the respective labels indicate the iso-levels of $M_{T'_s}$ in GeV. The exclusion area qualitatively agrees with the one found in \cite{Baek:2016lnv}, however its lower edge corresponding to lower $M_{T'_s}$ is slightly extended in our study, since in addition to \verb|ATLAS_CONF_2016_050|~\cite{ATLAS:2016ljb} we  are  using results of \verb|ATLAS_1604_07773|~\cite{Aaboud:2016tnv}, which  are more sensitive to a smaller mass gap between DM and the $T'_s$. Figure~\ref{fig:DMTP-parspace1} also holds the exclusion region from DM direct detection based on the latest Xenon 1 Ton data~\cite{Aprile:2017iyp} for $\lambda_{\phi T'_s t}=10$, shown as grey-shaded area.
One should note that for  smaller values of  $\lambda_{\phi T'_s t}=0.3$, 0.5 and 1.0, which are chosen as other benchmarks for this plot, Xenon 1 Ton does not have any sensitivity to the parameter space yet, since the cross section for DM scattering off nuclei scales quadratically with $\lambda_{\phi T'_s t}$, limiting the experiment to probe only large values of $\lambda_{\phi T'_s t}$ at the moment.

Also in figure~\ref{fig:DMTP-parspace1}, the  blue, red, grey and yellow contours (for $\lambda_{\phi T'_s t}=0.3$, 0.5, 1 and 10, respectively) show the parameter values which reproduce a relic density of $\Omega h^2=0.12$, corresponding to the value observed by PLANCK~\cite{Ade:2013zuv,Planck:2015xua}. The parameter space above these lines (for the respective value of $\lambda_{\phi T'_s t}$) yields too large relic densities and is thus excluded. 
From figure~\ref{fig:DMTP-parspace1}, one can see that the LHC plays an important and complementary role to DM DD and relic density constraints in covering the $m_\phi > m_t$ region, which is not fully constrained by non-collider experiments, especially for not-so-small values of $\lambda_{\phi T'_s t}$. 
\begin{figure}[htbp]
\centering
\includegraphics[width=0.8\textwidth]{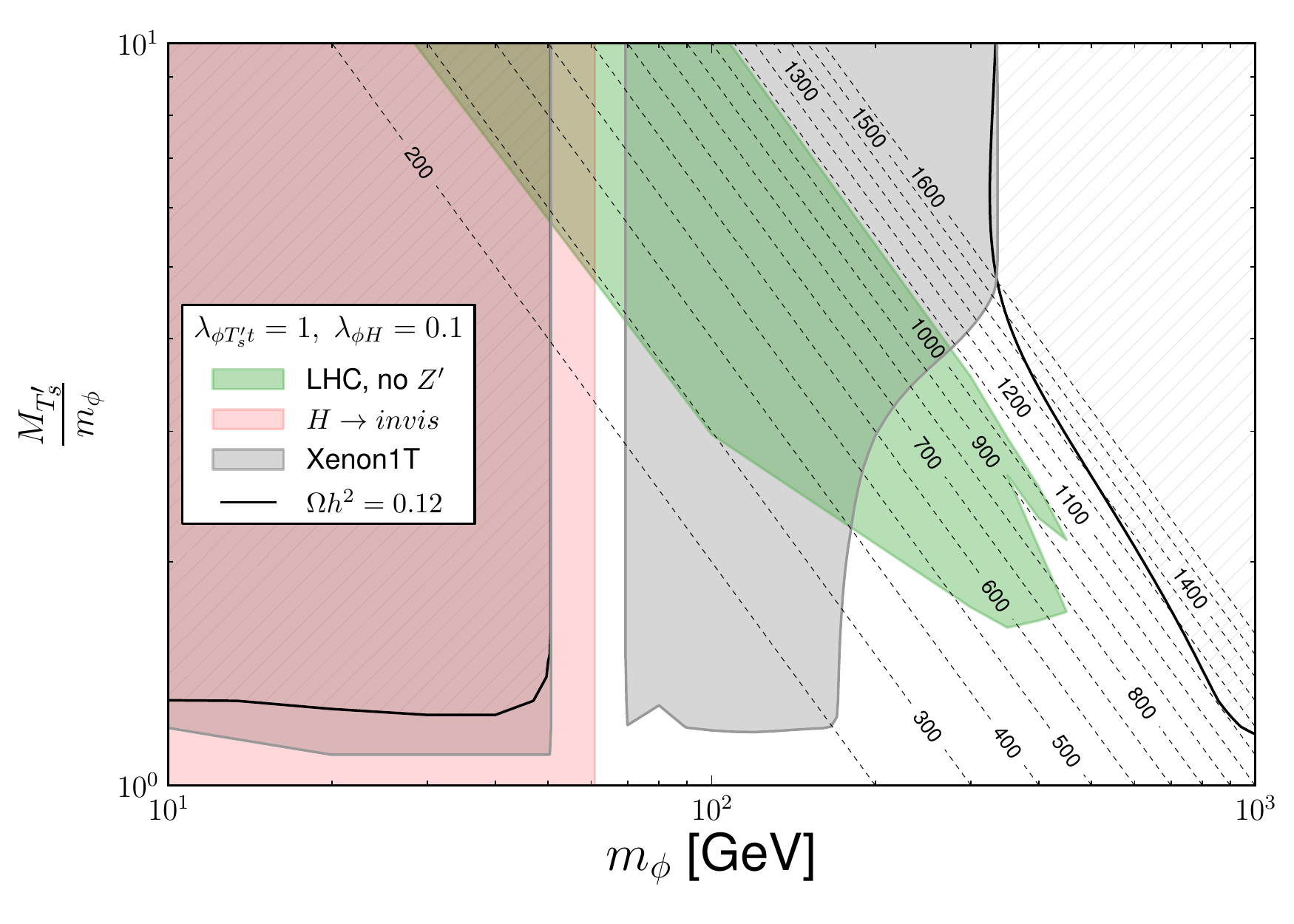}\\
\vskip -0.3cm
\includegraphics[width=0.8\textwidth]{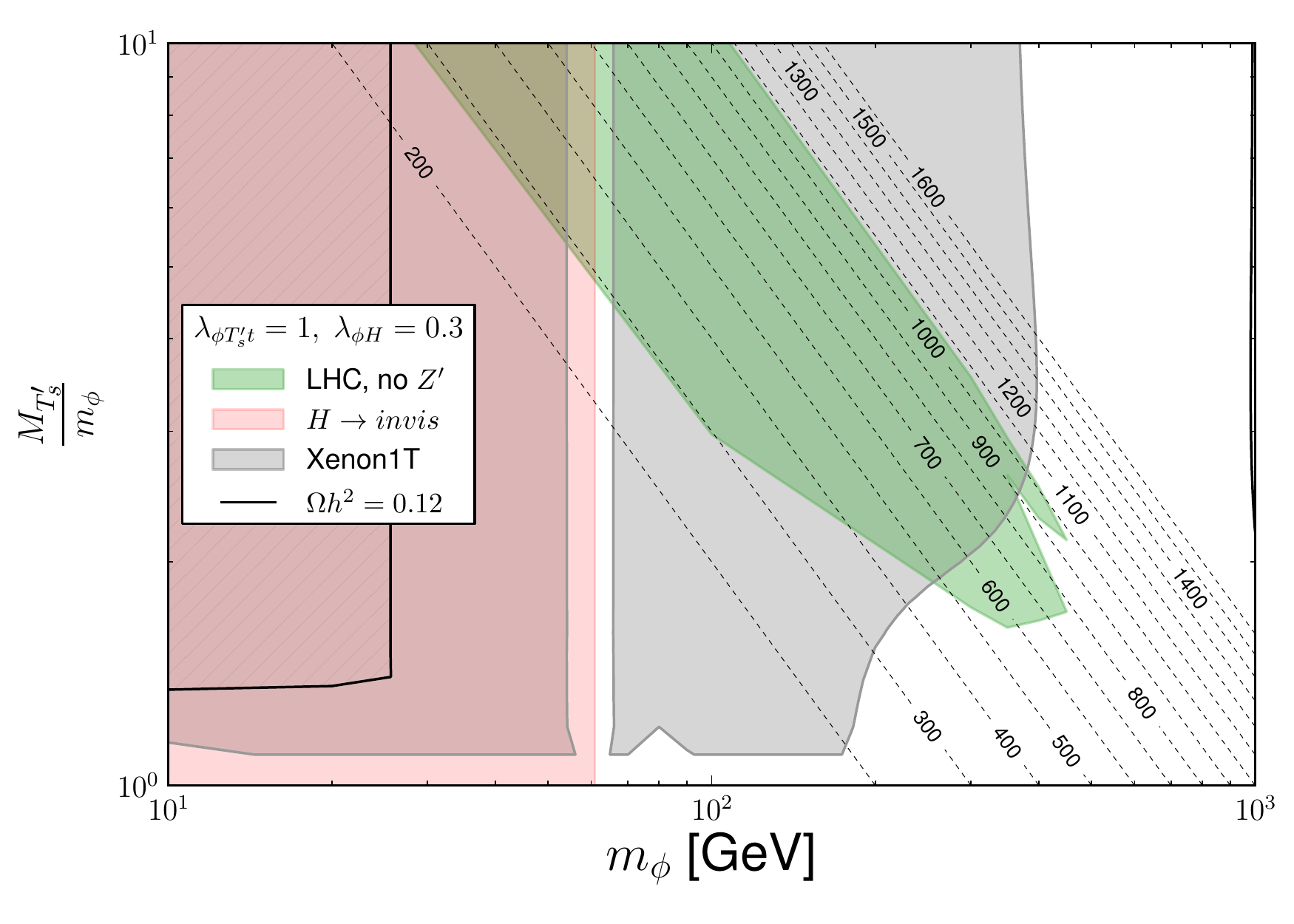}%
\vskip -0.5cm
  \caption{LHC, DM Direct Detection and relic density constraints on the parameter space of the ZP-TP-DM  model
  in the $\(\frac{M_{T'_s}}{m_\phi}\text{ , }m_\phi\)$ plane for $\lambda_{\phi T'_s t}=1$ and two values of 
  $\lambda_{\phi H} = 0.1$ (top) and  0.3 (bottom):
  a) the green-shaded area indicates the current LHC exclusion region from the process $pp \to T'\overline{T'} \to t\bar{t}\phi\bar{\phi}$ without $Z'$ exchange;
  b) the grey-shaded area indicates the exclusion region from DM DD from  the latest Xenon 1 Ton data~\cite{Aprile:2017iyp};
  c) the hatched parameter space is excluded by relic density constraints;
  d)  the pink-shaded area indicates the limit from the invisible Higgs decay searches.
  The thin dashed lines indicate the iso-levels of $M_{T'_s}$ in GeV.}
  \label{fig:DMTP-parspace2}
\end{figure}
One should also note that the allowed parameter space in the $\(\frac{M_{T'_s}}{m_\phi}\text{ , }m_\phi\)$ plane
can be affected by non-zero $\lambda_{\phi H}$. The relic abundance can be strongly altered in this case, because DM can annihilate through Higgs interactions instead of co-annihilation with the top partner. At the same time, direct detection bounds are modified as the DM can now interact with nuclei via Higgs exchange. Finally, the Higgs can decay into DM if the DM is sufficiently light, which yields additional bounds from LHC Higgs measurements. In  figure~\ref{fig:DMTP-parspace2}, we show the resulting bounds for fixed $\lambda_{\phi T'_s t}=1$ and two cases of  $\lambda_{\phi H} = 0.1$ (top) and 0.3 (bottom).
One can see that also for non-zero $\lambda_{\phi T'_s t}$, the very narrow region for $m_\phi$ around $M_H/2$ is allowed by non-collider searches, because in this region, the relic density is strongly reduced due to resonant DM annihilation into a Higgs boson, while DM DD rates, which are rescaled with the relic density, are also suppressed. At the same time, this region can be effectively probed by the LHC. 
As an illustration, the pink-shaded area indicates the limit from the invisible Higgs decay searches from ATLAS~\cite{Aad:2015txa}, which exclude BR$(H\to \text{invisible}) < 28\ \%$ at 95 \% CL. Eventually, this limit is relevant for $m_\phi < M_H/2$ and sufficiently large ($\simeq 0.015$) values of  $\lambda_{\phi H}$.

As mentioned above, we present the non-collider constraints for illustration purposes only. Since the $Z'$ does not affect the non-collider DM phenomenology for this model, we refer the reader to Ref.~\cite{Baek:2016lnv,Giacchino:2015hvk} for a detailed exploration of the DM direct and indirect collider constraints, where an analogous model, but without the $Z'$, was studied. We also  refer the reader to other works on scalar singlet DM (see e.g. Ref.~\cite{Athron:2017kgt} and the references therein), which are relevant for the case when heavy top partners are decoupled.

Finally, we would like to note that there is a special region of the parameter space, where the mass gap between $m_\phi$ and $M_{T'_s}$ is small. In  this region, where the relic density is equal or below the PLANCK constraint, the $T'_s$ decays to DM, soft b-jets and light jets or leptons (coming from virtual top quark decays). This case is very similar to the case of SUSY with degenerate stops and neutralinos and requires a dedicated analysis beyond the scope of this work, where we focus on the role of the $Z'$ boson to extend the LHC reach of the ZP-TP-DM parameter space for the case where $M_{T'_s} - m_\phi > m_{t}$.
            
\section{\texorpdfstring{Analysis of $pp \rightarrow Z'\rightarrow T'_s\overline{T'_s}\rightarrow t\bar{t}\phi\phi$ for the LHC}{Analysis of pp to Z' to T'T' to ttbar DM DM for the LHC}}
\label{sec:results}

\begin{figure}[ht] 
  \centering
  \includegraphics[scale=.28]{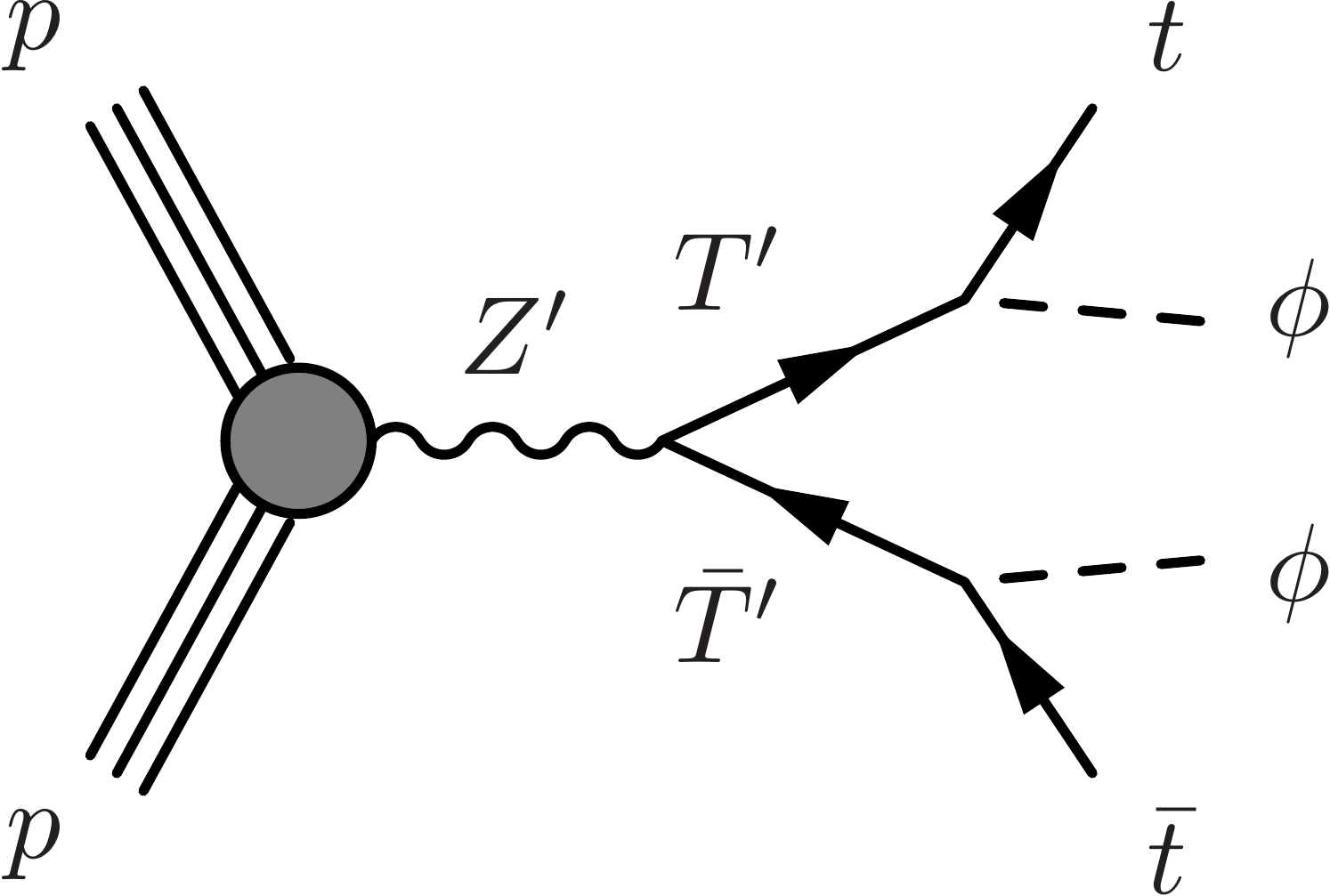}\hfill
  \includegraphics[scale=.28]{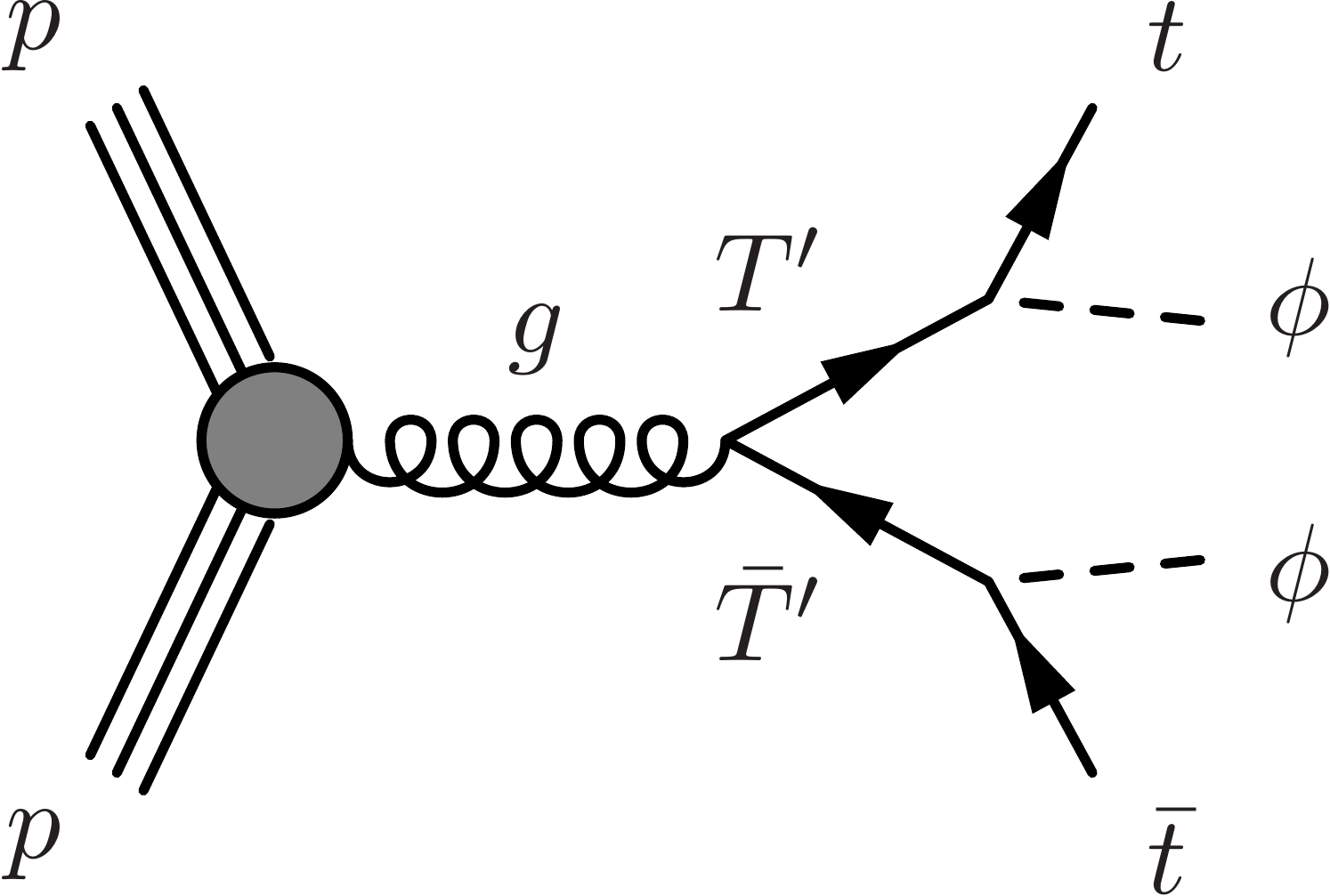}\hfill
  \includegraphics[scale=.28]{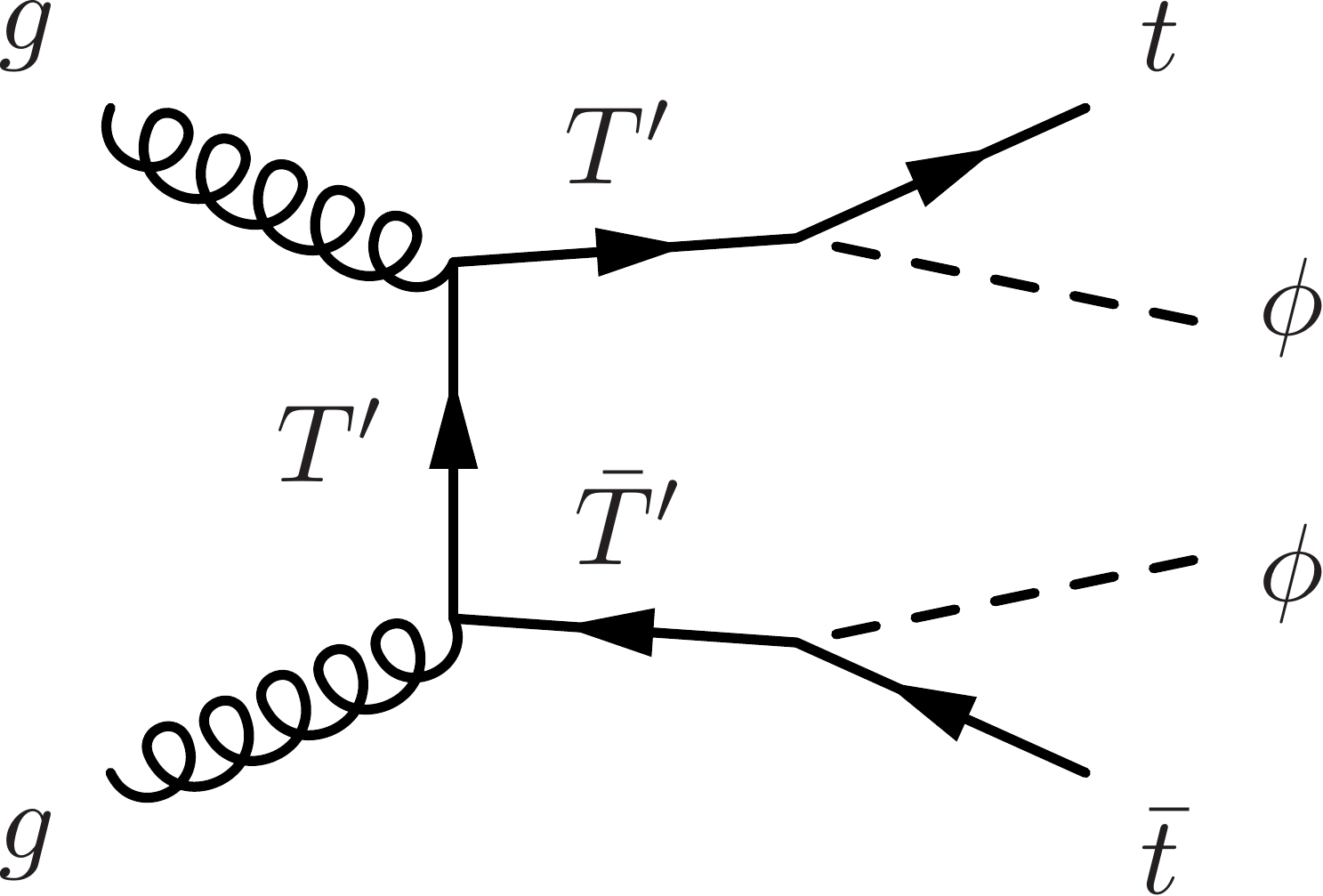}
  \caption{Feynman diagrams for $t \bar{t} \, \phi \phi$ production via $T'_s$ decays from $Z'$ bosons (left) and gluons (center and right).}
  \label{fig:Zp-prod-Feynman-diagram}
\end{figure}

In this study, we focus on $Z'$ production, where the $Z'$ then decays to a $T'_s$-pair which further decays into the final state consisting of two top quarks and DM, i.e. $t \bar{t} \, \phi \phi$. The same final state also arises from QCD pair production of $T'_s\overline{T'_s}$. The corresponding Feynman diagrams are shown in figure~\ref{fig:Zp-prod-Feynman-diagram}.
%

We start our investigation with several pre-studies and checks. In section~\ref{sec:PLA}, we investigate the dependence of kinematic distributions on the chirality of the couplings involved in the $Z'$ production and its subsequent cascade-decay. We will find that the chirality of the couplings $\lambda_{Z'q\bar{q}}$, $\lambda_{Z'T'_{s}\overline{T'}_{s}}$ and $\lambda_{\phi'T'_{s}\bar{t}}$ have a minor influence on the kinematics at parton level. This justifies to pick one specific set of chiralities for further studies, which we choose as $\lambda_{Z'q\bar{q},L}$ and $\lambda_{Z'T'_{s}\overline{T'_s}_{,L}}$, and we consider an $SU(2)$ singlet top partner (implying a coupling to $\phi$ and $t_R$). Similar minor differences in the kinematics can also be expected in the $T'_d$ case.
 
As two further pre-studies, in section \ref{sec:interference-effects} we explicitly check that interference between the $Z'$ produced and QCD produced $T'_s\overline{T'_s}$-pair is very small and not relevant to our studies, while in section \ref{sec:NWA-validity} we quantify finite-$Z'$-width effects. 

To determine the constraints from LHC searches on the model parameter space, we first determine the constraints on QCD-only $T'\overline{T'_s}$-pair production in section~\ref{sec:QCD-Tp}. The details of the  LHC searches used are specified at the end of that subsection. In section~\ref{sec:dijet-dilep} we determine the bounds on the couplings $\lambda_{Z'q\bar{q}}$ and  $\lambda_{Z'T'_{s}\overline{T'_s}}$ from di-lepton and di-jet resonance searches, which arise due to the $Z'$ being allowed to also decay into $q\bar{q}$ and $\ell^+\ell^-$.
In section~\ref{sec:ttMET}, we determine the improved bounds on the  $\(\lambda_{Z'q\bar{q}}, \lambda_{Z'T'_{s}\overline{T'_s}}\)$ parameter space when LHC SUSY search bounds are applied for the process $pp \to Z' \to T'_s\overline{T'_s} \to t\bar{t}\;\phi\phi$. Section~\ref{sec:DBA} contains a detailed benchmark analysis.

\subsection{Pre-study I: Impact of Chiral Couplings on Kinematical Distributions}
\label{sec:PLA}

\begin{figure}[htbp]
  \includegraphics[width=0.65\textwidth]{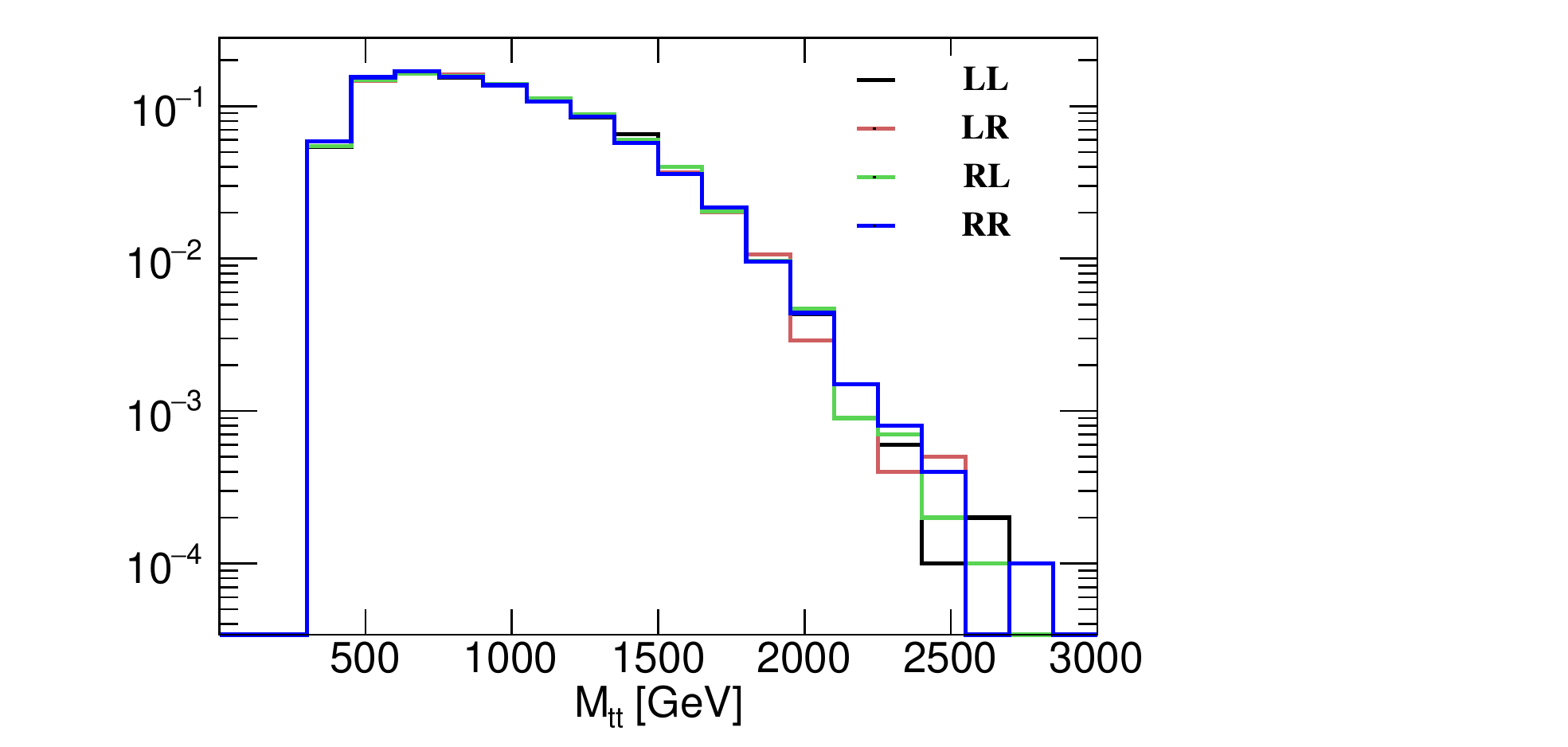}
  \hspace{-2 cm}
  \includegraphics[width=0.65\textwidth]{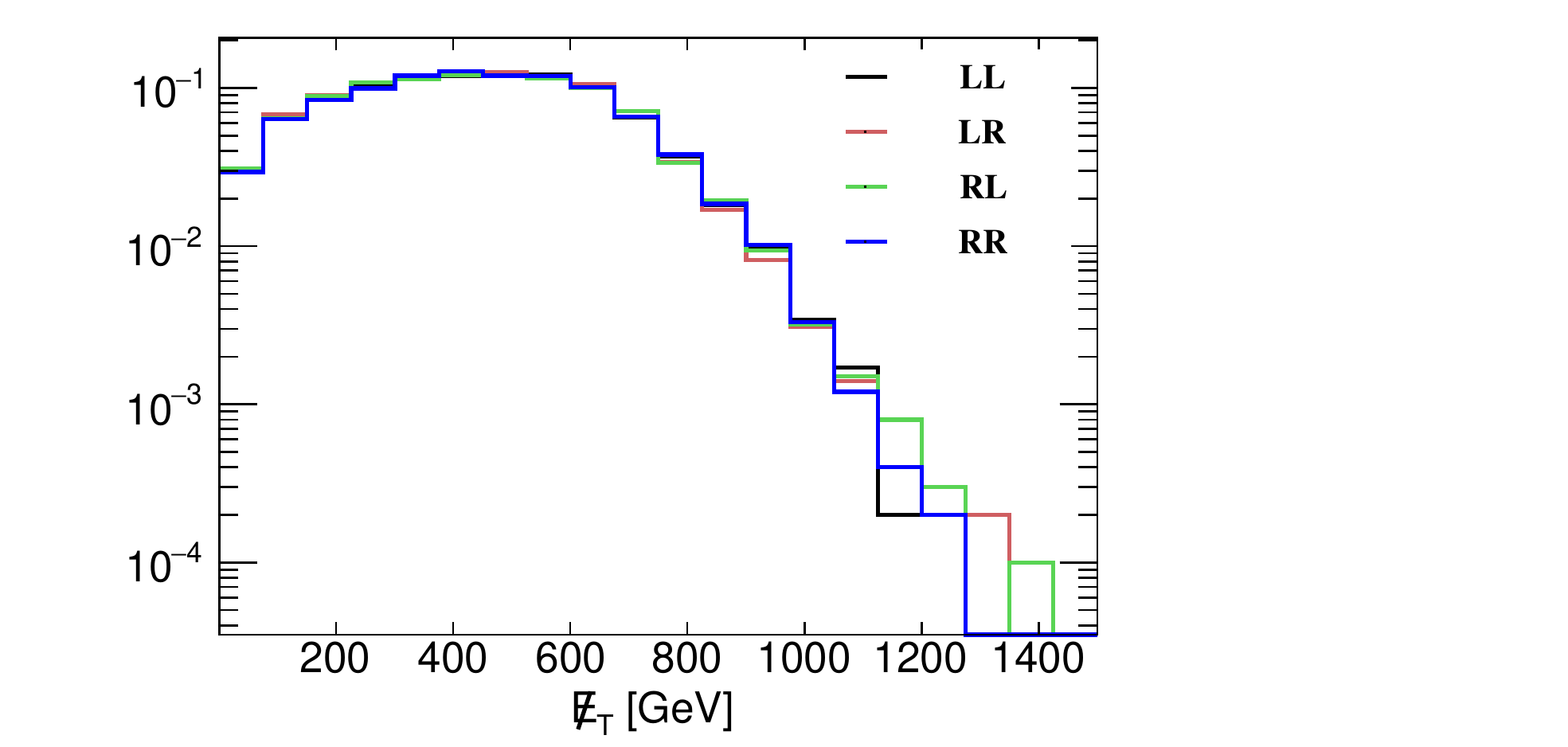}\\
  \includegraphics[width=0.65\textwidth]{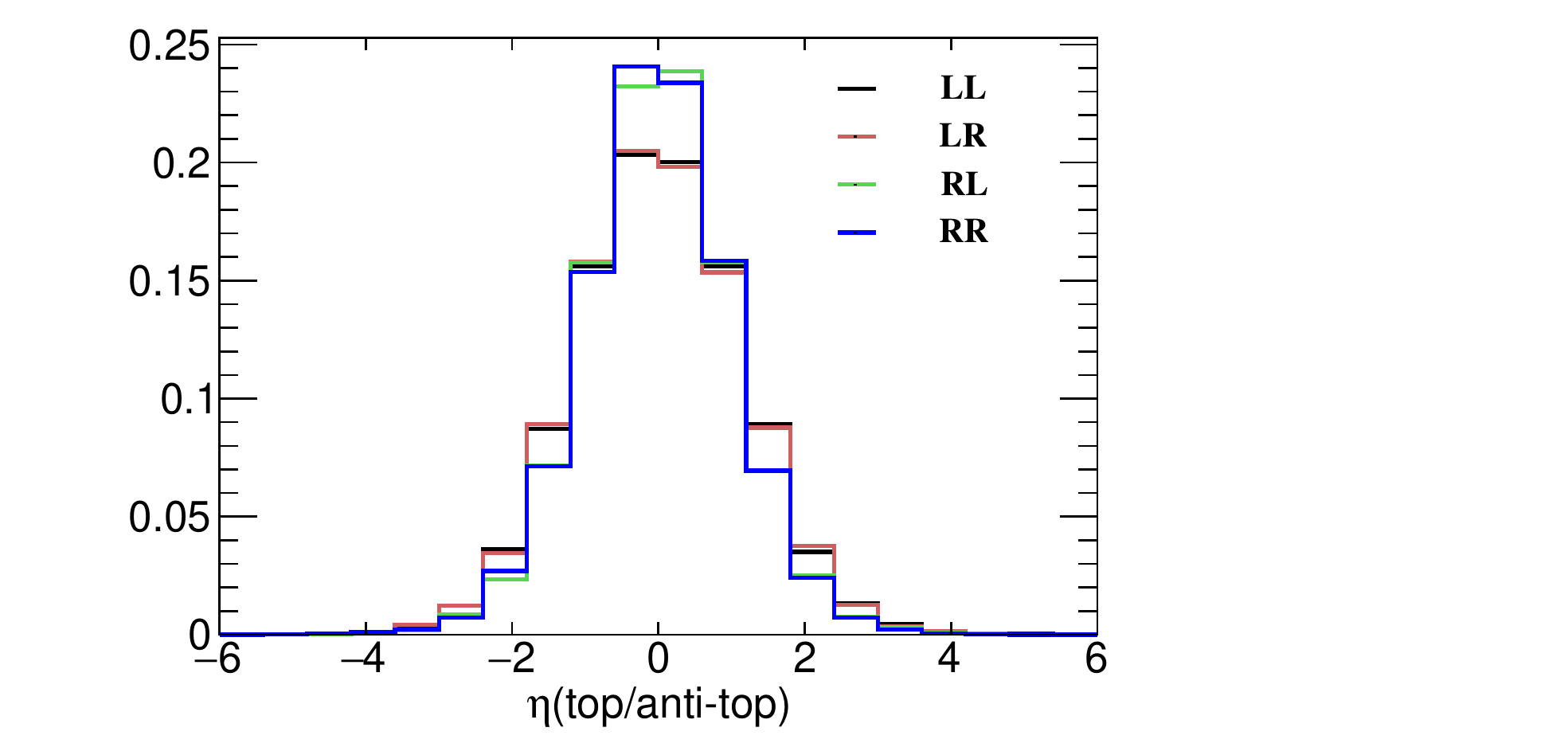}%
  \hspace{-2 cm}
  \includegraphics[width=0.65\textwidth]{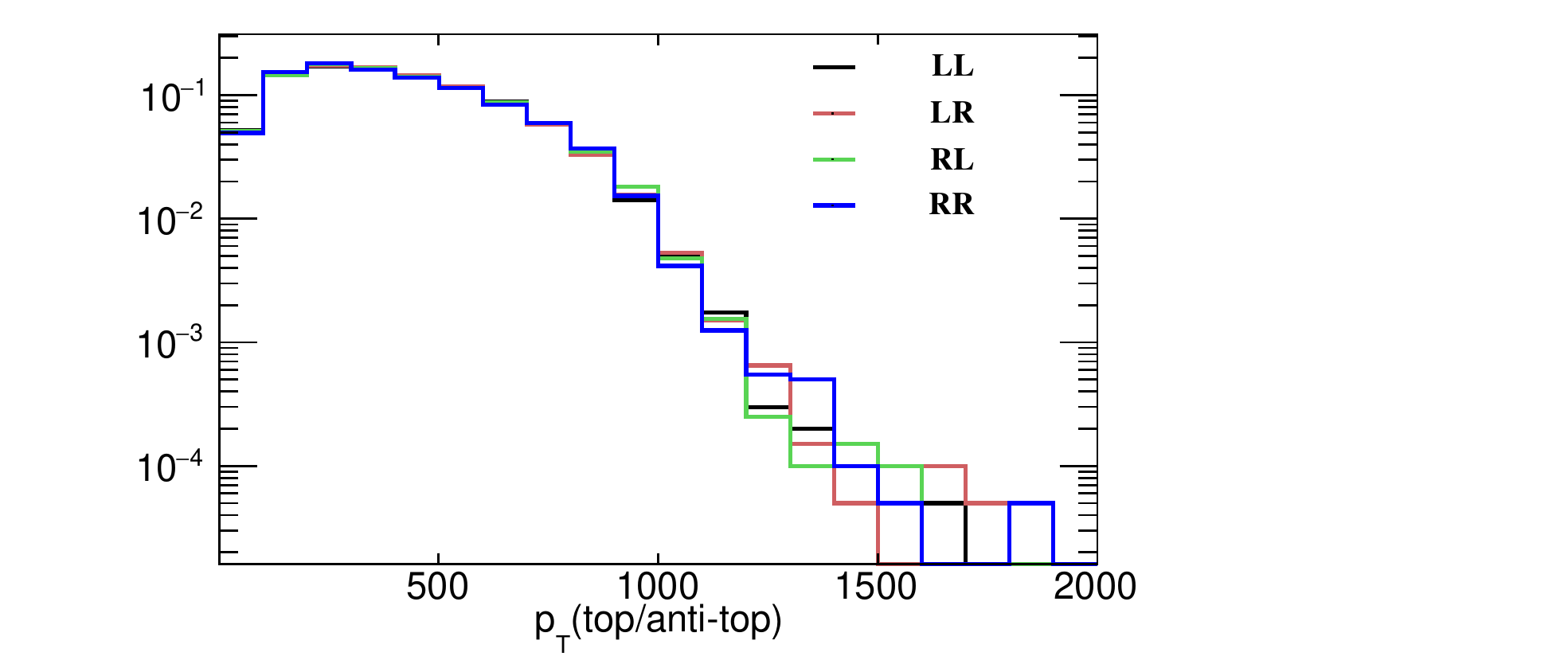}\\
  \caption{\label{fig:dist_tops} 
  Differential distributions (normalised to one) for different chiral choices of $\lambda_{Z' q\bar{q},L/R}$ (first letter) and $\lambda_{Z'T'_s\overline{T'_s},L/R}$ (second letter), when top partner production only via $Z'$ bosons is considered. The produced top partners decay to top quarks and dark matter. LL is shown in black,  LR in  red, RL in green, and RR in blue. The BSM particle masses are chosen as $M_{Z'}= 3$~TeV, $M_{T'_s}= 1$~TeV and $m_{\phi}= 500$~GeV.}
\end{figure}

\begin{figure}[htbp]
  \includegraphics[width=0.65\textwidth]{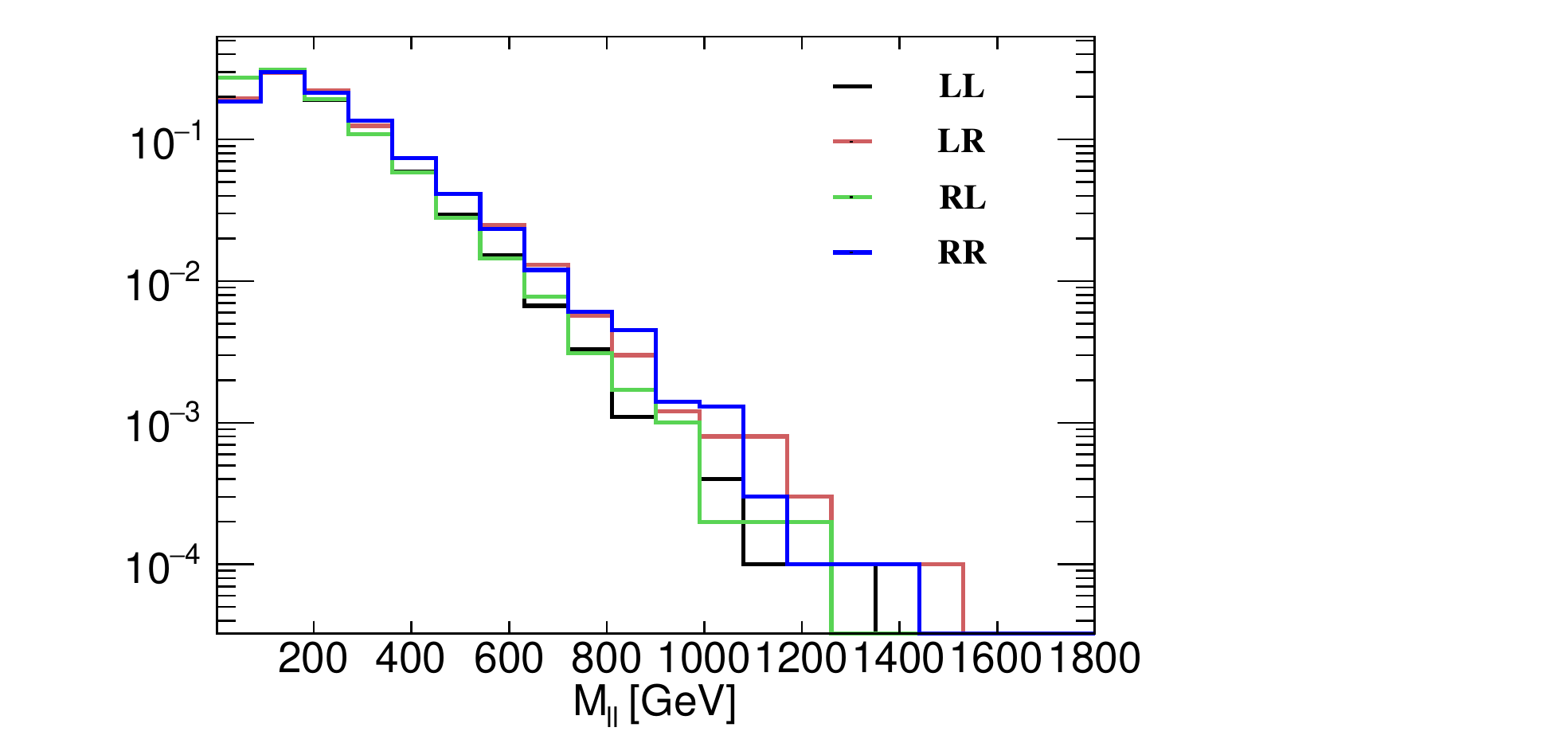}
  \hspace{-2 cm}
  \includegraphics[width=0.65\textwidth]{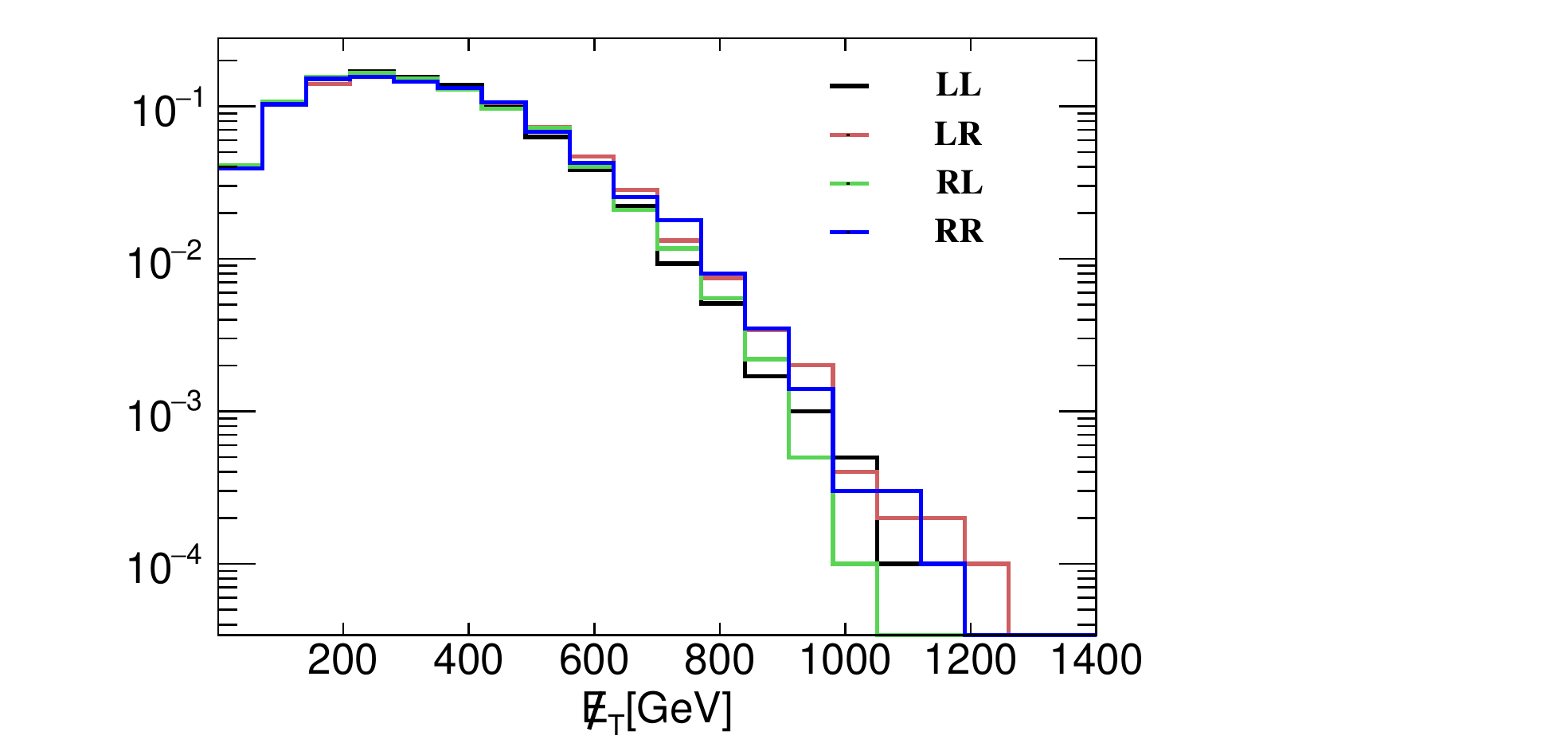}\\
  \includegraphics[width=0.65\textwidth]{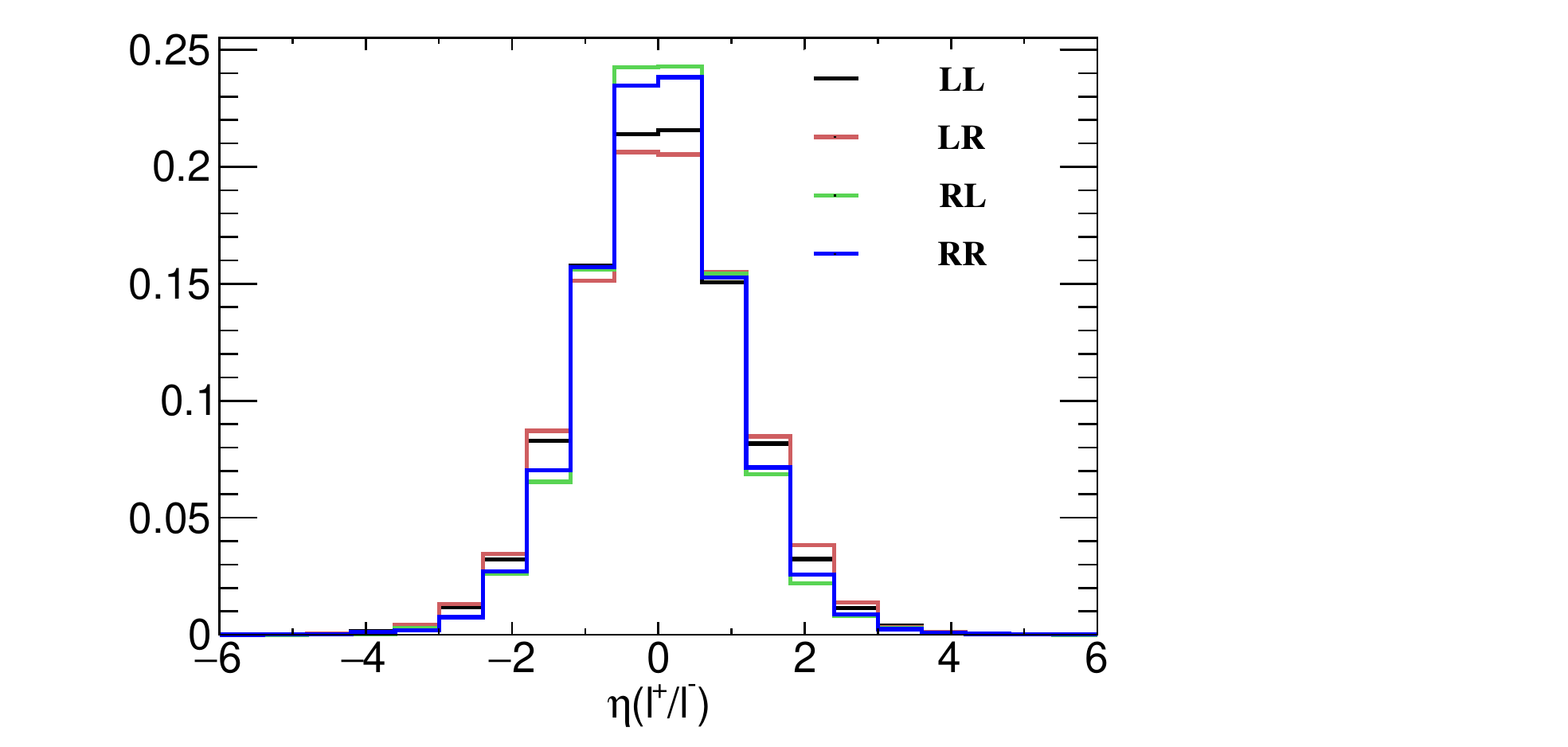}%
  \hspace{-2 cm}
  \includegraphics[width=0.65\textwidth]{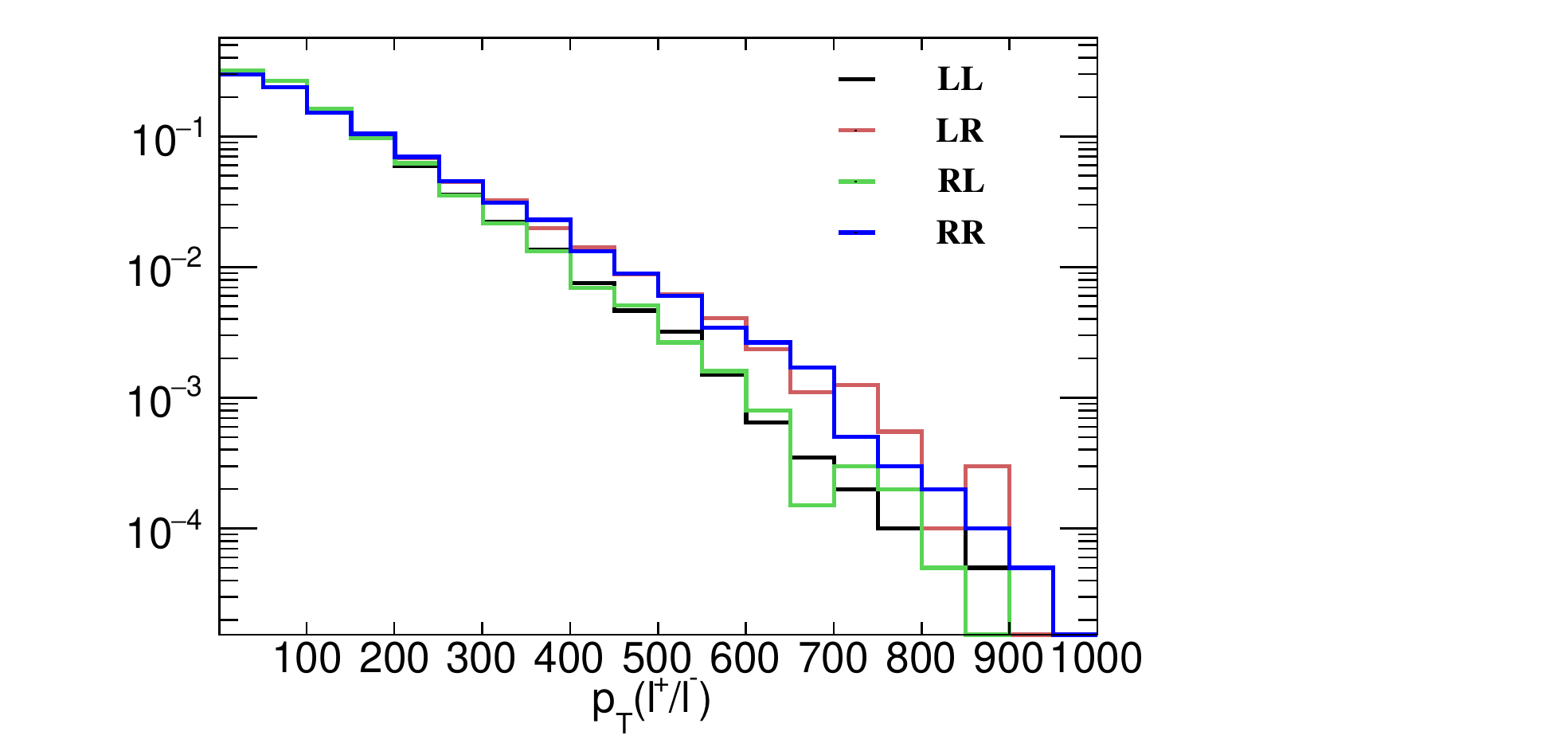}\\
  \caption{\label{fig:dist_leptons} Differential distributions (normalised to one) for different chiral choices of $\lambda_{Z' q\bar{q},L/R}$ (first letter) and $\lambda_{Z' T'_s \overline{T'_s},L/R}$ (second letter) when top partner production only via $Z'$ bosons is considered. The produced top partners decay to tops and dark matter, with the  top quarks decaying further into $b\;W_{\rm{lep}}$. LL is shown in black,  LR in  red, RL in green, and   RR in blue.  The BSM particle masses are chosen as $M_{Z'}= 3$~TeV, $M_{T'_s}= 1$~TeV and $m_{\phi}= 500$~GeV. }
\end{figure}

In order to understand the parameter space and the effect from different $Z'$ coupling combinations on the kinematical properties of the signature under study, we explore several parton level distributions  shown in figures~\ref{fig:dist_tops} and \ref{fig:dist_leptons}. These distributions have been obtained using the \texttt{MadGraph5\_aMC\@NLO 2.3.3} framework in conjunction with \texttt{MadAnalysis}. 
We present results for  the four different chiral combinations: ``LL'', ``LR'', ``RL'', ``RR'', where the first letter indicates the chirality of $\lambda_{Z' q \bar{q}}$  and the second letter indicates the chirality of $\lambda_{Z'T'_s\overline{T'_s}}$. The couplings themselves are fixed to $\lambda_{Z'q\bar{q}} = 0.25$ and $\lambda_{Z'T'_s\overline{T'_s}} = 2.5$, while the different masses are set to $M_{Z'} = 3$ TeV, $M_{T'_s} = 1$ TeV and $m_\phi = 500$ GeV. All events are generated for the process where $T'_s$ are pair-produced via $Z'$ bosons only. From figure~\ref{fig:dist_tops}~(top panel), one can see that the $\MET$ spectrum (evaluated from DM momentum only) and the invariant mass distributions of the di-tops barely depend on the choice of a chiral combination. Minor deviations only occur in the high energy tails. In section~\ref{sec:DBA} below, we show that the difference between the cutflow efficiencies for the two extreme cases ``LL'' and ``RR'' is at the level of 1--2 \%, which quantitatively proves our point to choose just one chiral combination and significantly reduce the model parameter space.

The pseudorapidity ($\eta$) distributions of the top and anti-top quark are very similar for the LL and LR chiral combinations.\footnote{To be precise, we plot the $\eta$ distribution of the top quark added to the $\eta$ distribution of the anti-top.} Likewise, the top/anti-top pseudorapidity distributions (figure~\ref{fig:dist_tops}, bottom left) for the RL and RR combination are close to each other, but slightly wider compared to the LL and LR distributions. The transverse momentum for top and anti-top quarks changes marginally in the high $p_T$ tail (for $p_{T} \gtrsim 1$ TeV), when the chirality of $\lambda_{Z'q\bar{q}}$ changes.   

The above (not directly observable) distributions define kinematics of the top quark decay products, which we present next in figure~\ref{fig:dist_leptons}.
For a leptonic $W$ decay, one can see that the RR and LR combinations have the same $p_T$ distributions of the leptons. The same is true for the LL and RL combinations. For RR and LR, however, this distribution has a slightly higher tail (slightly harder) in comparison to LL and RL. This behavior has been observed previously (see Ref.~\cite{Kraml:2016eti}) and occurs due to the influence of the top polarisation on the $p_{T}$ of the decay products. This difference in lepton $p_T$ distributions occurs for high values of the lepton $p_T$ and does not visibly affect the efficiency of the cuts for the signature under study. The same holds for the $M_{\ell\ell}$, $\eta_\ell$ and $\MET$ distributions. One can note the slight difference in $\eta_\ell$ is correlated with the slight difference in $\eta_t$ for different chiral combinations.  Also, the $\MET$ shape before and after a top quark decays are very similar.  For simplicity, we work with the case where both $\lambda_{Z'q\bar{q}}$ and $\lambda_{Z'T'_s\overline{T'_s}}$ are left-handed (LL). This choice yields a (marginally) softer lepton $p_T$ and therefore slightly lower cut-efficiencies compared to LR and RR, making the LL configuration a conservative choice.

\begin{figure}[htbp]
  \includegraphics[width=0.55\textwidth]{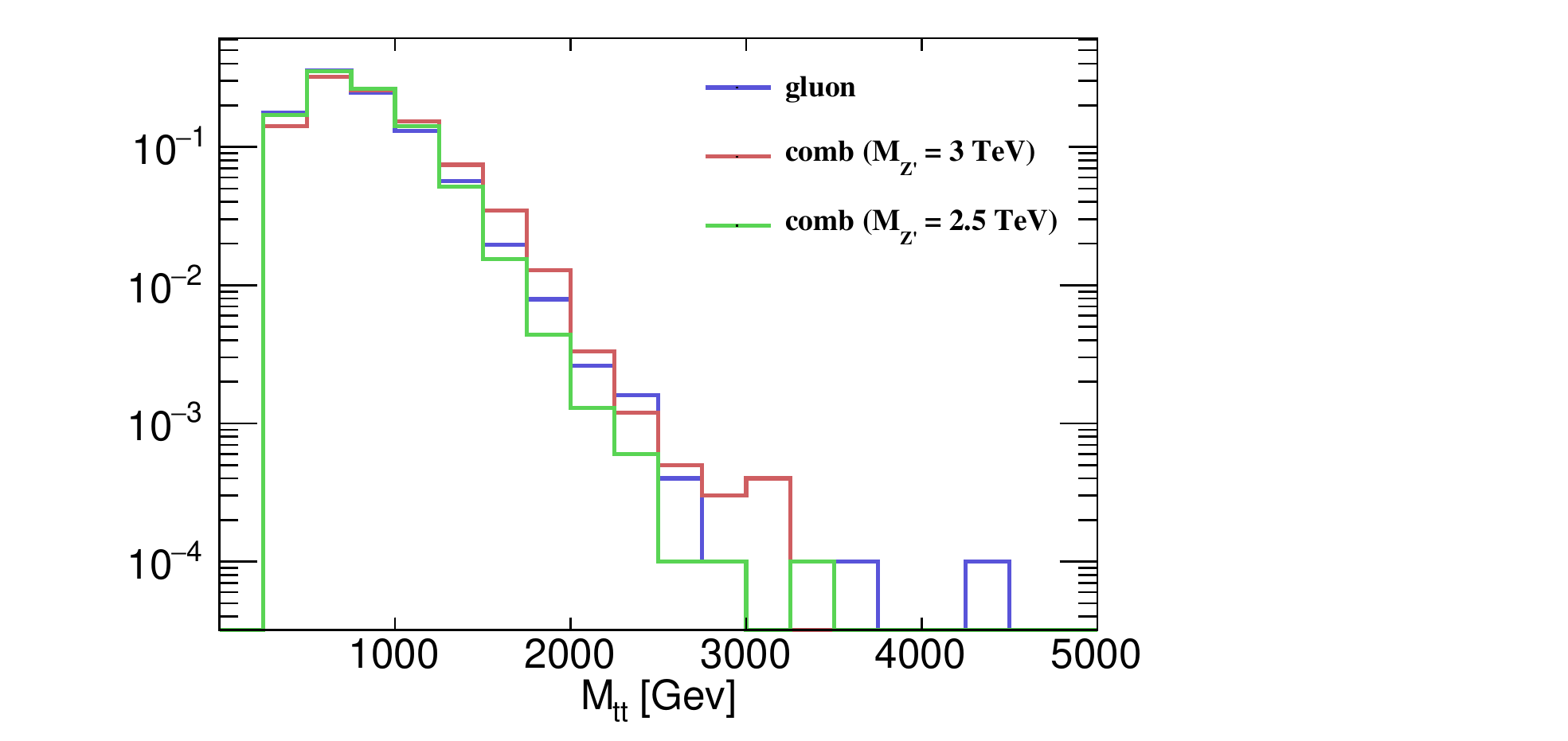}%
  \includegraphics[width=0.55\textwidth]{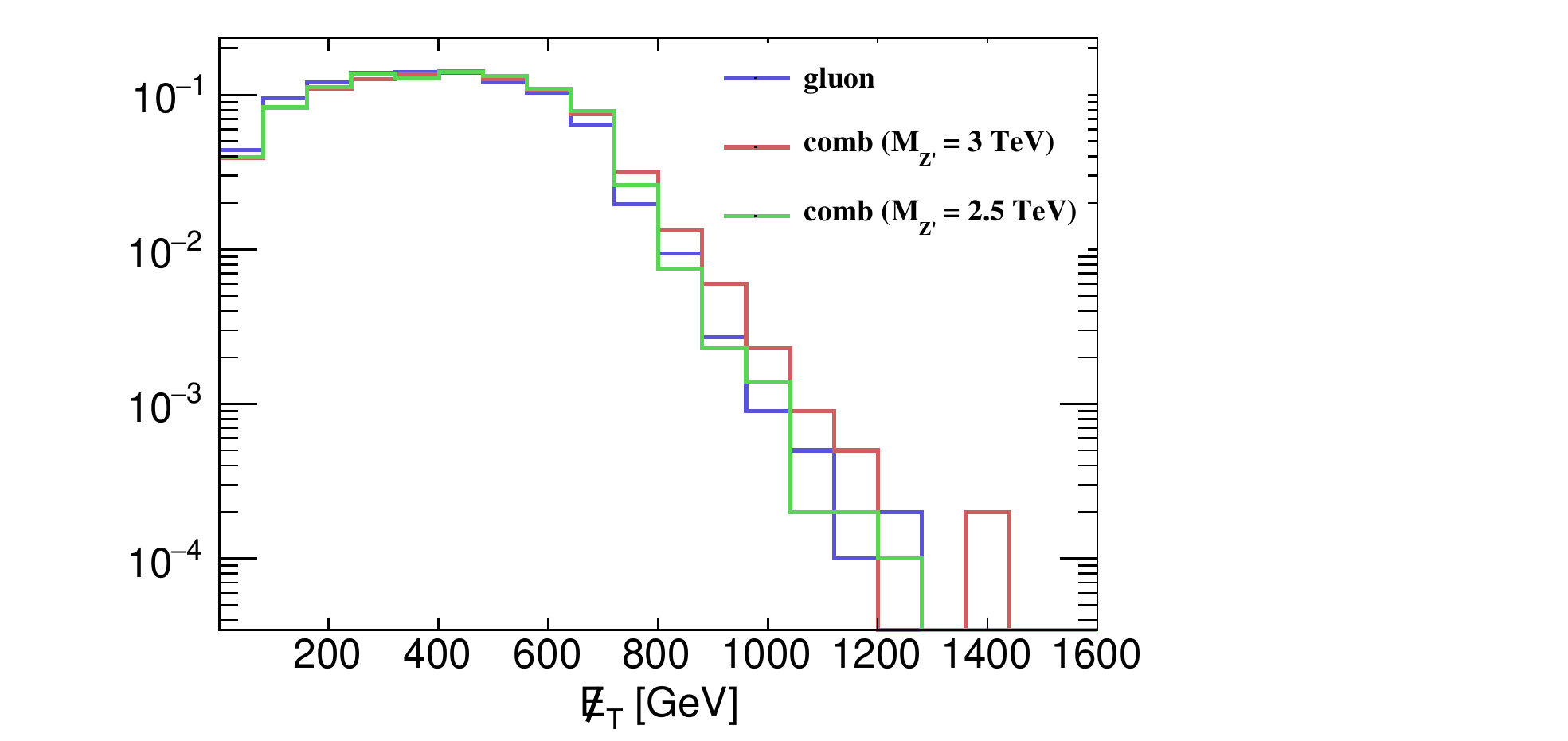}\\
  \includegraphics[width=0.55\textwidth]{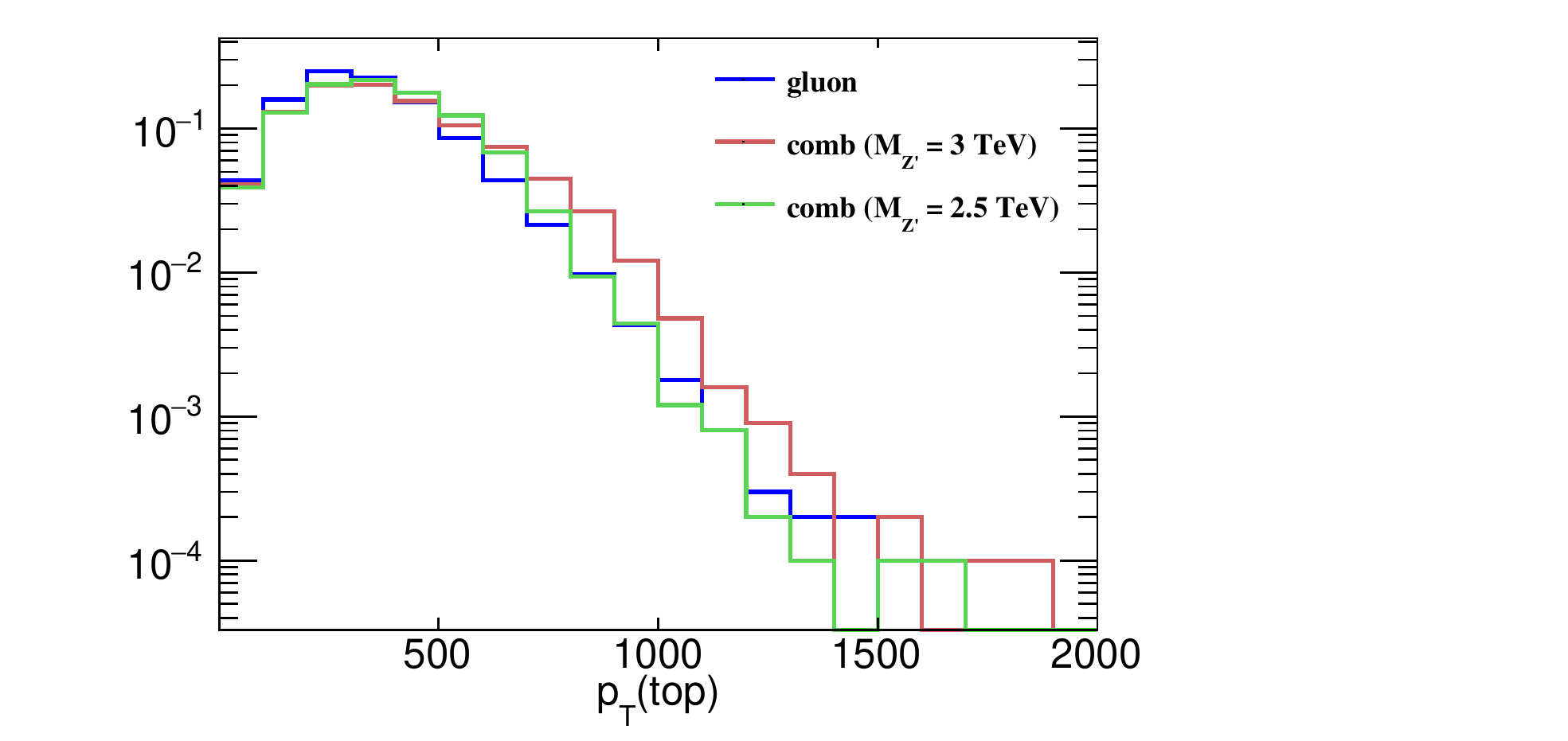}%
  \includegraphics[width=0.55\textwidth]{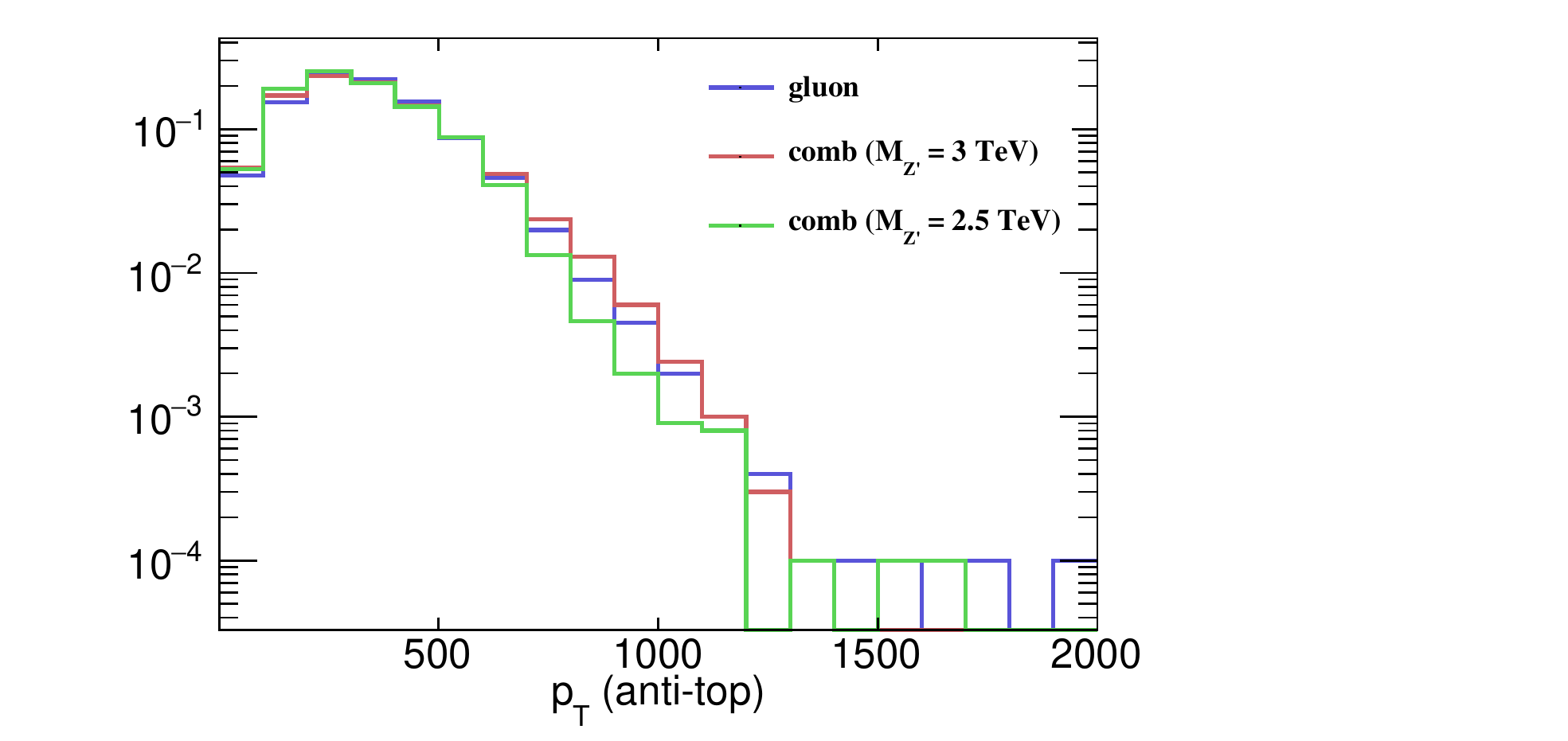}\\
  \includegraphics[width=0.55\textwidth]{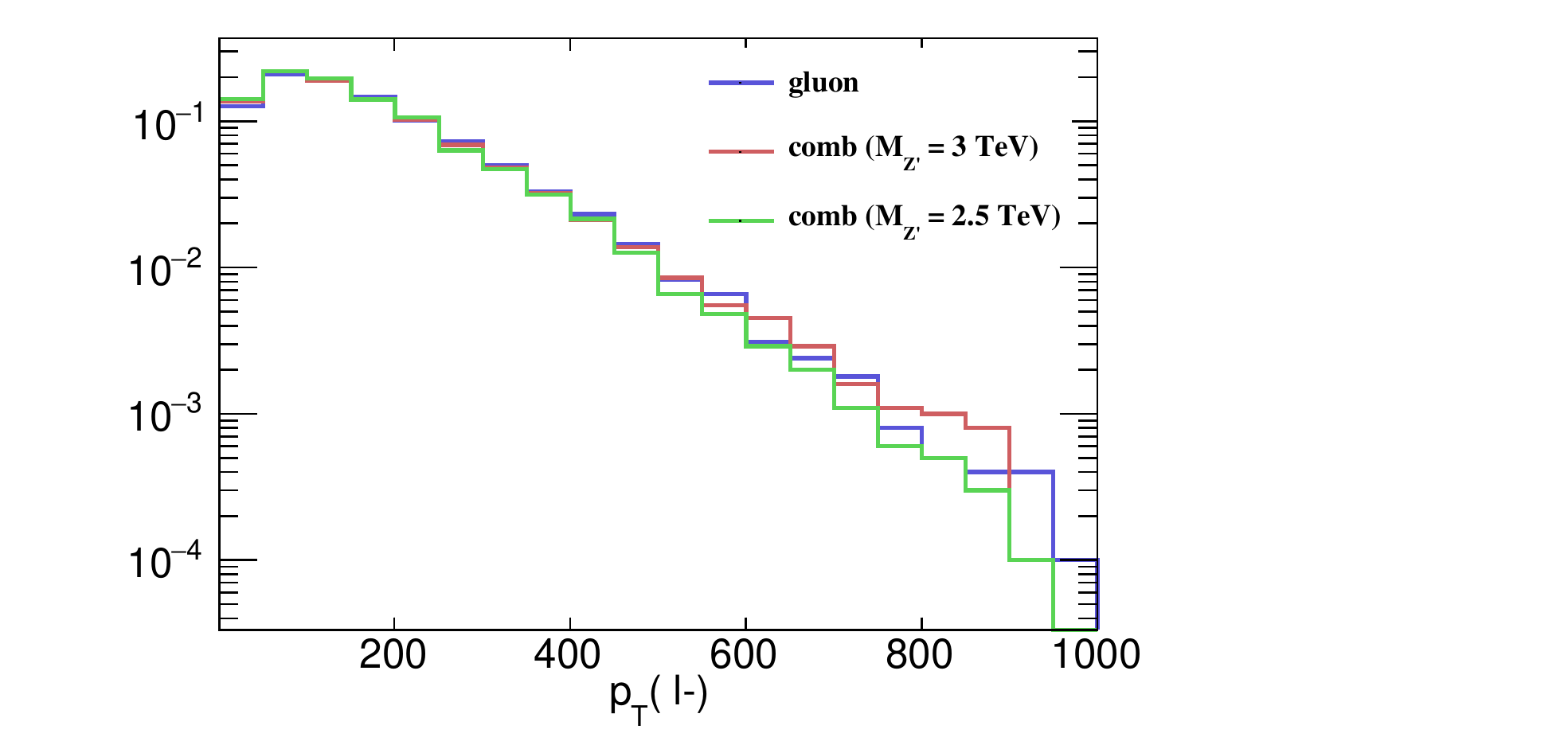}%
   \includegraphics[width=0.55\textwidth]{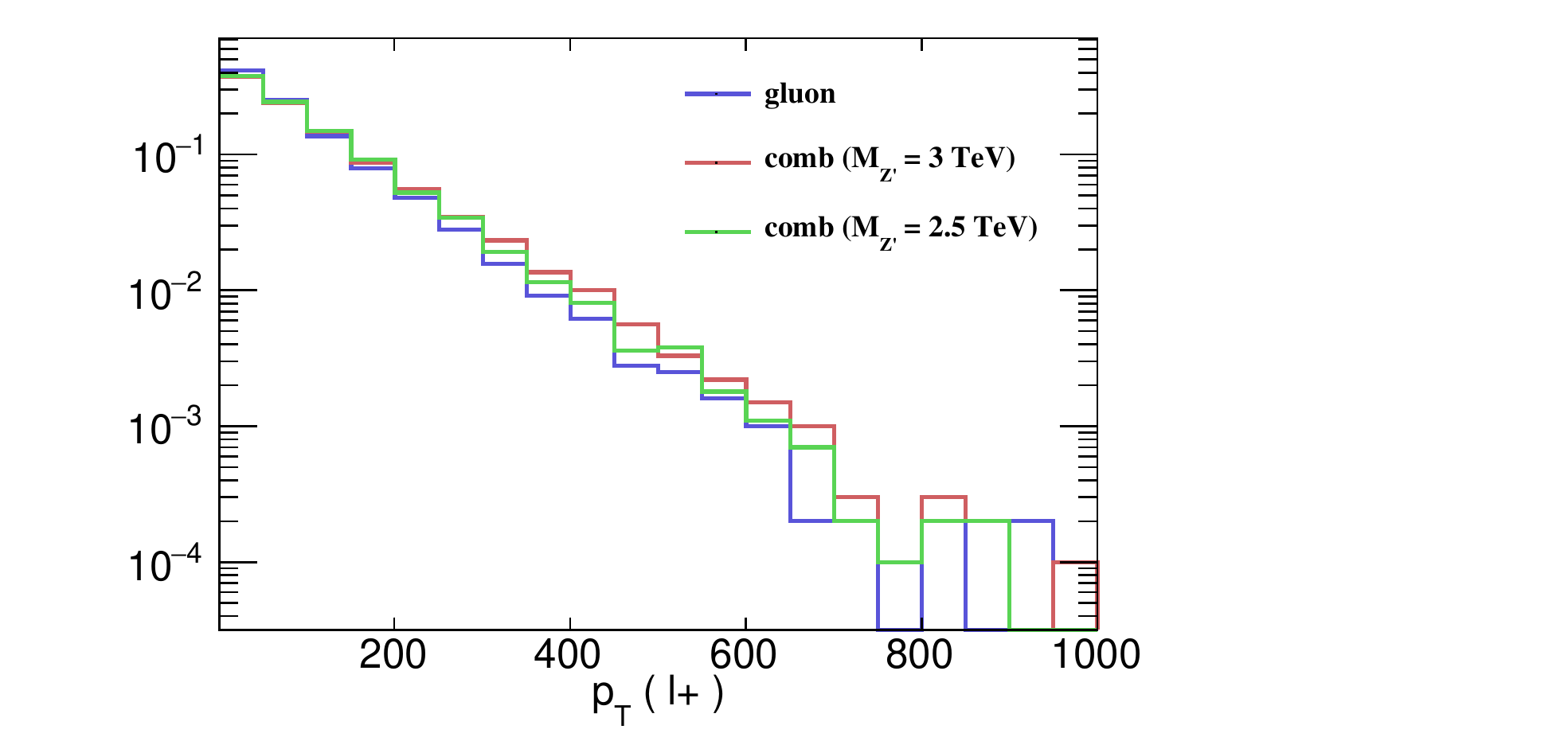}\\
  \includegraphics[width=0.55\textwidth]{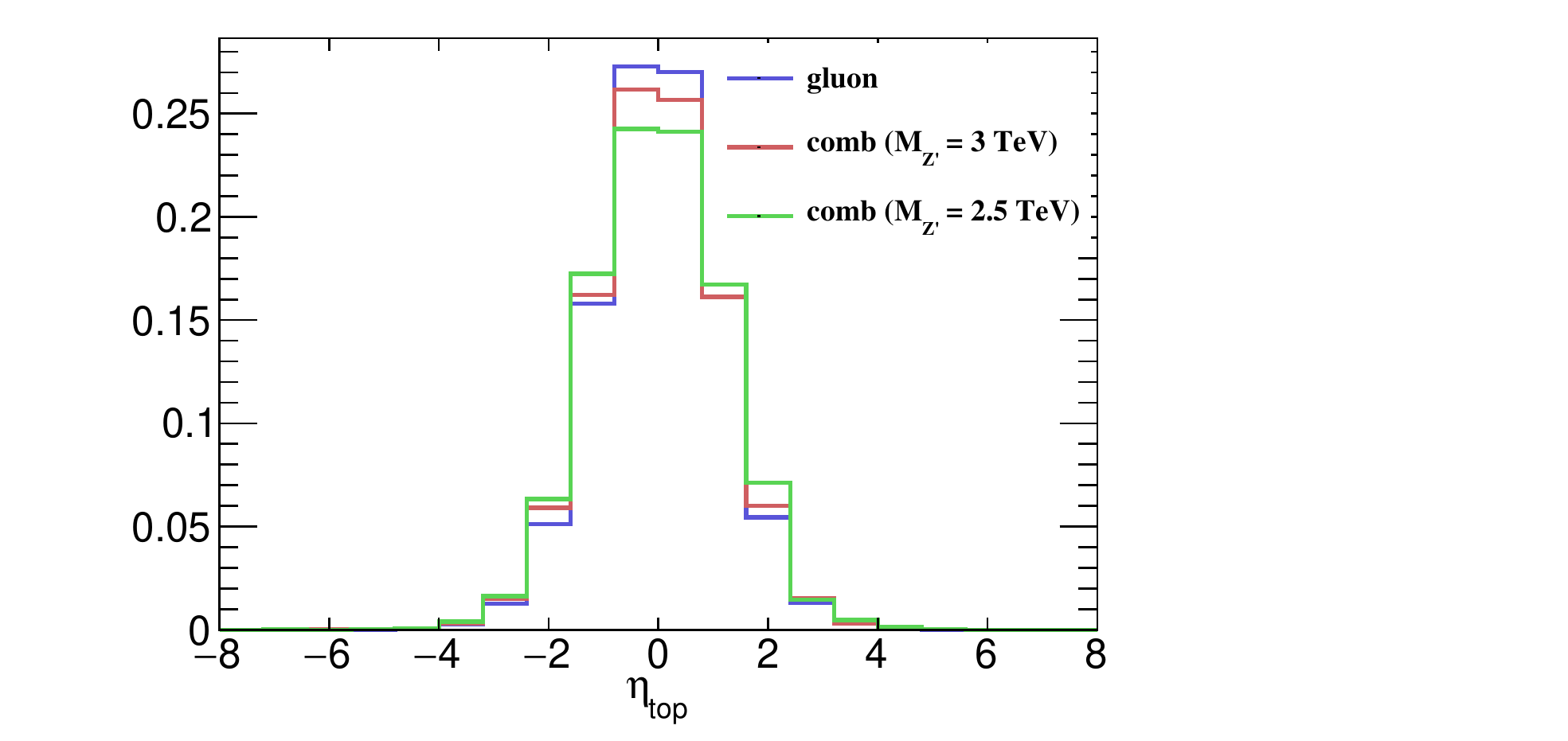}%
  \includegraphics[width=0.55\textwidth]{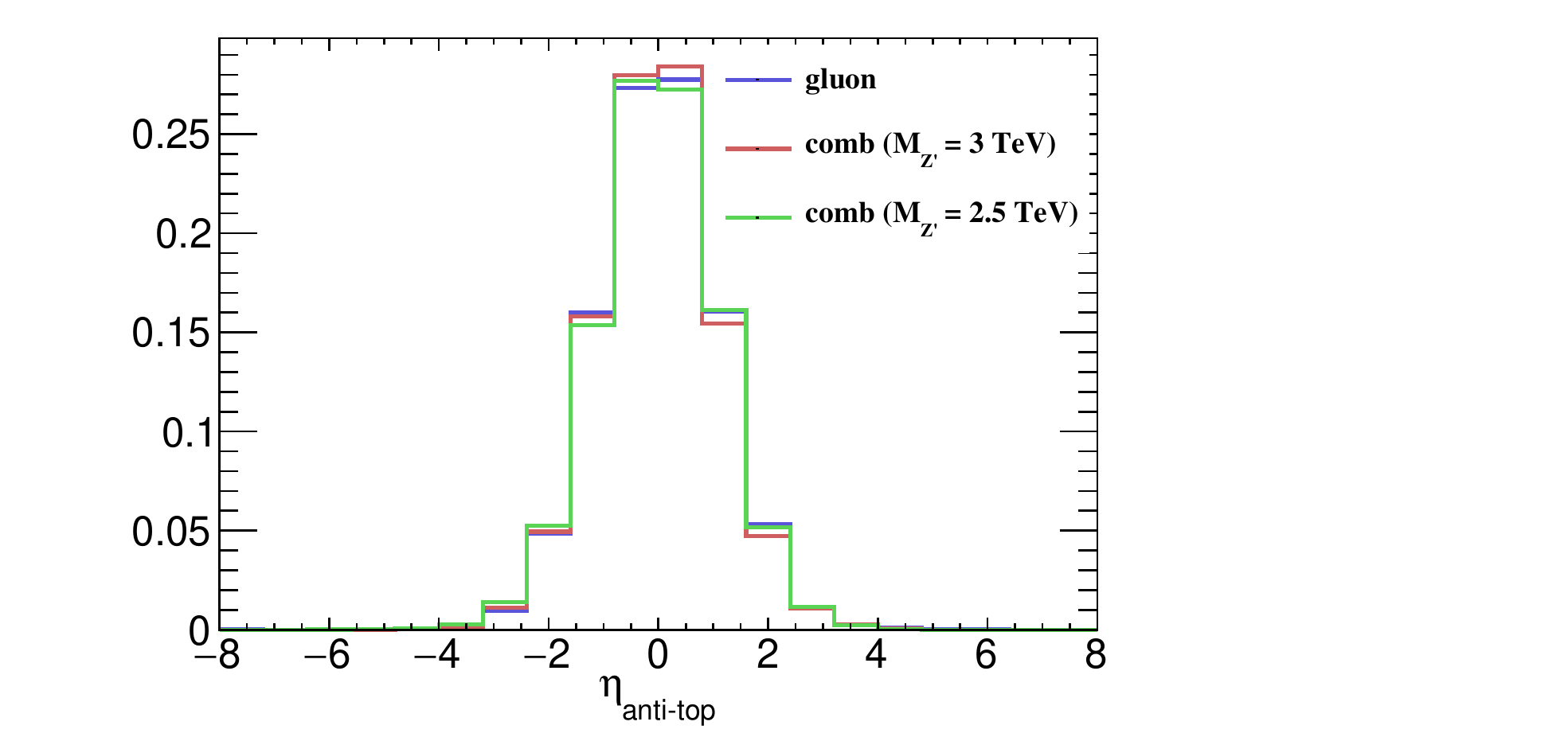}\\
  \caption{\label{fig:dist_int} Differential distributions (normalised to one) specifically for the chiral choice LL, when top partner-pair production occurs via QCD and for the combined production, i.e QCD$+Z'$, for $M_{Z'} = 2.5$ and 3 TeV. The produced top partners are allowed to decay to top quarks and dark matter. Distributions from QCD-only production are shown in green, the combined production for $M_{Z'} = 2.5$~TeV in blue and for $M_{Z'} = 3 $~TeV in red. Here, $\lambda_{Z' T'_s \overline{T'_s}_{,L}} = 2.5$, $\lambda_{Z' q \bar{q},L } = 0.25$, $M_{T'_s}= 1$~TeV and $m_\phi = 500$~GeV.}
\end{figure}
It is also instructive to explore the difference in kinematics between the setup where top partners are produced via gluons and for the combined production involving both $Z'$ bosons and gluon mediation. This difference is directly related  to the main point of our paper --- the role of the $Z'$ boson in exploring the $t\bar{t}+\MET$ signature at the LHC. In figure~\ref{fig:dist_int}, we present kinematical distributions for two points with $M_{Z'} = 2.5$ TeV and 3 TeV with $\lambda_{Z'q\bar{q}}$ and $\lambda_{Z'T'_s\overline{T'_s}}$ taken to be left-handed and kept at $\lambda_{Z'q\bar{q}} = 0.25$ and $\lambda_{Z' T'_s \bar{T'_s}} = 2.5$. One can see that the lepton $p_T$ distributions are similar for gluon exchange alone and $g+Z'$ exchange for both $Z'$ masses. However, one can also observe that the transverse momentum of the $\ell^+$ as well as $\MET$ are systematically harder in the $g+Z'$ exchange case, especially for larger $p_T$ or $\MET$. As we will see later from the fast detector simulation studies, these differences lead to a non-negligible difference of the final selection efficiencies, which are higher for $g+Z'$ exchange in comparison to just gluon exchange alone.

\subsection{Pre-study II: Interference Effects}
\label{sec:interference-effects}

Besides $T'_s$-pair production via $Z'$ bosons, $T'_s$-pairs are also produced through QCD interactions, as shown in figure \ref{fig:Zp-prod-Feynman-diagram}. To quantify possible interference effects, we study the three cases where $T'_s$-pairs are produced only via $Z'$ bosons, only via gluons, and for the combined production involving both $Z'$ bosons and gluons using \texttt{MadGraph5 2.3.3}. Additionally, we try to maximise the interference effects by choosing the couplings and masses such that the $Z'$-mediated and gluon-only cross sections are nearly identical. Because the $Z'$ width can also affect interference, we study two parameter points: one with a narrow $Z'$ width ($M_{Z'} = 2.5$ TeV, $M_{T'_s} = 1$ TeV, $\lambda_{Z'q\bar{q}} = 0.3$, $\lambda_{Z'T'_s\bar{T'_s}} = 0.58$) and one with a very broad $Z'$ width ($M_{Z'} = 2.5$ TeV, $M_{T'_s} = 1$ TeV, $\lambda_{Z'q\bar{q}} = 0.21$, $\lambda_{Z'T'_s\bar{T'_s}} = 4.7$). The results are summarised in table~\ref{tab:interferences}.
\begin{table}[ht]
  \centering
  \begin{tabular}{crcc}
    \toprule
    $\Gamma_{Z'}$ [GeV]   & production channel & $\sigma$ [fb] & difference [fb] (rel. difference [\%])  \\
    \midrule
                          & $Z'$               & 30.9          &                  \\ \cline{2-3}
    70.5                  & QCD                & 32.4          & +1.6 (+2.5 \%)   \\ \cline{2-3}
                          & combined           & 64.9          &                  \\
    \midrule
                          & $Z'$               & 31.3          &                  \\ \cline{2-3}
    1134                  & QCD                & 32.5          & +2.4 (+3.6 \%)   \\ \cline{2-3}
                          & combined           & 66.2          &                  \\
    \bottomrule
  \end{tabular}
  \caption{Cross sections for $T'_s$ pair production for different production channels with $M_{Z'} = 2.5$ TeV. The difference is computed as ``combined - ($Z'$ + gluon)'' and the relative difference as ``$1 - \frac{Z' + \text{gluon}}{\text{combined}}$''.}
  \label{tab:interferences}
\end{table}

We find the interference to be constructive and ranging from +2.5 \% gain in cross section in the narrow width scenario to +3.6 \% in the broad width scenario.

\subsection{Pre-study III: Narrow Width Approximation and Corrections}
\label{sec:NWA-validity}

\begin{figure}[ht] 
  \centering
  \includegraphics[width=0.77\linewidth]{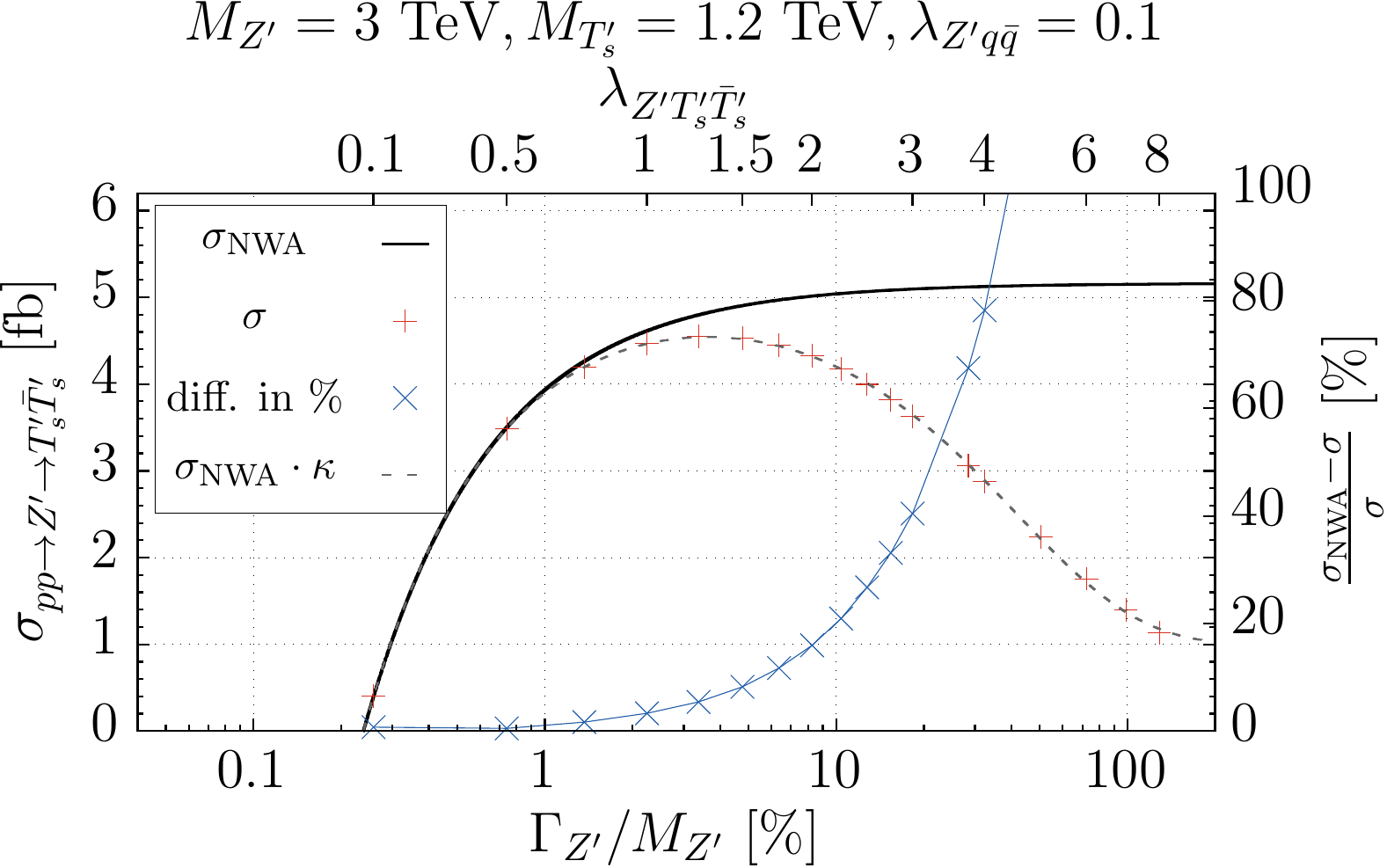}
  \caption{Comparison of $pp \to Z' \to T'_s\overline{T'_s}$ in the NWA (black line) and without (red '+') for $M_{Z'} = 3$ TeV. The blue crosses show the difference in \% between the two based on $\sigma$. The blue curve shows the fitting function eq.~(\ref{eq:NWA-fit-def-val}) with fit parameters given in eq.~(\ref{eq:NWA-fit-def-val-coeff}).}
  \label{fig:NWA-check}
\end{figure}

The narrow width approximation (NWA) enables us to easily estimate and scale cross sections for variable model parameters, such as couplings and masses. However, as the NWA becomes less accurate with an increasing width of the decaying particle, it is important to study and estimate corrections to it. In figure~\ref{fig:NWA-check}, we show the cross sections for $pp \to Z' \to T'_s\overline{T'_s}$\footnote{For simplicity, we do not consider finite width effects arising from the $T'_s$ decays.} in the NWA (black line) and computed with \texttt{CalcHEP} (red '+') for $M_{Z'} = 3$ TeV, $M_{T'_s} = 1.2$ TeV, $\lambda_{Z'q\bar{q}} = 0.1$, $\lambda_{Z'\ell^+\ell^-} = 0$ and varying $\lambda_{Z'T'_s\overline{T'_s}}$. The relative  difference between the two (normalised to the \texttt{CalcHEP} result) is shown as blue crosses.
For very small $\Gamma_{Z'}/M_{Z'}$ up to approximately 1 \%, the NWA perfectly estimates the cross section without any approximation $\sigma$. Increasing $\Gamma_{Z'}/M_{Z'}$ to $\approx5 \%$ yields a difference in cross section of roughly 10 \%. Further increasing $\Gamma_{Z'}/M_{Z'}$ leads to relative differences of 40 \% or more, therefore making the NWA very inaccurate in that region. Also note that the NWA for this particular choice of $M_{Z'}$ and $M_{T'_s}$ always overestimates the actual cross section, which would lead to over-optimistic limits and constraints. To correct for finite width effects, we use our simulation results in figure~\ref{fig:NWA-check} to define a correction factor $\kappa\(\lambda^2_{Z'T'_s\overline{T'_s}}\)$ as 
\beq
   \kappa\(\lambda^2_{Z'T'_s\overline{T'_s}}\) \equiv \frac{\sigma}{\sigma_{\text{NWA}}}.
  \label{eq:NWA-fit-kappa}
\eeq
With the ansatz function
\beq
  \kappa\(\lambda^2_{Z'T'_s\overline{T'_s}}\) = c_0 + c_1 \cdot \text{exp}\(-c_2 \, \lambda^2_{Z'T'_s\overline{T'_s}}\)
  \label{eq:NWA-fit-def-val}
\eeq
we obtain
\beq
  c_0 = 0.193(4) \,, \quad c_1 = 0.812(4) \,, \quad c_2 = 0.049(1)
  \label{eq:NWA-fit-def-val-coeff}
\eeq 
for $M_{Z'} = 3$ TeV, $M_{T'_s} = 1.2$ TeV (cf. the grey dashed curve in figure~\ref{fig:NWA-check} for the quality of the fit). For different mass choices, these fitting coefficients will vary due to cutoff effects appearing for large $\Gamma_{Z'}$ and due to PDF effects. For a fixed mass pair, however, $\kappa\(\lambda^2_{Z'T'_s\overline{T'_s}}\)$ can be used universally in the $\(\lambda_{Z'T'_s\overline{T'_s}}, \lambda_{Z'q\bar{q}}\)$ coupling space. Therefore, whenever the NWA is used in the following chapters, we rescale
\beq
  \sigma_{\text{NWA}} \mapsto \sigma_{\text{NWA}} \cdot \kappa_{(M_{Z'},M_{T'_s})}\(\lambda^2_{Z'T'_s\overline{T'_s}}\) \,,
  \label{eq:NWA-rescaling}
\eeq
with the fitting function $\kappa_{(M_{Z'},M_{T'_s})}\(\lambda^2_{Z'T'_s\overline{T'_s}}\)$, determined as shown above.

\subsection{\texorpdfstring{QCD-only $T'_s$ Pair Production}{QCD-only T' Pair Production}}
\label{sec:QCD-Tp}

With the results of our pre-studies in place, we turn to constraints on the simplified model from current LHC searches at 13 TeV. For each parameter point (with the parameter grid to be defined below), we simulate 50000 events which are used to compare against \texttt{CheckMATE} implemented  ATLAS and CMS searches \cite{Aad:2016tuk,Aaboud:2016uro,Aaboud:2016tnv,Aaboud:2016zdn,Aad:2016qqk,Aad:2016eki,Aaboud:2016lwz,TheATLAScollaboration:2015nxu,TheATLAScollaboration:2016gxs,ATLAS:2016xcm,ATLAS:2016ljb,CMS:2015bsf}.\footnote{These searches focus mainly on SUSY-like final states with $\MET$ and are thus expected to yield relevant bounds on the $t\bar{t}\, \phi\phi$ (with $\phi$ representing $\MET$) final state which we investigate.}
 
\begin{figure}[ht]
  \centering
  \includegraphics[width=0.9\textwidth]{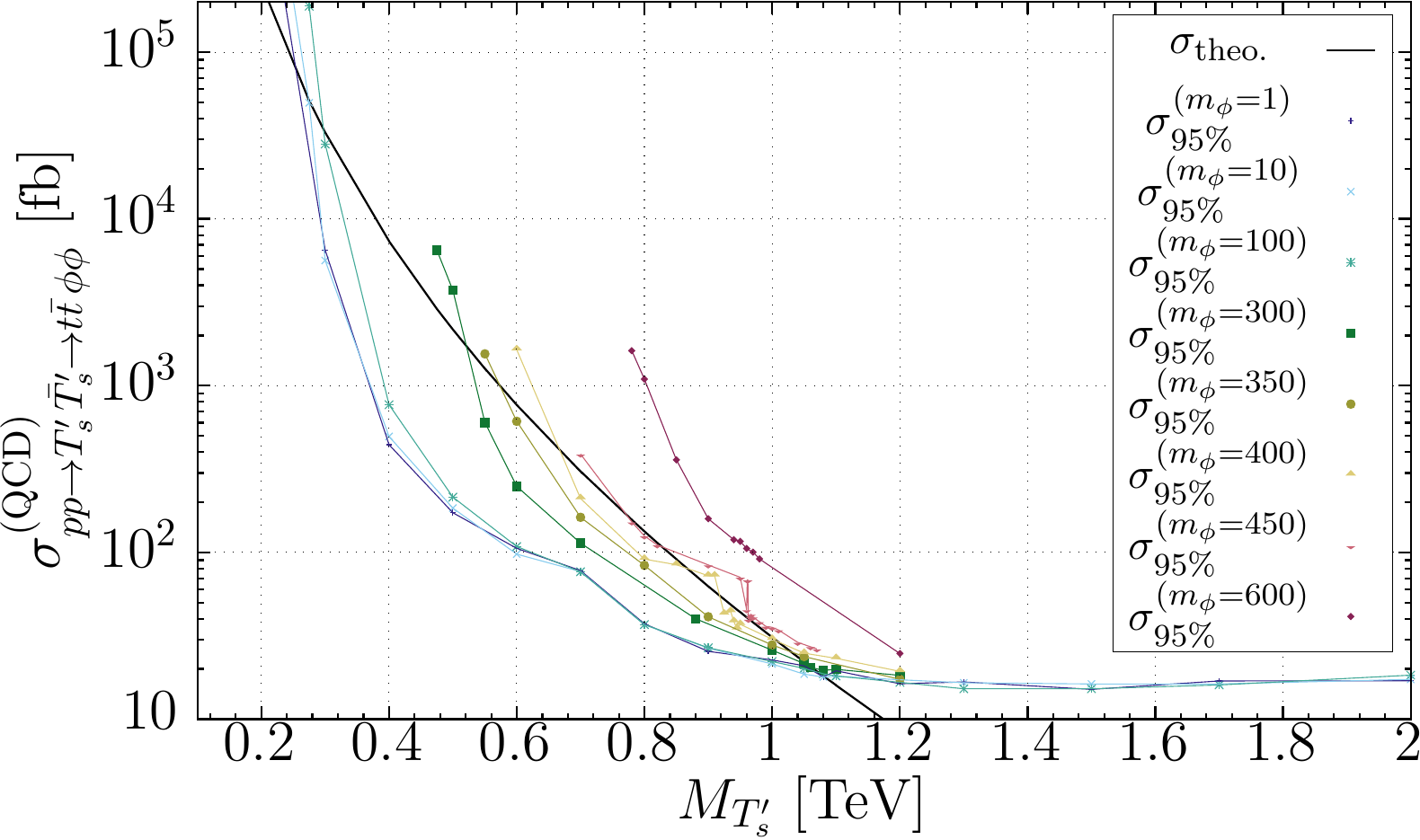}%
  \caption{Theoretical (black) and experimental (coloured) cross sections for $pp \to T'_s\overline{T'_s} \to t \bar{t} \, \phi \phi$ in fb without $Z'$ mediation in dependence of $M_{T'_s}$ and $m_\phi$. $m_{\phi}$ is given in GeV.}
  \label{fig:crossx-1D-gluon}
\end{figure}

We first determine the limits for QCD-only $T'_s$-pair production (see figure~\ref{fig:Zp-prod-Feynman-diagram}, centre and right), which only depend on $M_{T'_s}$ and $m_\phi$, since the kinematics and the rate for the process are completely fixed by these two masses. In figure~\ref{fig:crossx-1D-gluon}, we show the production cross section (simulated at leading order) for the QCD-only $T'_s$-pair production as a function of $M_{T'_s}$ together with the experimental limits for  different dark matter masses $m_\phi$. Using \texttt{CheckMATE}, we have found the strongest observed bound at 95 \% confidence level out of all implemented analyses with the $r$-value given as \cite{Dercks:2016npn}
\beq
  r = \frac{S - 1.64 \cdot \Delta S}{S95}\,, \notag
  \label{eq:r-value}
\eeq
where $S$ is the number of predicted signal events with its uncertainty $\Delta S$ and $S95$ is the experimental 95 \% upper limit on the number of signal events.

It turns out that almost all limits are coming from the analysis \texttt{ATLAS\_CONF\_2016\_050} \cite{ATLAS:2016ljb}, a search aimed at top squarks in final states together with one isolated lepton, jets and $\slashed{E}_T$ at $\sqrt{s} = 13$ TeV. The most sensitive signal regions are \texttt{tN\_high} and \texttt{SR1}, where the first (latter) one is aimed at high (low) mass splittings between $\tilde{t}_1$ and $\tilde{\chi}^0_1$. A large mass split results in highly boosted top quarks, whereas a small mass split is responsible for all decay products to be fully resolved \cite{ATLAS:2016ljb}.
For a very small mass split between $M_{T'_s}$ and $m_\phi$ just above the top quark threshold, i.e. when $M_{T'_s} \gtrsim m_t + m_\phi$, the analysis \texttt{ATLAS\_1604\_07773} \cite{Aaboud:2016tnv} yields the best limits.
For heavy $T'_s$ between 1.2 and 2 TeV, the experimental limits are almost constant for any $m_\phi$, as the top quarks will now always be heavily boosted. The theoretically predicted signal, however, stays well below the exclusion range due to the high suppression of cross section coming from the heavy $T'_s$-pair. 
For $M_{T'_s} \approx 1.08$ TeV, the signal is excluded for all $m_\phi$ up to $\approx 300$ GeV and stays excluded up to a lower bound, where $M_{T'_s} \gtrsim m_t + m_\phi$. Reducing $M_{T'_s}$ further results in off-shell top quarks with highly different kinematical distributions not studied in this work.
Increasing $m_\phi$ beyond 300 GeV shrinks the excluded $M_{T'_s}$ region and eventually closes it for $m_\phi \gtrsim 450$ GeV, leaving all $M_{T'_s}$ allowed.

It is also worth noticing that for $m_\phi = 400$ and 450 GeV, there are small regions of non-excluded $M_{T'_s}$ in the otherwise excluded area (e.g. at $M_{T'_s} \approx 900$ GeV). These are not actually physical, but rather correspond to a gap in the regions of parameter space the signal regions \texttt{tN\_high} and \texttt{SR1} are able to cover. E.g. for $m_\phi = 400$ GeV and $M_{T'_s} \lesssim 920$ GeV, \texttt{ATLAS\_CONF\_2016\_050} \cite{ATLAS:2016ljb} is sensitive to the signal in \texttt{SR1}, whereas it is most sensitive in \texttt{tN\_high} for $M_{T'_s} \gtrsim 920$ GeV. As both signal regions do not overlap entirely, a gap in form of a kink or step is seen in the experimental cross section. 

With these $Z'$-independent constraints on $T'_s$-pair production, we now investigate the current experimental di-jet and di-lepton limits in the $\(\lambda_{Z'T'_s\overline{T'_s}}, \lambda_{Z'q\bar{q}}\)$ plane for $M_{T'_s} \geq 1.1$ TeV, which are generally safe for any $m_\phi$.

\subsection{Di-jet and Di-Lepton Constraints}
\label{sec:dijet-dilep}

\begin{figure}[ht] 
  \centering
  \includegraphics[scale=.45]{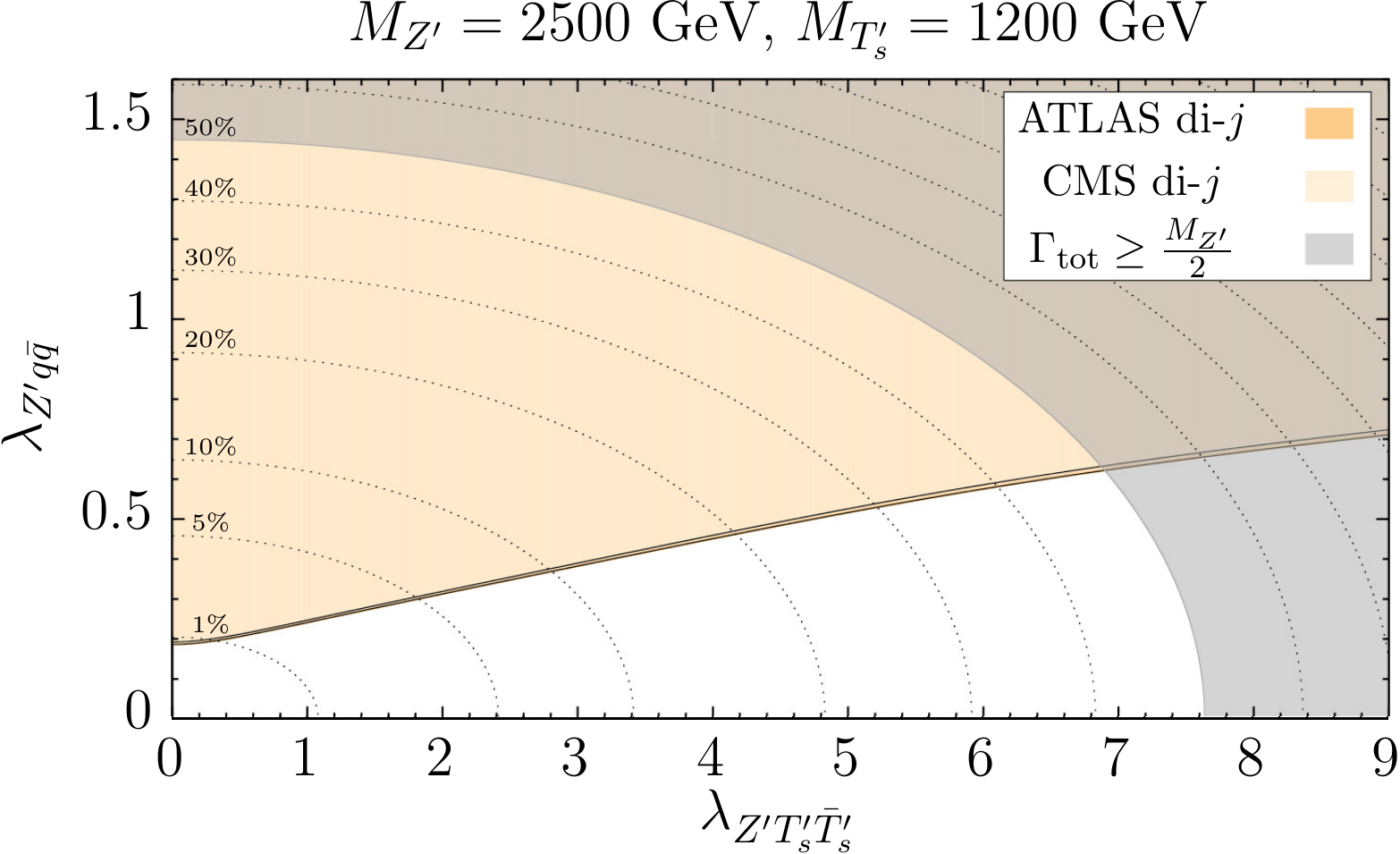}\hfill
  \includegraphics[scale=.45]{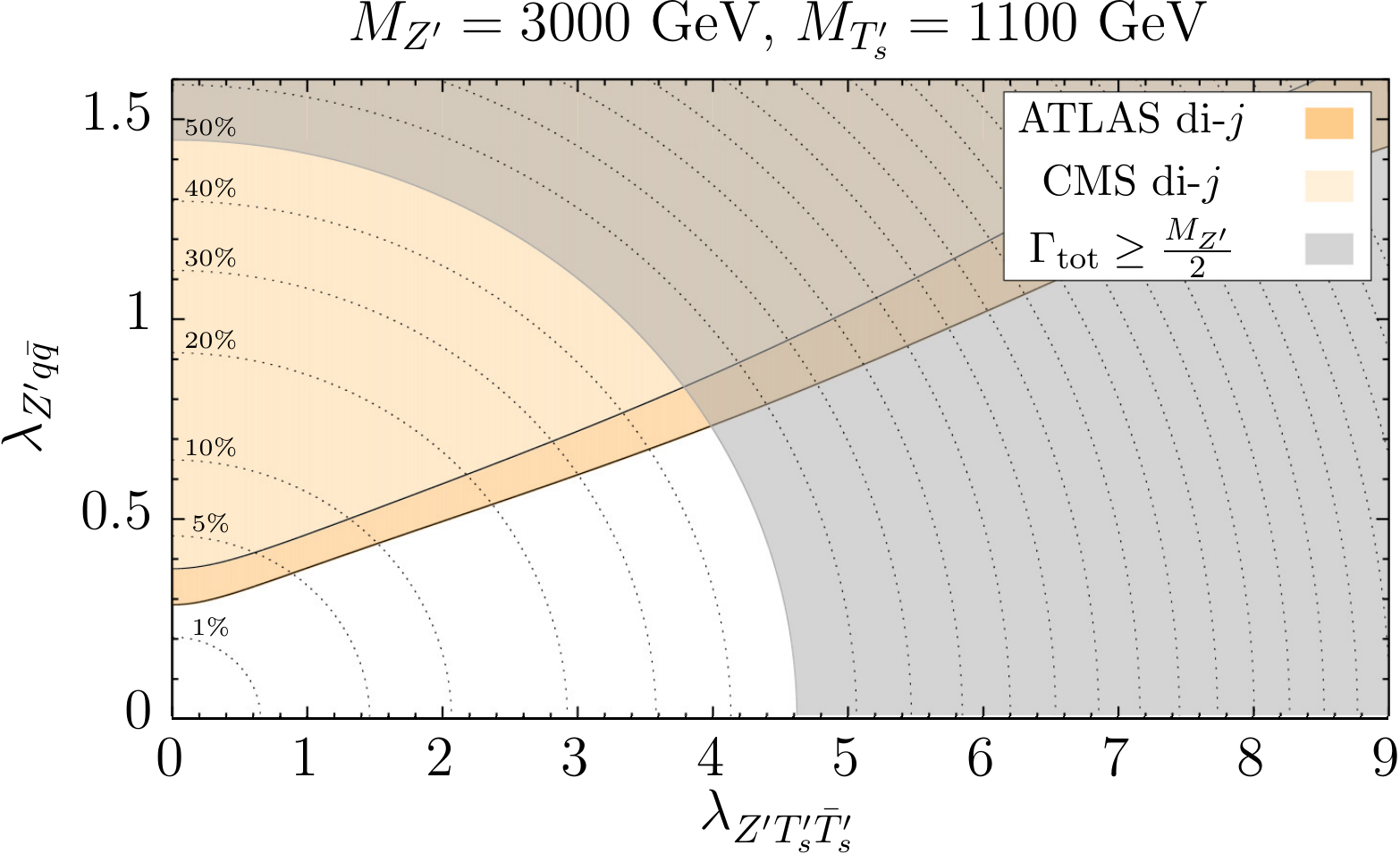}\\[0.2cm]
  \includegraphics[scale=.45]{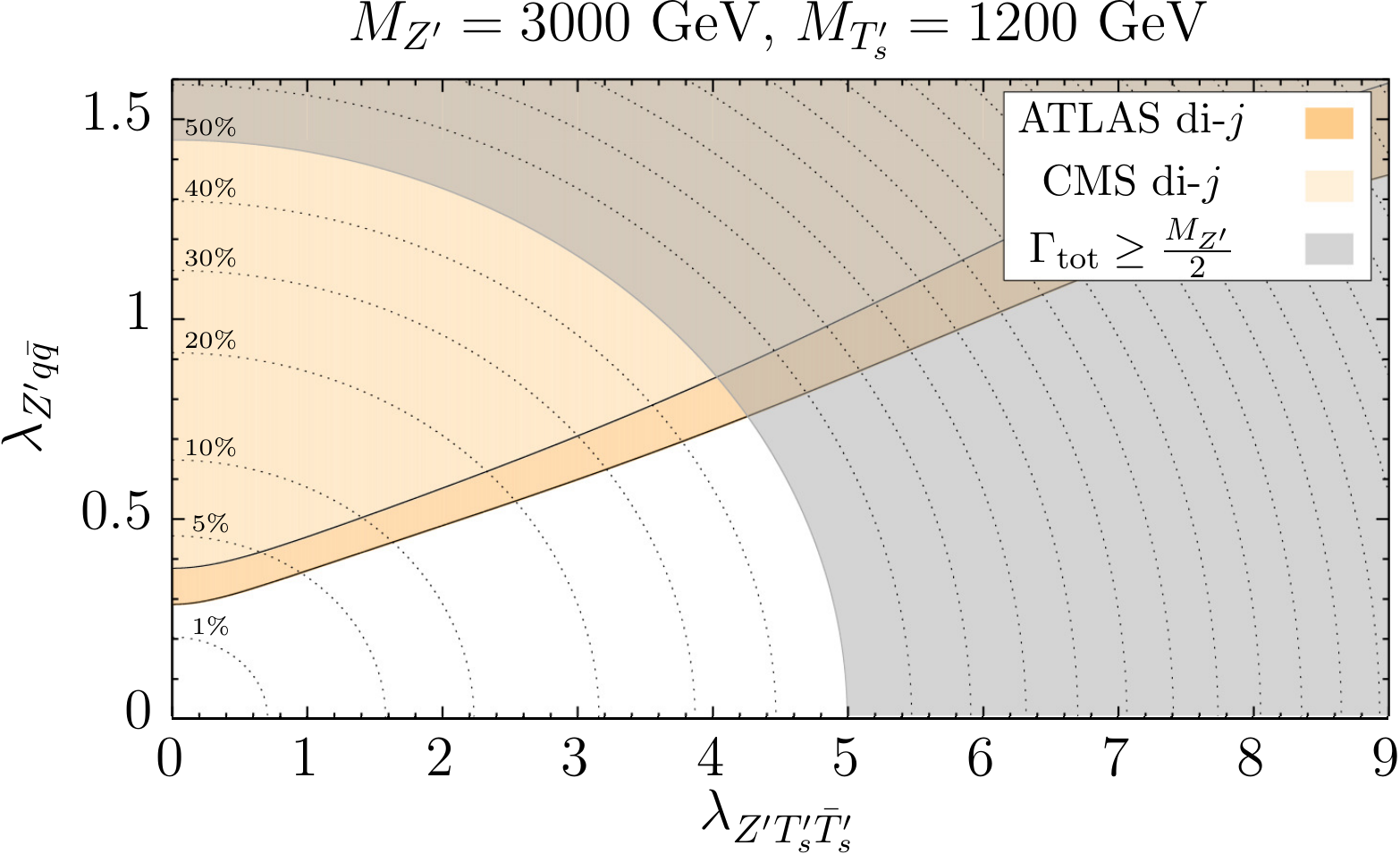}\hfill
  \includegraphics[scale=.45]{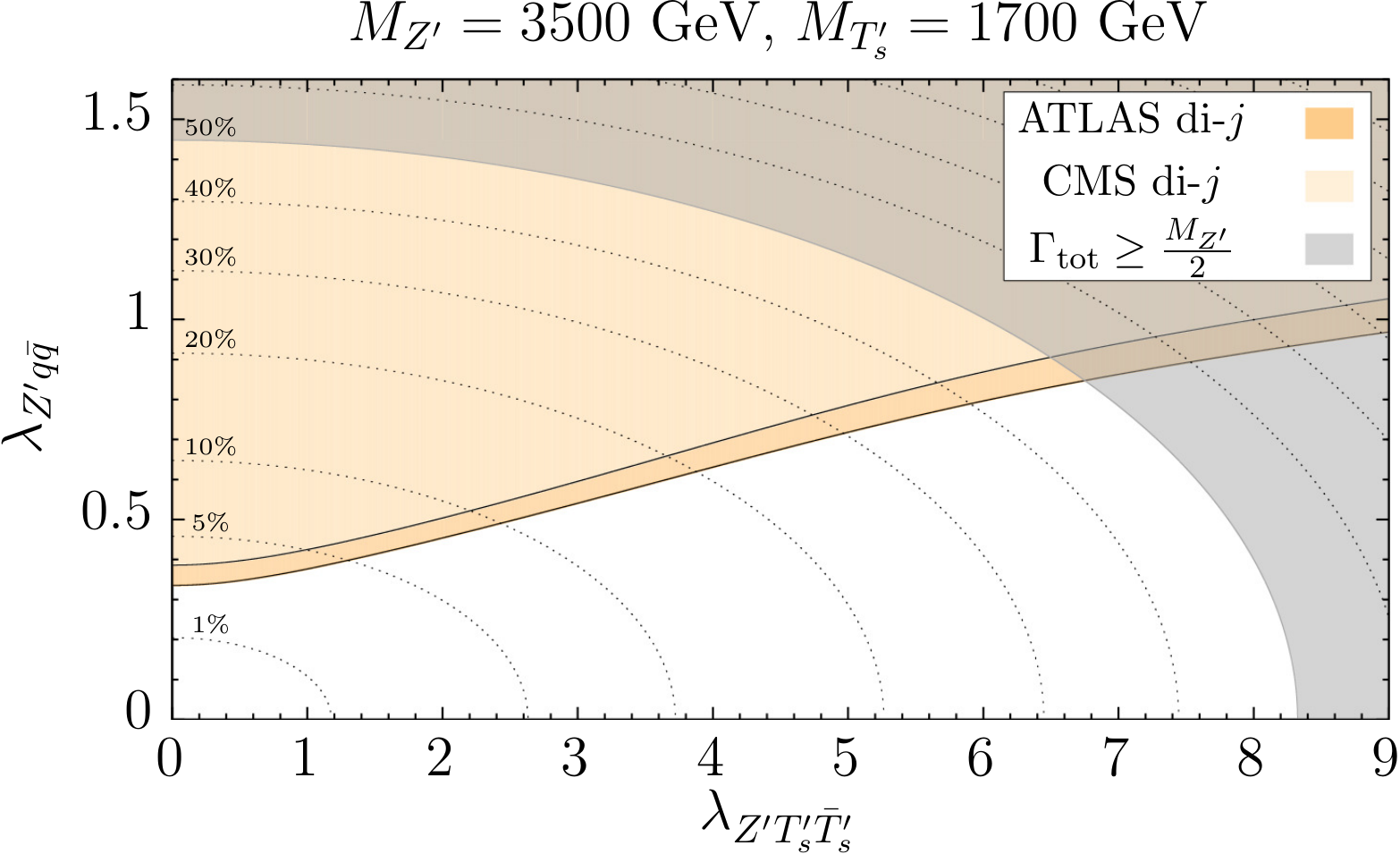}
  \caption{$\(\lambda_{Z'T'_s\overline{T'_s}}, \lambda_{Z'q\bar{q}}\)$ parameter space for $\lambda_{Z'\ell^+\ell^-} = 0$ and different $M_{Z'}$ and $M_{T'_s}$ with di-jet and di-lepton bounds. The dotted lines from bottom to top show when the $Z'$ width is $(1, 5, 10, 20, \ldots)$\% of its mass.}
  \label{fig:parspace-gl0}
\end{figure}

\begin{figure}[ht] 
  \centering
  \includegraphics[scale=.45]{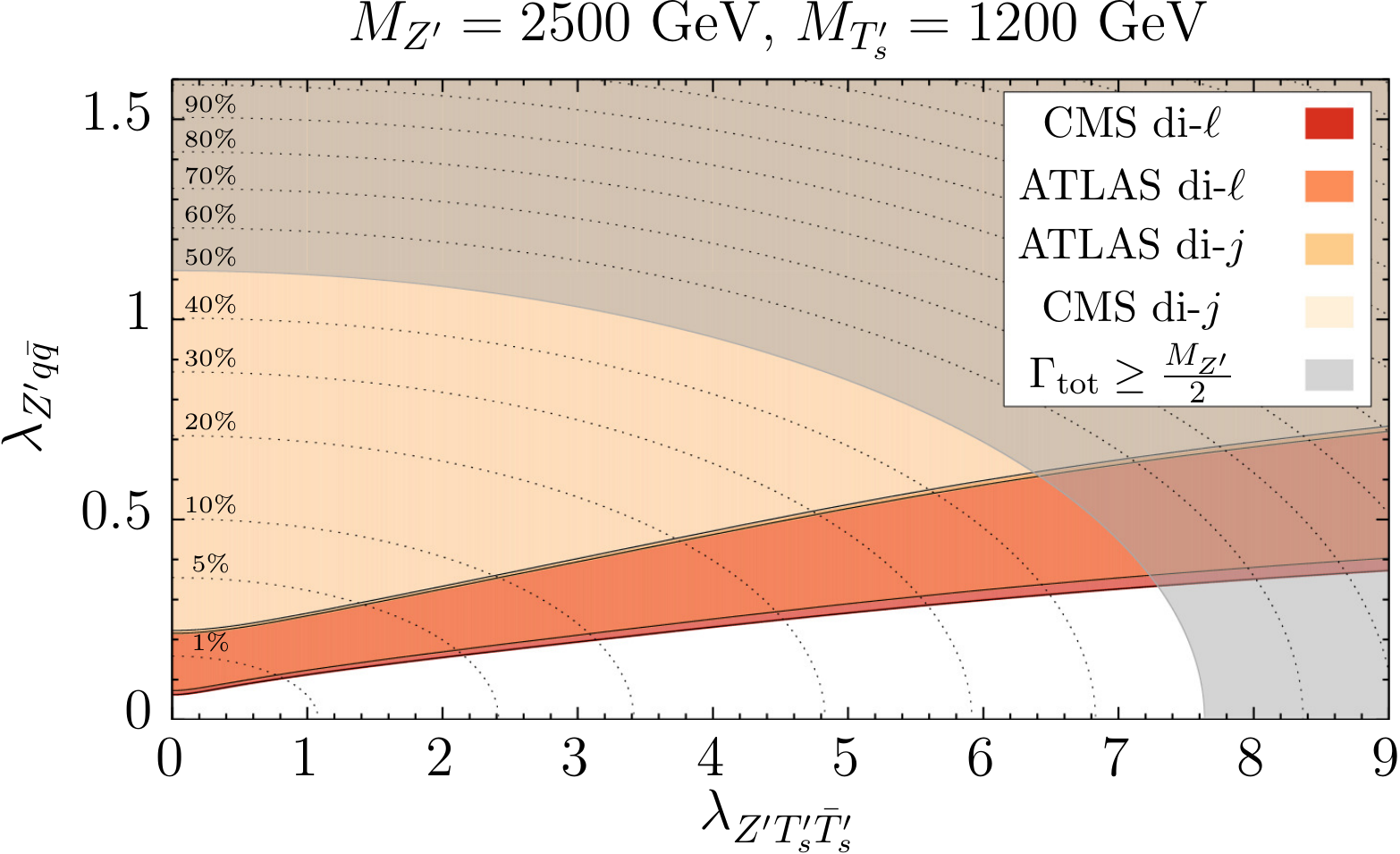}\hfill
  \includegraphics[scale=.45]{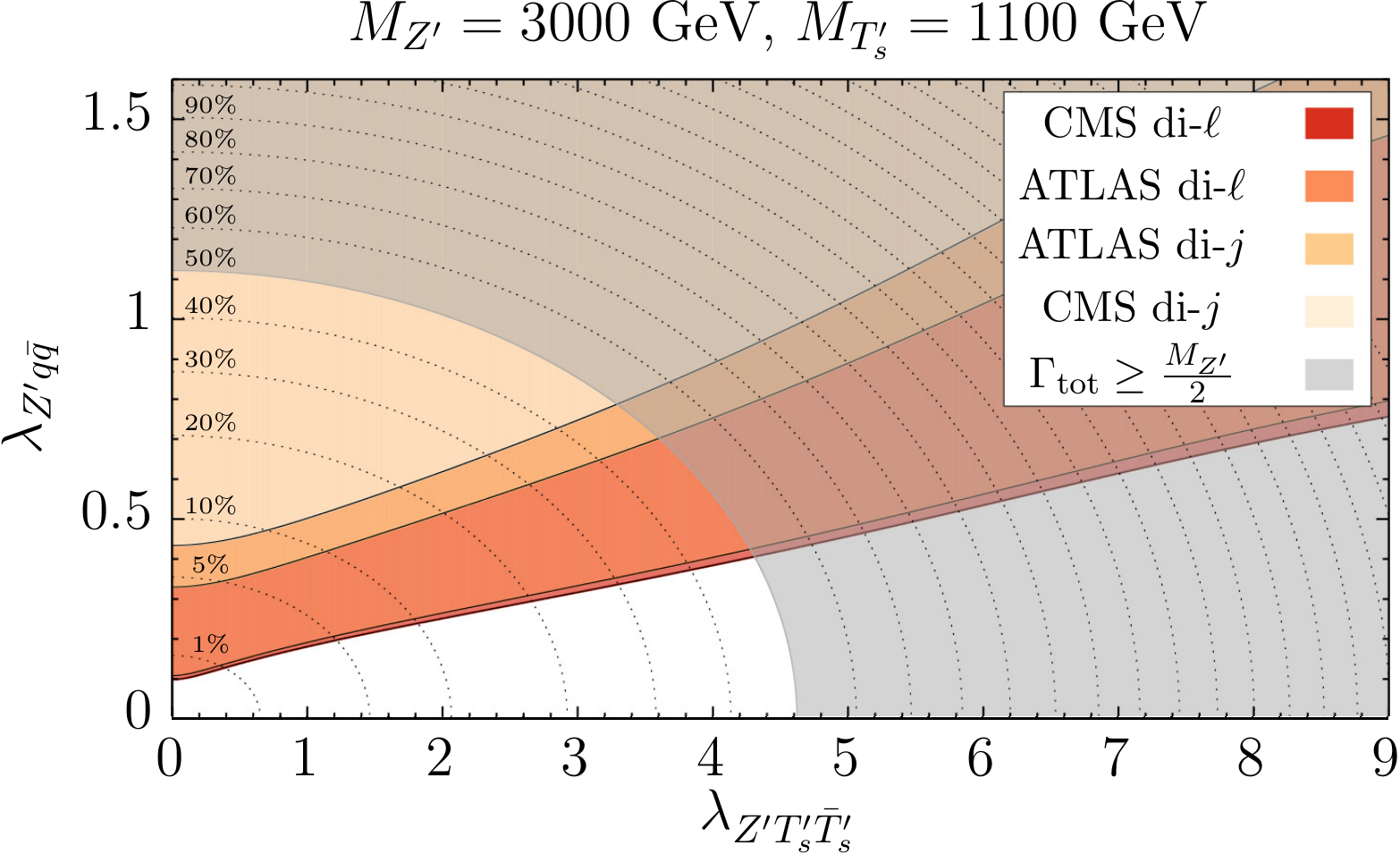}\\[0.2cm]
  \includegraphics[scale=.45]{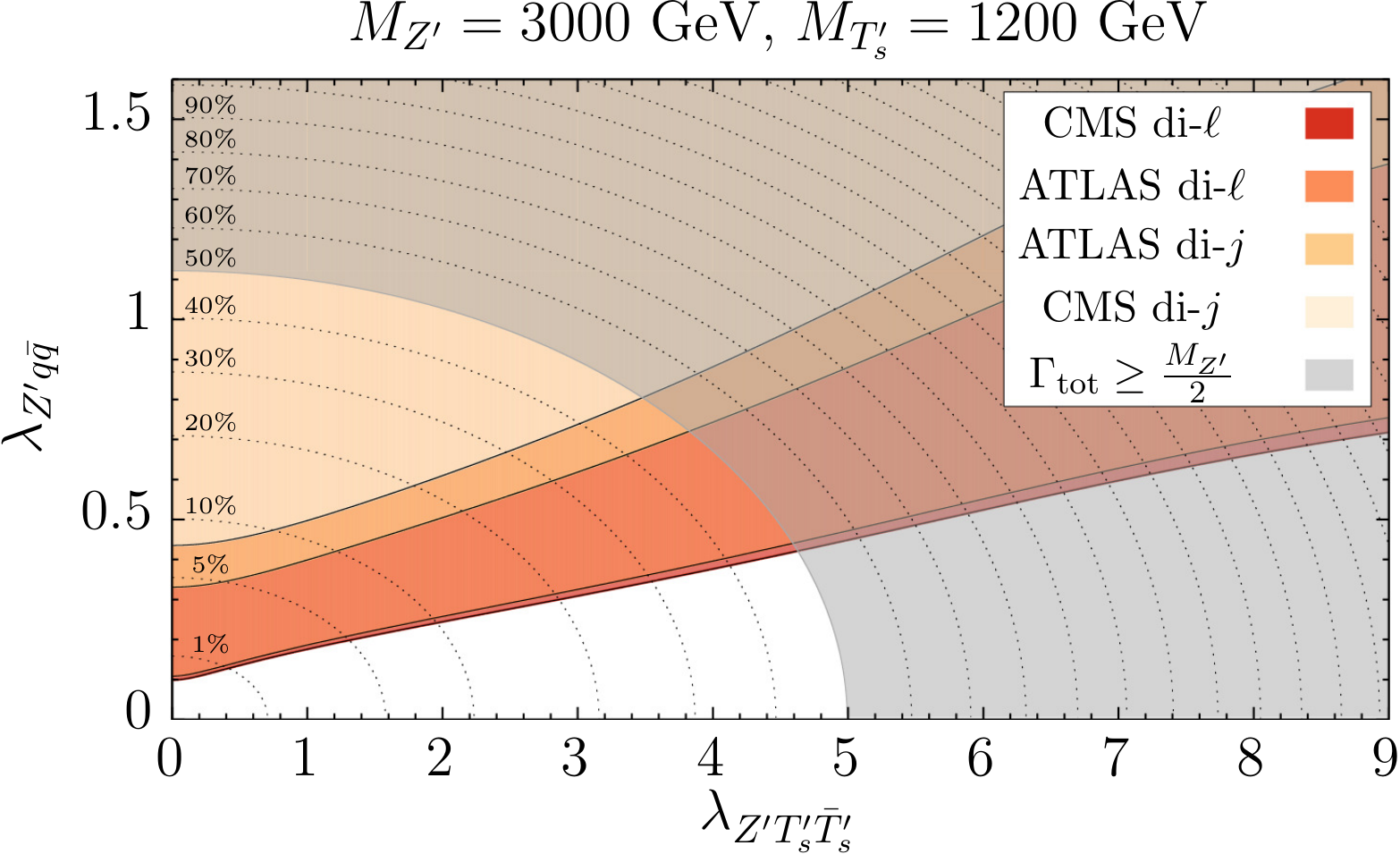}\hfill
  \includegraphics[scale=.45]{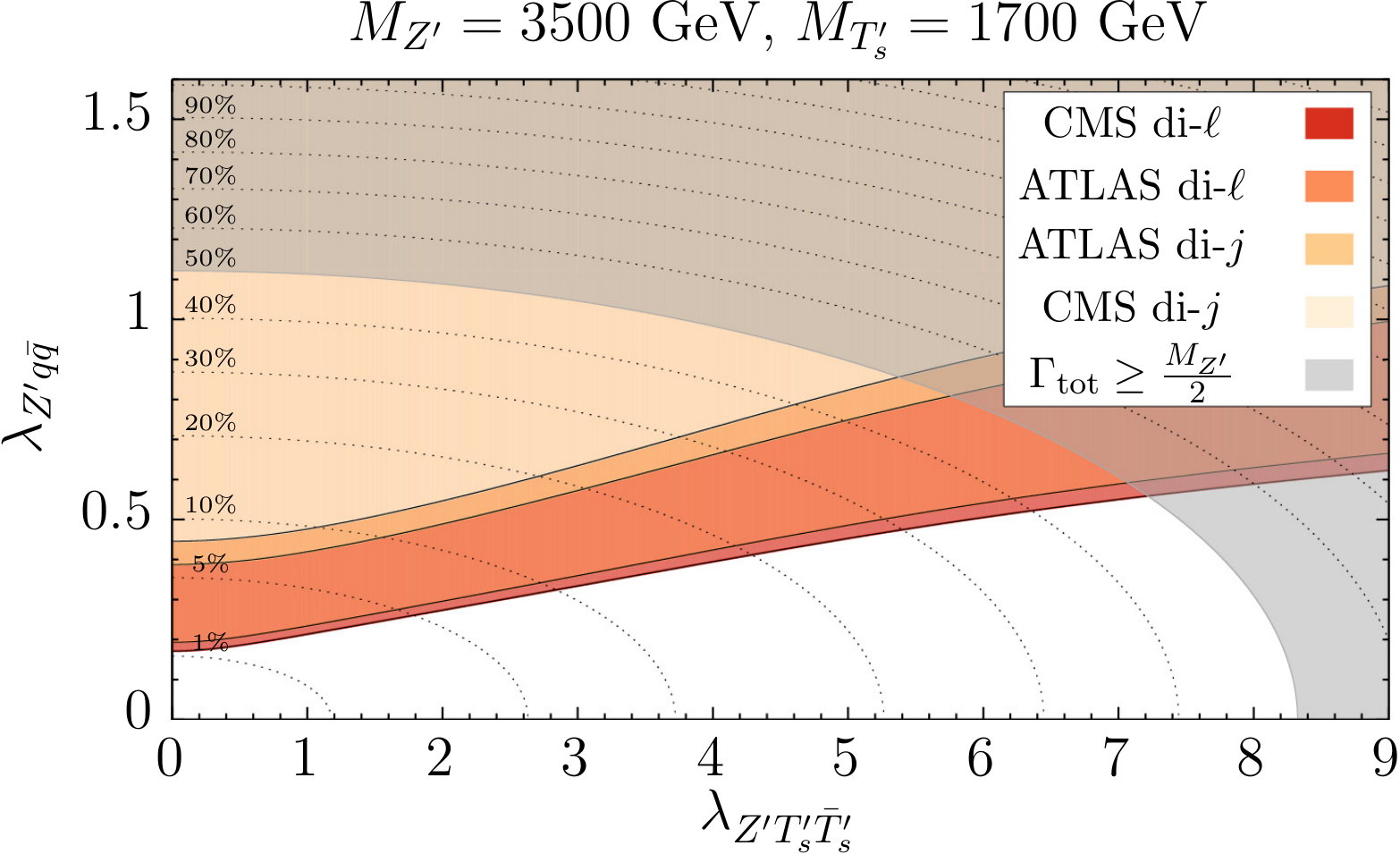}
  \caption{$\(\lambda_{Z'T'_s\overline{T'_s}}, \lambda_{Z'q\bar{q}}\)$ parameter space for $\lambda_{Z'\ell^+\ell^-} = \lambda_{Z'q\bar{q}}$ and different $M_{Z'}$ and $M_{T'_s}$ with di-jet and di-lepton bounds. The dotted lines from bottom to top show when the $Z'$ width is $(1, 5, 10, 20, \ldots)$\% of its mass.}
  \label{fig:parspace-gl-EQ-gq}
\end{figure}

Before further examining the model, we  check which parts of the $\(\lambda_{Z'T'_s\overline{T'_s}}, \lambda_{Z'q\bar{q}}\)$ coupling space are already excluded by current experimental di-jet and di-lepton limits. We also require the width of the $Z'$ boson to be not excessively large, such that
\beq
  \Gamma_{\text{tot}}\(Z'\) < \frac{M_{Z'}}{2}\,,
  \label{eq:pert_bound-1}
\eeq
which also ensures that $Z'$ couplings to fermions are perturbative and one can trust our tree-level study.
The total two-body $Z'$ decay width is:
\beq
  \Gamma_{\text{tot}}\(Z'\) = \frac{1}{8\,\pi\, M^2_{Z'}}\, \sum\limits_{\text{final states}} \!\!\!\!\! |\mathcal{M}|^2 \,|p_1|
  \label{eq:Zp_tot_width}
\eeq
with the 4-momentum of the first final-state particle
\beq
  \notag
  |p_1| = \frac{\sqrt{\[M^2_{Z'}-\(m_1+m_2\)^2\] \[M^2_{Z'}-\(m_1-m_2\)^2\]}}{2\,M_{Z'}} = 
  \begin{cases}
    \dfrac{\sqrt{M^2_{Z'} - 4m^2}}{2} &, m_1=m_2 \equiv m 
    \\
    \dfrac{M_{Z'}}{2} &, m_1=m_2=0
  \end{cases}
  \label{eq:Zp-tot-width-p1}
\eeq
and already having integrated over the solid angle of the first final-state particle $\int \d\Omega = \iint \sin(\theta_1)\,\d\theta_1 \, \d\phi_1  = 4\pi$.
The squared matrix element for a $Z'$ decaying to quark, $T'_s$ and lepton pairs times $|p_1|$ then reads
\begin{align}
  \sum\limits_{\text{final states}} \!\!\!\!\! |\mathcal{M}|^2 \, |p_1| &= \underbrace{2\[\sum\limits_{\{q\}} \(M^2_{Z'} - m^2_q\) \cdot \left|p_1(m_q)\right|\]}_{\equiv A_q} \, \lambda^2_{Z'q\bar{q}} \notag \\
                                                                        &+ \, \underbrace{2 \(M^2_{Z'} - M^2_{T'_s}\) \cdot \left|p_1(M_{T'_s})\right|}_{\equiv A_{T'_s}} \, \lambda^2_{Z'T'_s\overline{T'_s}}   \label{eq:sqMZp} \\
                                                                        &+ \, \underbrace{\frac23\[\sum\limits_{\{\ell\}} \(M^2_{Z'} - m^2_\ell\) \cdot \left|p_1(m_{\ell})\right|\]}_{\equiv A_{\ell}} \, \lambda^2_{Z'\ell^+\ell^-} \,. \notag
  \label{eq:Msq}
\end{align}
Plugging eq.~(\ref{eq:sqMZp}) into eq.~(\ref{eq:Zp_tot_width}) and (\ref{eq:pert_bound-1}) yields a combined upper bound on $\lambda_{Z'q\bar{q}}$, $\lambda_{Z'T'_s\overline{T'_s}}$ and $\lambda_{Z'\ell^+\ell^-}$. This bound together with the experimental limits from ATLAS and CMS for di-jet \cite{ATLAS:2016lvi,Sirunyan:2016iap} and di-lepton \cite{ATLAS:2016cyf,CMS:2016abv} searches is shown in figures~\ref{fig:parspace-gl0}  ($\lambda_{Z'\ell^+\ell^-} = 0$) and \ref{fig:parspace-gl-EQ-gq} ($\lambda_{Z'\ell^+\ell^-} = \lambda_{Z'q\bar{q}}$) in detail for several combinations of $M_{Z'}$ and $M_{T'_s}$, where the limits for $\lambda_{Z'T'_s\overline{T'_s}} > 0$ were found using the NWA
\begin{subequations}
  \begin{align}
    \sigma_{pp \to Z' \to jj}\(\lambda_{Z'q\bar{q}}\(0\)\) &= \sigma_{pp \to Z'}\(\lambda_{Z'q\bar{q}}\(\lambda_{Z'T'_s\overline{T'_s}}\)\) \cdot \text{BR}\(Z' \to q\bar{q}\) \,, \\
    \sigma_{pp \to Z' \to \ell^+\ell^-}\(\lambda_{Z'q\bar{q}}\(0\)\) &= \sigma_{pp \to Z'}\(\lambda_{Z'q\bar{q}}\(\lambda_{Z'T'_s\overline{T'_s}}\)\) \cdot \text{BR}\(Z' \to \ell^+\ell^-\) \,,
  \end{align}
  \label{eq:corr-limits}
\end{subequations}
with $\lambda_{Z'q\bar{q}}\(0\)$ being the experimental limit for $\lambda_{Z'T'_s\overline{T'_s}} = 0$.
%
%

While the di-jet bounds already set rather strong limits on $\lambda_{Z'q\bar{q}}$, the di-lepton bounds extend these even more, especially at small $M_{Z'}$. With increasing $M_{Z'}$, all bounds get substantially weaker, whereas an increase of $M_{T'_s}$ is affecting the bounds only slightly.
When $M_{Z'}$ is getting close to  $2 M_{T'_s}$ (see figure~\ref{fig:parspace-gl0} and \ref{fig:parspace-gl-EQ-gq}, top left and bottom right), the parameter space broadens in the $\lambda_{Z'T'_s\overline{T'_s}}$ direction due to $A_{T'_s}$ becoming small and thus allowing for larger $\lambda_{Z'T'_s\overline{T'_s}}$ to be realised. 

Eventually, we find that a significant fraction of parameter space, especially for large values of  $\lambda_{Z'T'_s\overline{T'_s}}$, is still available for study.

\subsection{\texorpdfstring{LHC Reach Including $Z'$ Bosons}{LHC Reach Including Z' Bosons}}
\label{sec:ttMET}

After establishing constrains on the $\(\lambda_{Z'T'_s\overline{T'_s}}, \lambda_{Z'q\bar{q}}\)$ parameter space from current di-jet and di-lepton bounds, we are ready to investigate the remaining parameter space for our $t\bar{t} + \slashed{E}_T$ signature. To do so, we first use \texttt{CheckMATE} to analyse 50000 events generated for the $Z'$-mediated part of figure~\ref{fig:Zp-prod-Feynman-diagram} (left graph) individually for all on-shell $(M_{Z'},M_{T'_s})$ combinations and $m_\phi = (10, 62.5, 100, 300, 600)$ GeV together with $\lambda_{Z'q\bar{q}} = 0.2$ and $\lambda_{Z'T'_s\overline{T'_s}} = 2$.

\begin{figure}[ht]
  \centering
  \includegraphics[width=0.5\textwidth]{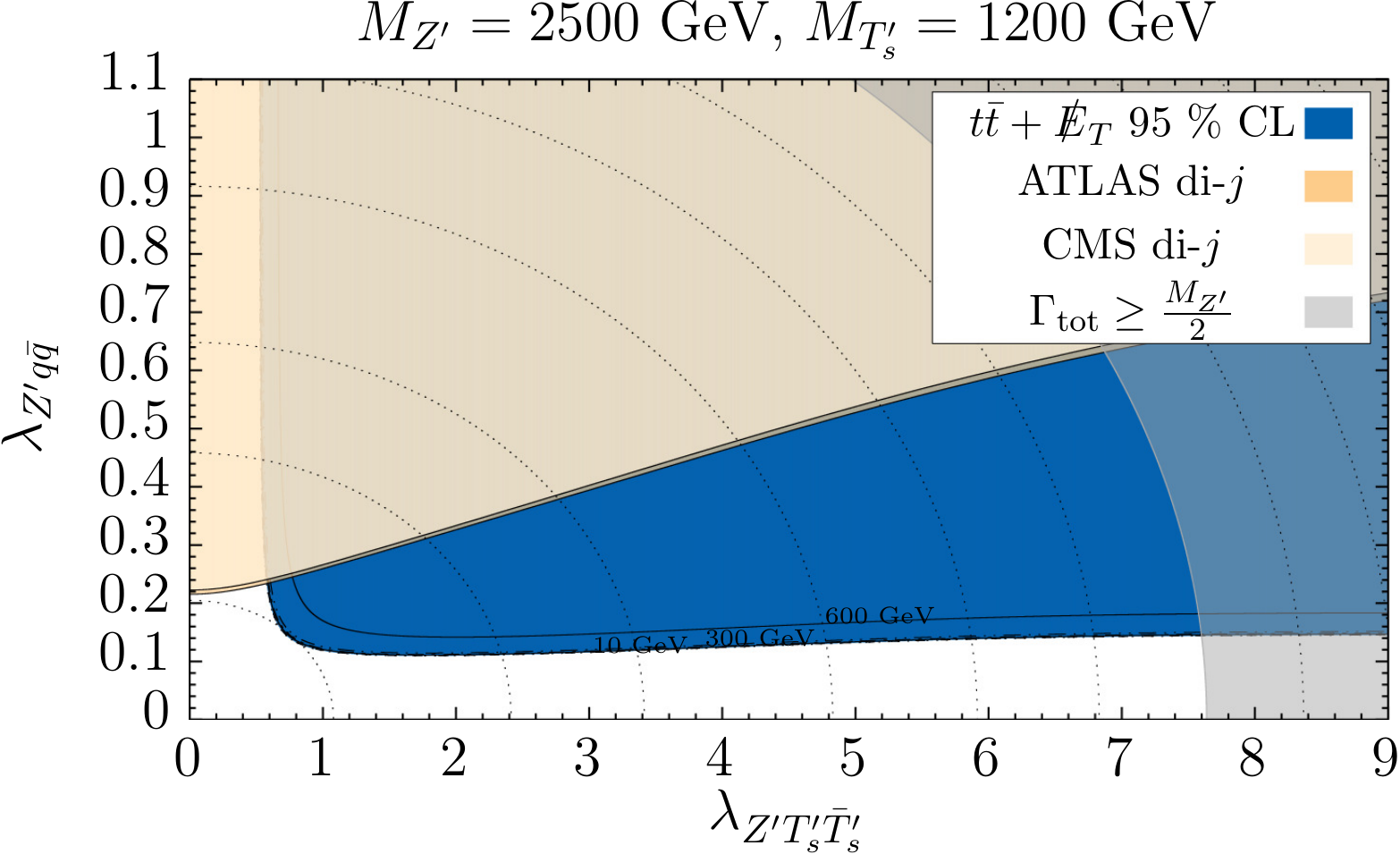}\hfill%
  \includegraphics[width=0.5\textwidth]{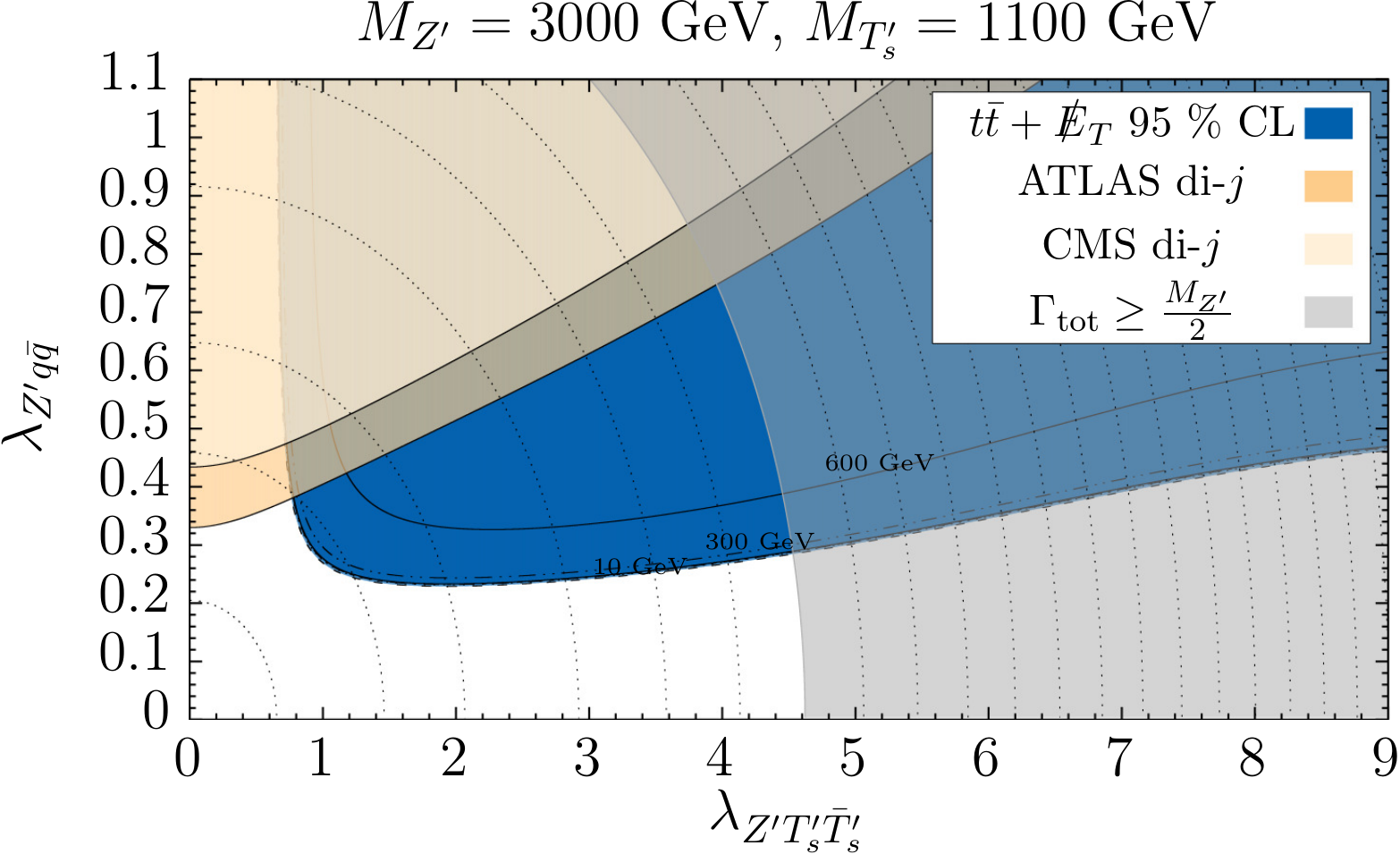}\\
  \includegraphics[width=0.5\textwidth]{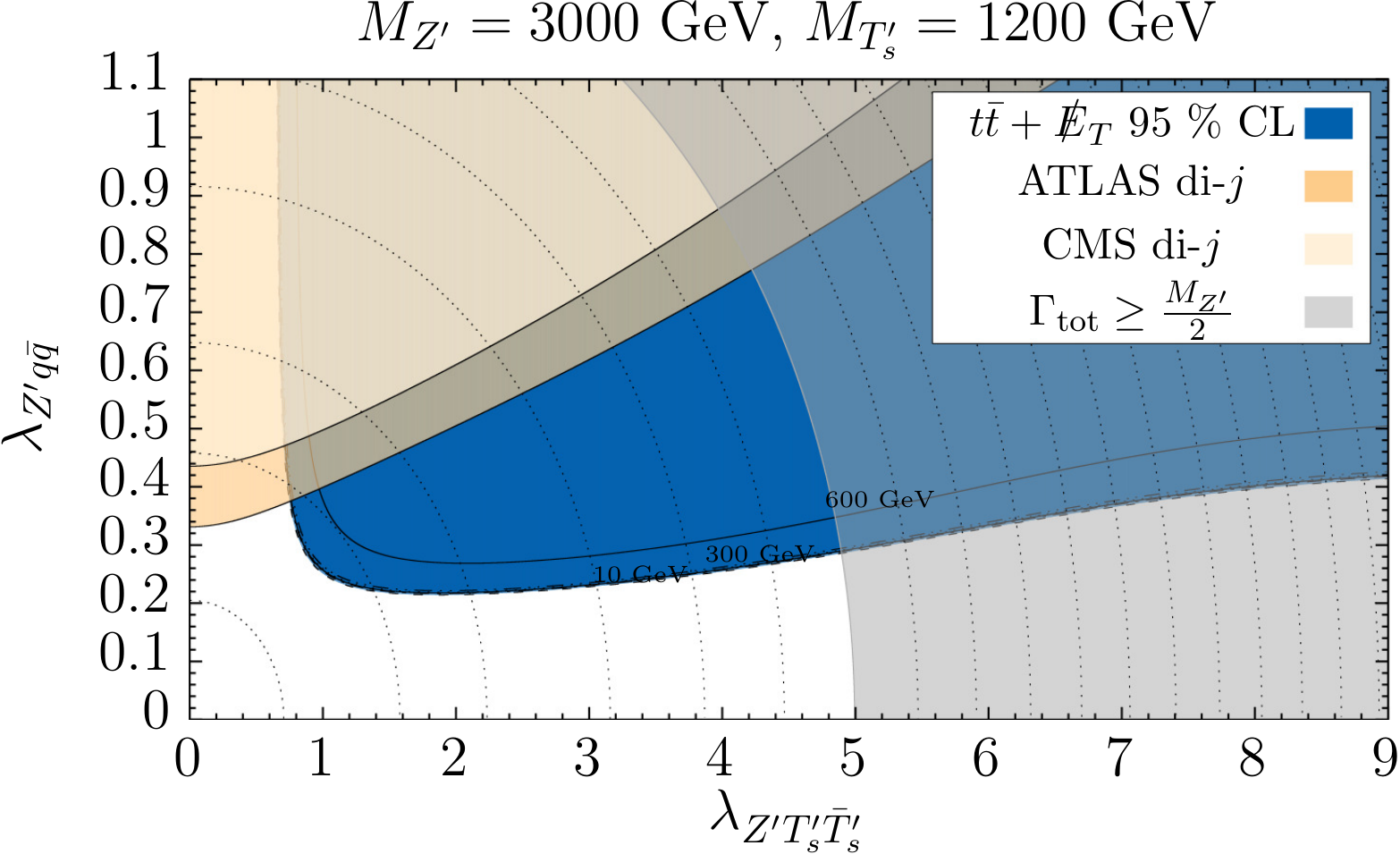}\hfill%
  \includegraphics[width=0.5\textwidth]{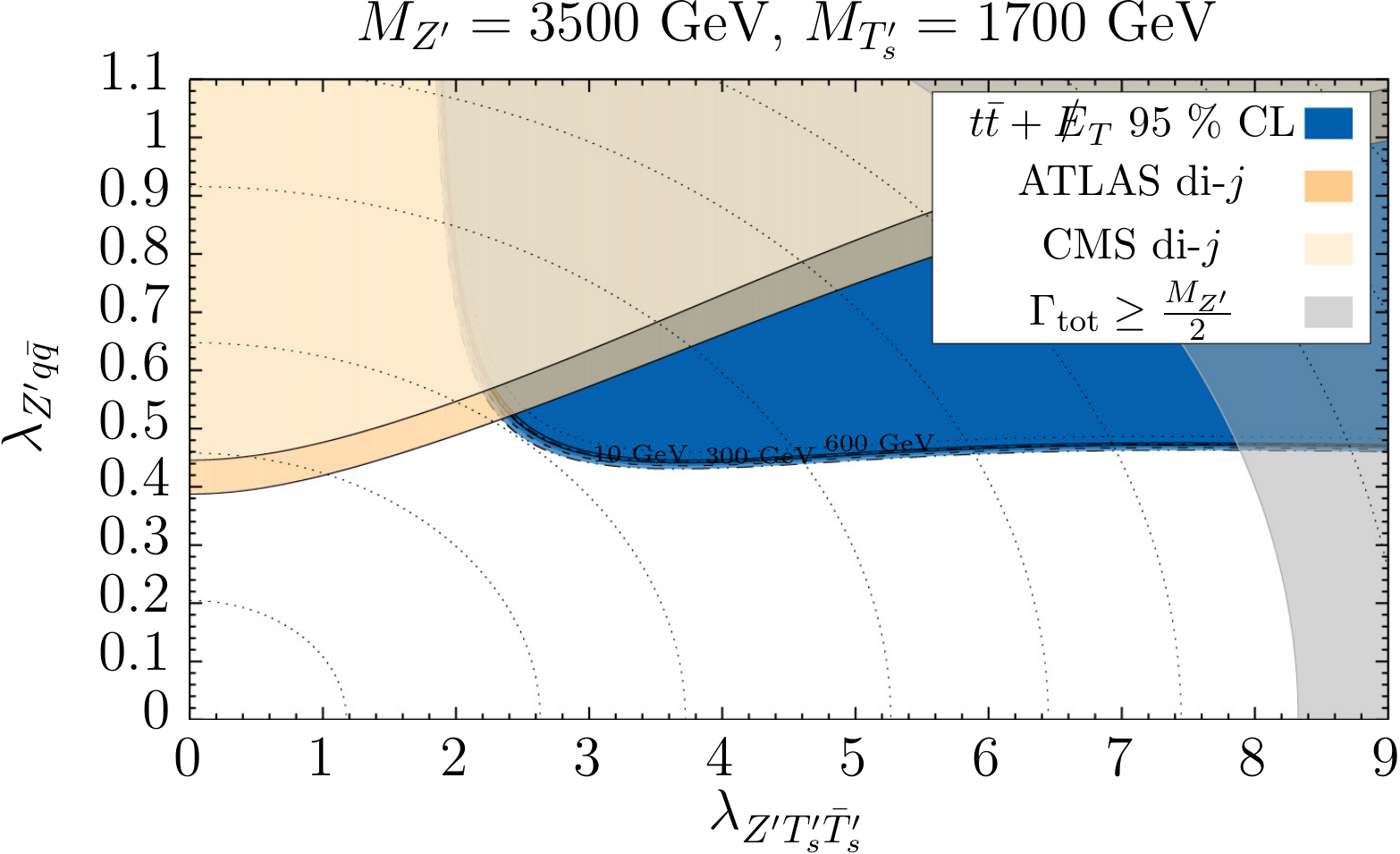}
  \vspace*{-5mm}
  \caption{Di-jet and di-lepton bounds together with the most constraining $t\bar{t} + \slashed{E}_T$ bounds coming from \texttt{ATLAS\_CONF\_2016\_050} for $\lambda_{Z'\ell^+\ell^-} = 0$. The parameter space below the coloured bands is not excluded and available for study. The labels "10", "300" and "600" on the black lines refer to $m_\phi$ in GeV for the blue $t\bar{t} + \slashed{E}_T$ bound.}
  \label{fig:ttMET-pspaces-0}
\end{figure}

\begin{figure}[ht]
  \centering
  \includegraphics[width=0.5\textwidth]{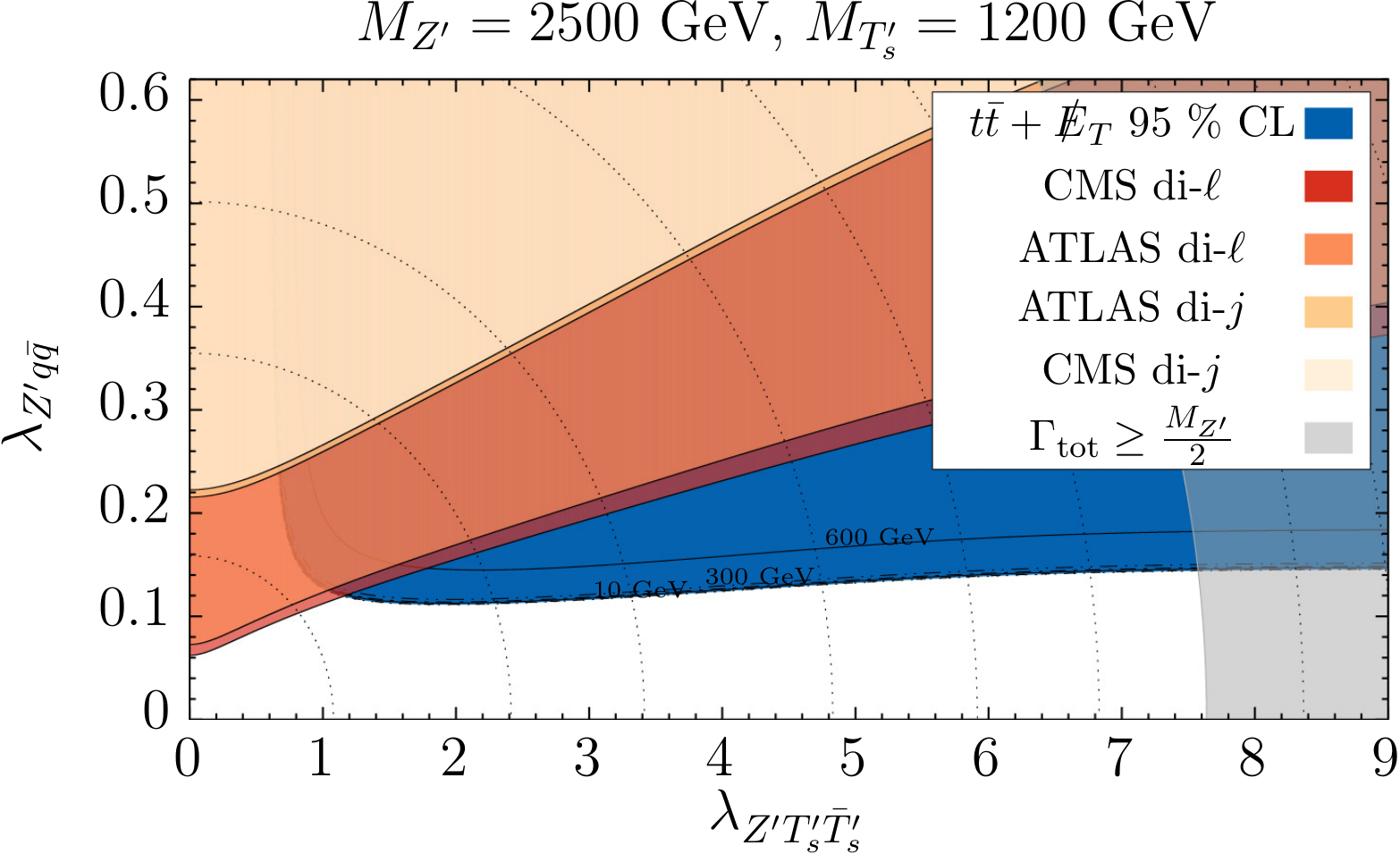}\hfill%
  \includegraphics[width=0.5\textwidth]{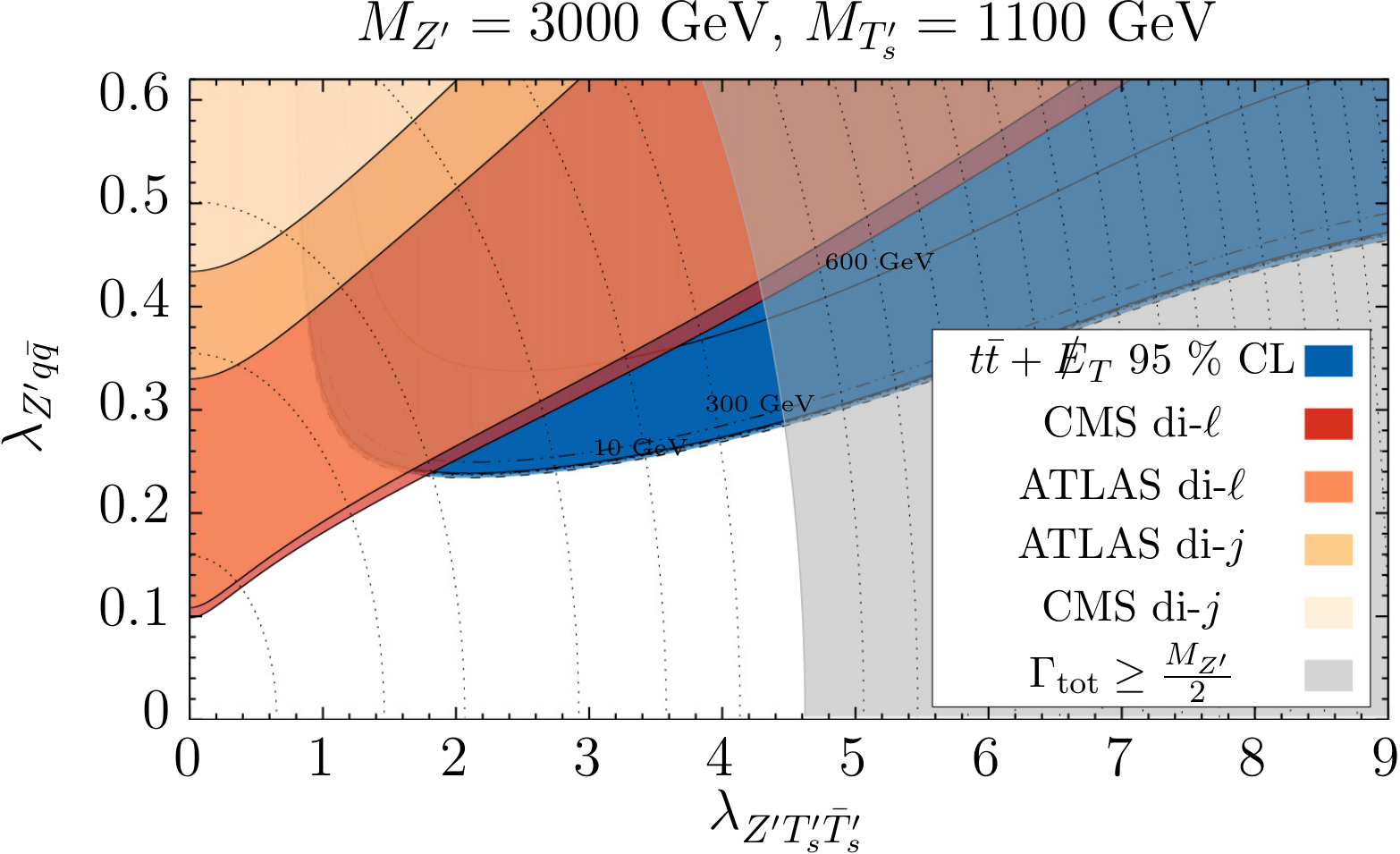}\\
  \includegraphics[width=0.5\textwidth]{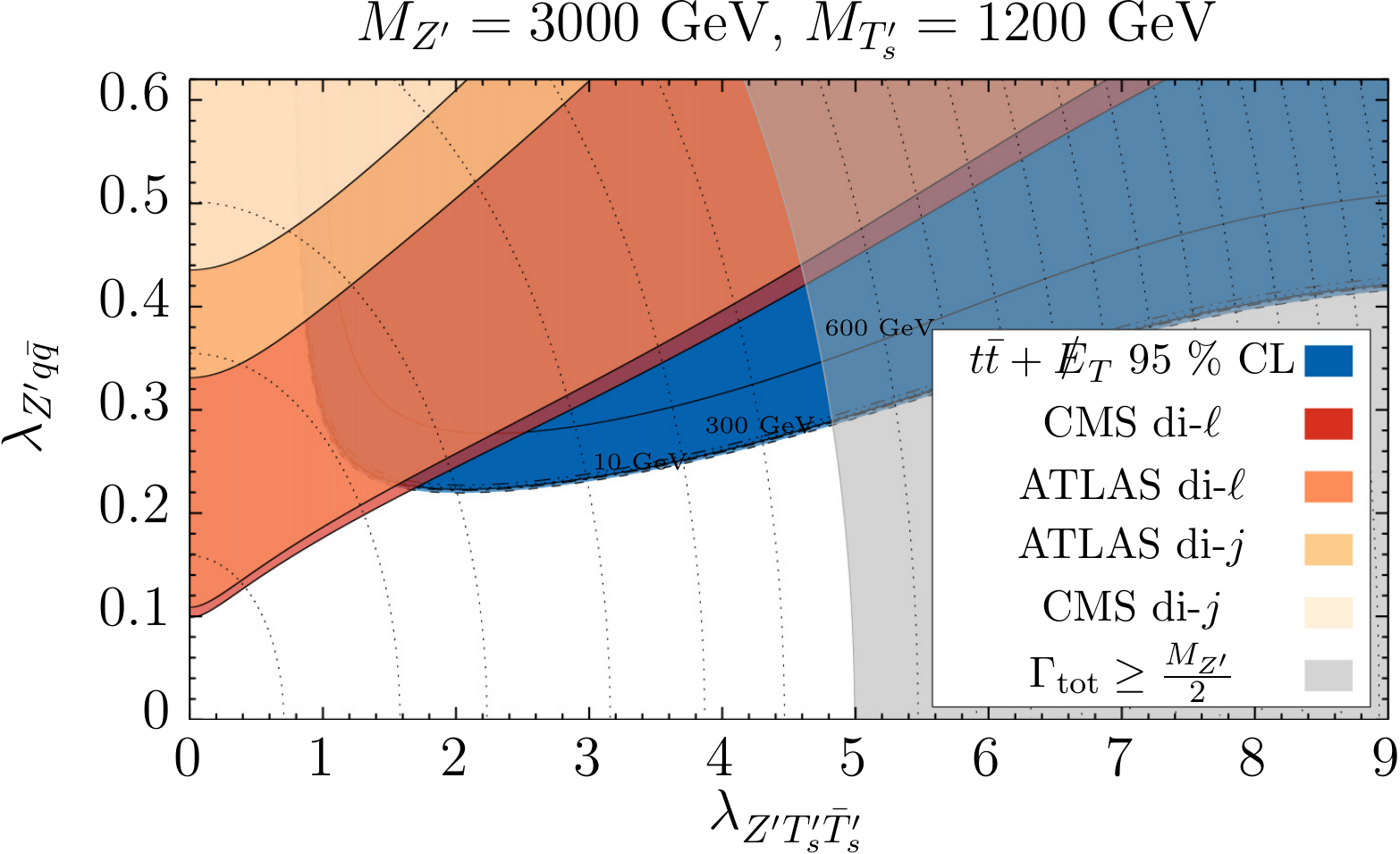}\hfill%
  \includegraphics[width=0.5\textwidth]{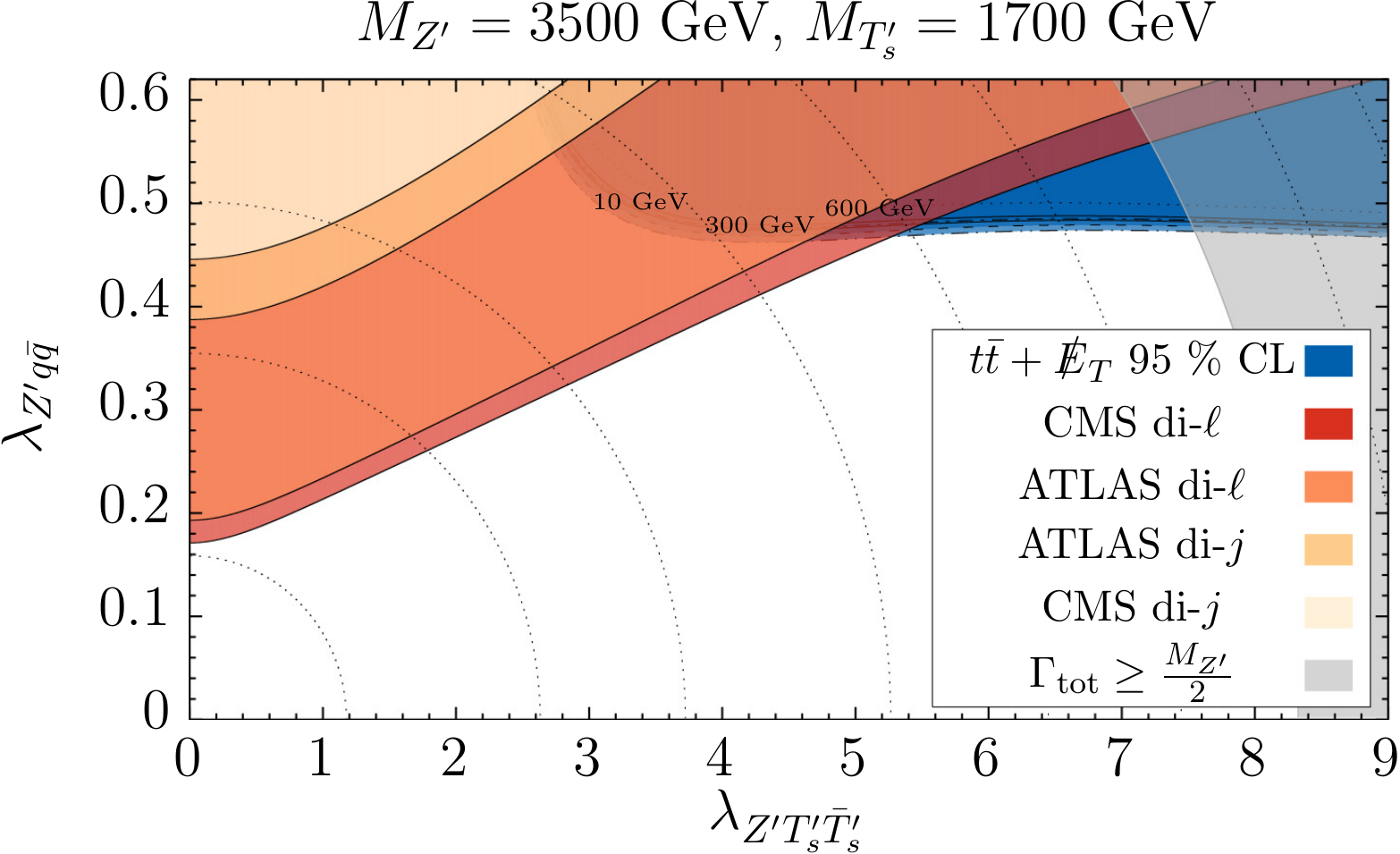}
  \vspace*{-5mm}
  \caption{Di-jet and di-lepton bounds together with the most constraining $t\bar{t} + \slashed{E}_T$ bounds coming from \texttt{ATLAS\_CONF\_2016\_050} for $\lambda_{Z'\ell^+\ell^-} = \lambda_{Z'q\bar{q}}$. The parameter space below the coloured bands is not excluded and available for study. The labels "10", "300" and "600" on the black lines refer to $m_\phi$ in GeV for the blue $t\bar{t} + \slashed{E}_T$ bound.}
  \label{fig:ttMET-pspaces}
\end{figure}

It turns out that the majority of points are constrained mostly by the \texttt{ATLAS\_CONF\_2016\_050} \cite{ATLAS:2016ljb} analysis in the \texttt{tN\_high} signal region. Only for very large $M_{Z'}$ and $M_{T'_s}$, \texttt{ATLAS\_1605\_03814} \cite{Aaboud:2016zdn} gives stronger limits in the \texttt{2jt} signal region (although being only slightly more constraining than the \texttt{ATLAS\_CONF\_2016\_050 tN\_high} limit). Therefore, we can use the NWA once more to rescale the detection/exclusion limits computed by \texttt{CheckMATE} via
\bea
  \sigma_{95\%}&\stackrel{!}{=}&\sigma_{\rm signal} = \sigma_{pp \to Z'} \cdot \text{BR}(Z' \to T'_s\overline{T'_s}) \notag \\
  &=&  \mathcal{P} \, \lambda^2_{Z'q\bar{q}} \cdot \frac{A_{T'_s} \lambda^2_{Z'T'_s\overline{T'_s}}}{A_{T'_s} \lambda^2_{Z'T'_s\overline{T'_s}} + A_q \lambda^2_{Z'q\bar{q}} + A_\ell \lambda^2_{Z'\ell^+\ell^-}} \,, 
  \label{eq:LHC-NWA}
\eea
where $\sigma_{95\%}$ indicates the bound (at 95\% C.L.) on the observed production cross section from the most constraining search and parameter region and $\mathcal{P} = \mathcal{P}(M_{Z'}) = \frac{\sigma_{pp \to Z'}}{\lambda^2_{Z'q\bar{q}}}$ is the prefactor of the $Z'$ production cross section and only depends on $M_{Z'}$. This allows us to find the excluded $\(\lambda_{Z'T'_s\overline{T'_s}}, \lambda_{Z'q\bar{q}}\)$ parameter space. An interesting set of these exclusion limits is shown in blue in figures~\ref{fig:ttMET-pspaces-0} ($\lambda_{Z'\ell^+\ell^-} = 0$) and \ref{fig:ttMET-pspaces} ($\lambda_{Z'\ell^+\ell^-} = \lambda_{Z'q\bar{q}}$) for various  values of $m_\phi$, indicated as black solid and dashed lines at the bottom of the blue band.
For large $\lambda_{Z'T'_s\overline{T'_s}}$, the $t\bar{t} + \slashed{E}_T$ limit is nearly saturated with respect to $\lambda_{Z'q\bar{q}}$. This is due to $\text{BR}(Z' \to T'_s\overline{T'_s}) \gg \text{BR}(Z' \to q\bar{q})$, ensuring a sufficient amount of $T'_s$-pairs surviving the experimental cuts is produced. However, as we rescale the NWA according to eq.~(\ref{eq:NWA-rescaling}), the limits also slightly decrease for increasing $\lambda_{Z'T'_s\overline{T'_s}}$ (see e.g. figure~\ref{fig:ttMET-pspaces-0} and \ref{fig:ttMET-pspaces}, top right and bottom left). On the contrary, for decreasing $\lambda_{Z'T'_s\overline{T'_s}}$, the $t\bar{t} + \slashed{E}_T$ limit gets maximal before abruptly vanishing soon after. This is due to $\text{BR}(Z' \to T'_s\overline{T'_s}) \lesssim \text{BR}(Z' \to q\bar{q})$, resulting in a suppression of $T'_s$-pair production. The slope or shape of that drop is mainly influenced by $\Delta M = M_{Z'} - 2M_{T'_s}$. The smaller $\Delta M$, the more rectangular the shape will become, whereas for larger $\Delta M$, the shape will turn smoother and rounder.

In comparison to the di-lepton and di-jet limits, the $t\bar{t} + \slashed{E}_T$ signature is able to cover large parts of the parameter space where $T'_s$-pair production gets more and more dominant (i.e. for large $\lambda_{Z'T'_s\overline{T'_s}}$), as long as $M_{Z'}$ is sufficiently small. With increasing $M_{Z'}$, the $t\bar{t} + \slashed{E}_T$ limit weakens substantially and gets even weaker than the di-lepton and di-jet bounds, once $M_{Z'} \geq 4$ TeV (not shown here). However, for $M_{Z'} \leq 3$ TeV, the $t\bar{t} + \slashed{E}_T$ limit is the most constraining one in terms of parameter space coverage, whereas the di-lepton and di-jet limits yield strong, additional constraints for very small $\lambda_{Z'T'_s\overline{T'_s}}$.

In summary, by taking into account both di-lepton and di-jet as well as the $t\bar{t} + \slashed{E}_T$ limits, we are able to efficiently constrain the $\(\lambda_{Z'T'_s\overline{T'_s}}, \lambda_{Z'q\bar{q}}\)$ parameter space for most values of $\lambda_{Z'q\bar{q}}$.

\subsection{Detailed Benchmark Analysis}
\label{sec:DBA}

\begin{figure}[ht]
  \centering
  \includegraphics[width=0.9\textwidth]{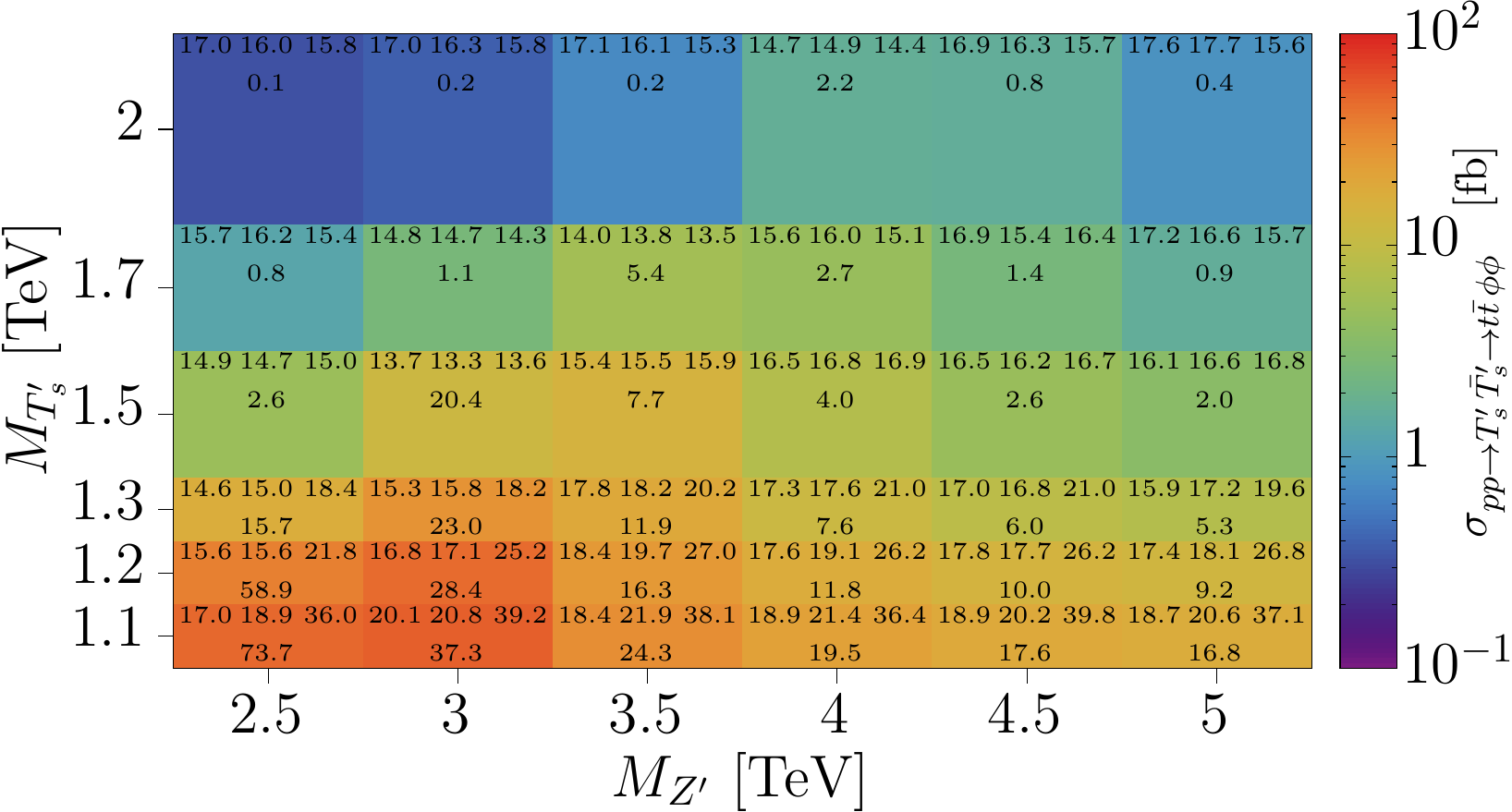}
  \caption{Predicted cross sections for $pp \to T'_s\overline{T'_s} \to t \bar{t} \phi \phi$ in fb in dependence of $M_{Z'}$ and $M_{T'_s}$. The top three numbers in a rectangle from left to right show the experimental limit on the cross section in fb for $m_\phi = 10, 300, 600$ GeV respectively, whereas the number below shows the theoretical prediction coinciding with the colour-coding. The couplings are chosen as $\lambda_{Z'q\bar{q}} = 0.25 = \lambda_{Z'\ell^+\ell^-}$ and $\lambda_{Z'T'_s\overline{T'_s}} = 2.5$.}
  \label{fig:benchmark-map}
\end{figure}

Based on the above studies, we choose a specific benchmark point in the  $\(\lambda_{Z'T'_s\overline{T'_s}}, \lambda_{Z'q\bar{q}}\)$ plane for further detailed investigation, followed by a qualitative discussion for other coupling choices. The benchmark we choose now covers all $T'_s\overline{T'_s}$ production processes shown in figure~\ref{fig:Zp-prod-Feynman-diagram} and covers all interesting states of exclusion, i.e. being excluded by all limits for small $M_{Z'}$, only excluded by the $t\bar{t} + \slashed{E}_T$ signature and not excluded at all. The benchmark comprises the following couplings
\begin{equation}
  \lambda_{Z'q\bar{q}} = 0.25\,, \qquad \qquad \lambda_{Z'T'_s\overline{T'_s}} = 2.5 \,, \qquad \qquad \lambda_{Z'\ell^+\ell^-} = \lambda_{Z'q\bar{q}}\,, 
  \label{eq:benchmark}
\end{equation}
and an overview over all related cross sections in the studied $M_{Z'}$-$M_{T'_s}$ range can be found in figure~\ref{fig:benchmark-map}.\footnote{For illustration, tables~\ref{tab:benchmarks} and \ref{tab:benchmarks-QCD} give more details on four benchmark points included in figure~\ref{fig:benchmark-map}.}
Therein, the top three numbers from left to right in each rectangle correspond to the experimental limits for $m_\phi = 10, 300, 600$ GeV respectively, whereas the number in the center corresponds to our signal cross section and coincides with the colour coding. By pinning down the benchmark-couplings in the $\(\lambda_{Z'T'_s\overline{T'_s}}, \lambda_{Z'q\bar{q}}\)$ plane from figures~\ref{fig:ttMET-pspaces-0} and \ref{fig:ttMET-pspaces} and then comparing with the numbers from figure~\ref{fig:benchmark-map}, it is clear that the mass pairs $(M_{Z'},M_{T'_s}) = (2.5, 1.2)$ TeV and $(3, 1.2)$ TeV are strongly excluded, even without the QCD contributions to $T'_s$-pair production. For $(3.5, 1.7)$ TeV, neither the full nor the $Z'$-only limit is able to probe that point, due to the predicted cross section being too small because of the large $M_{Z'}$ and $M_{T'_s}$.

\begin{figure}[ht]
  \centering
  \includegraphics[width=0.5\textwidth]{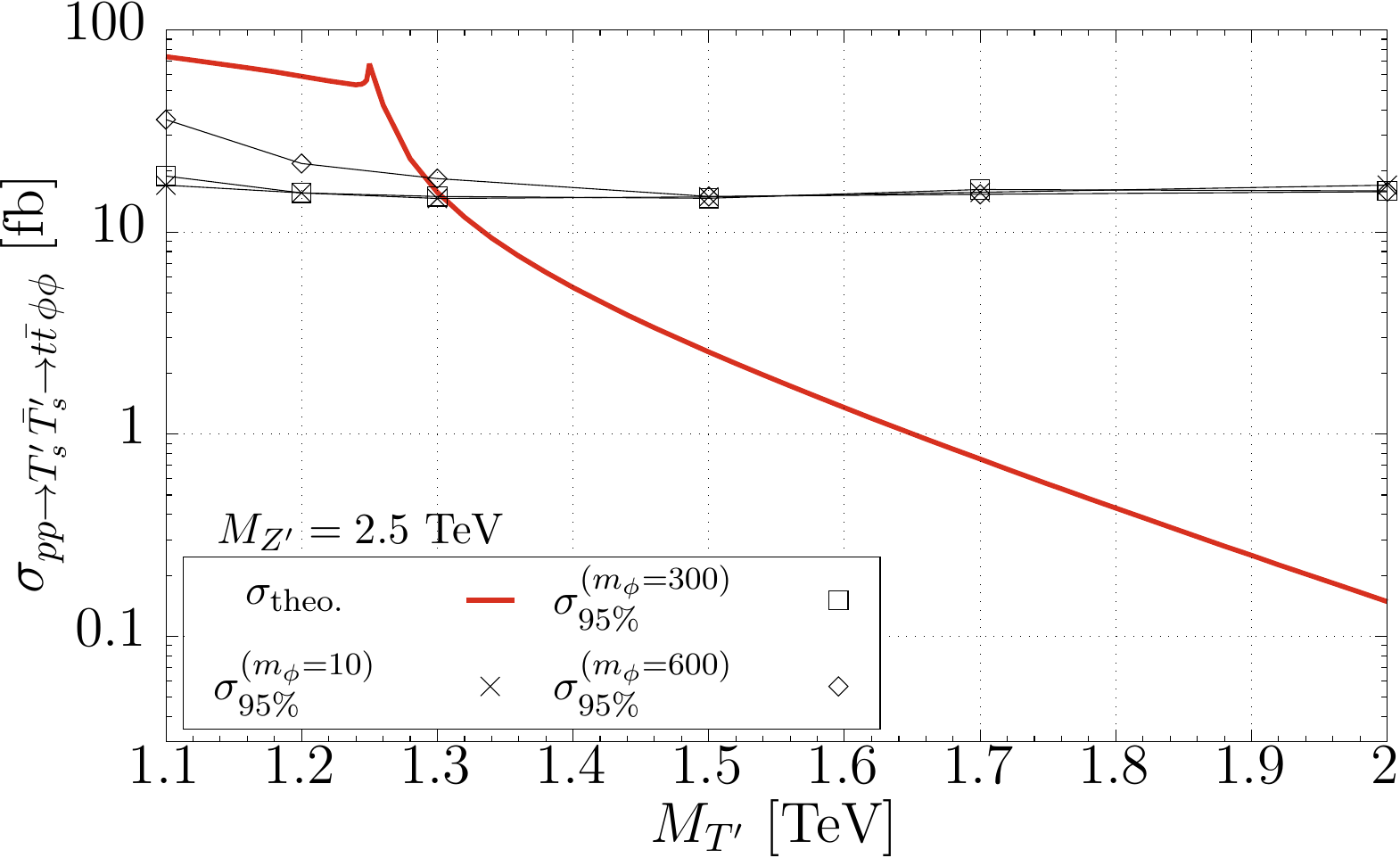}\hfill%
  \includegraphics[width=0.5\textwidth]{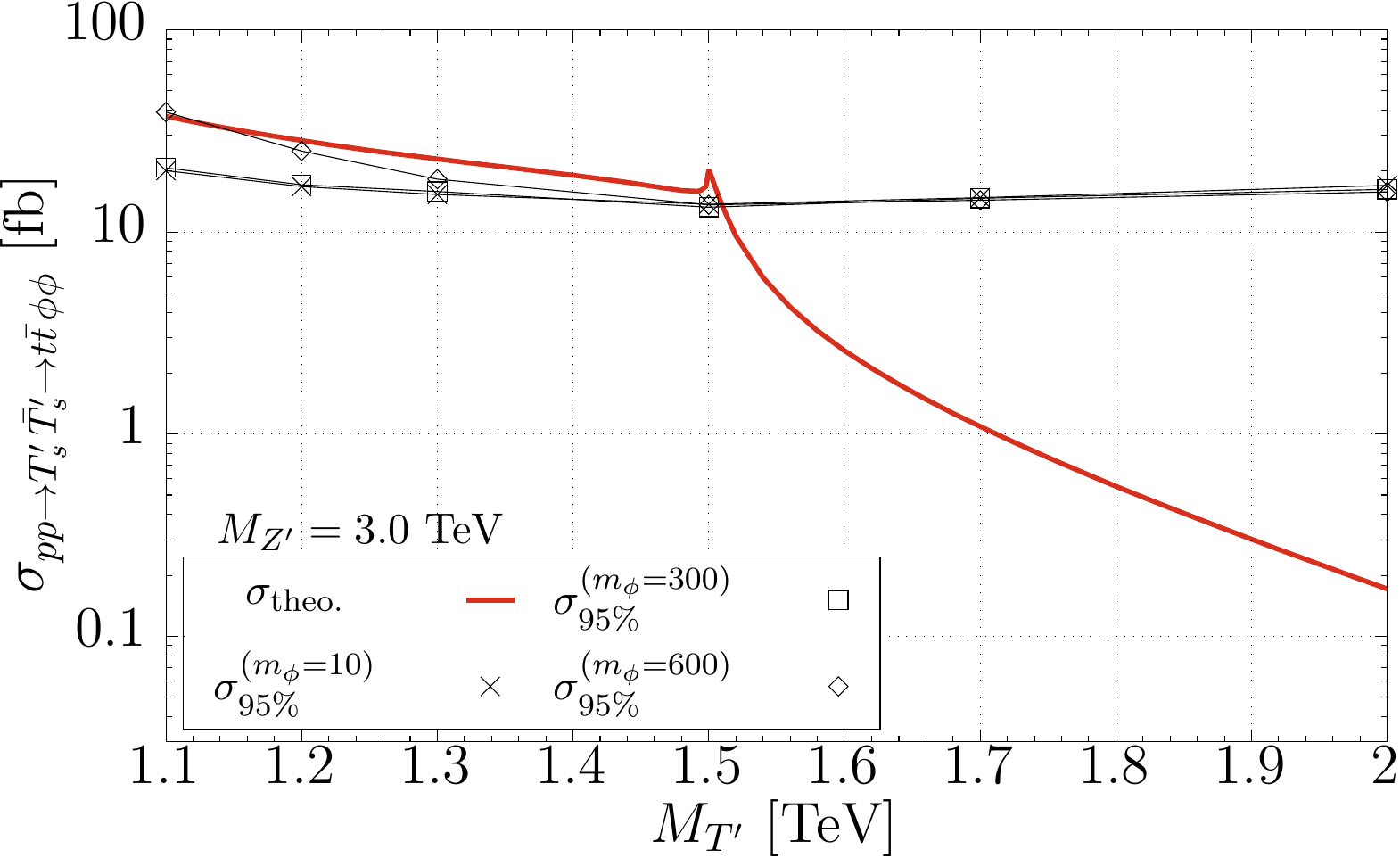}\\
  \includegraphics[width=0.5\textwidth]{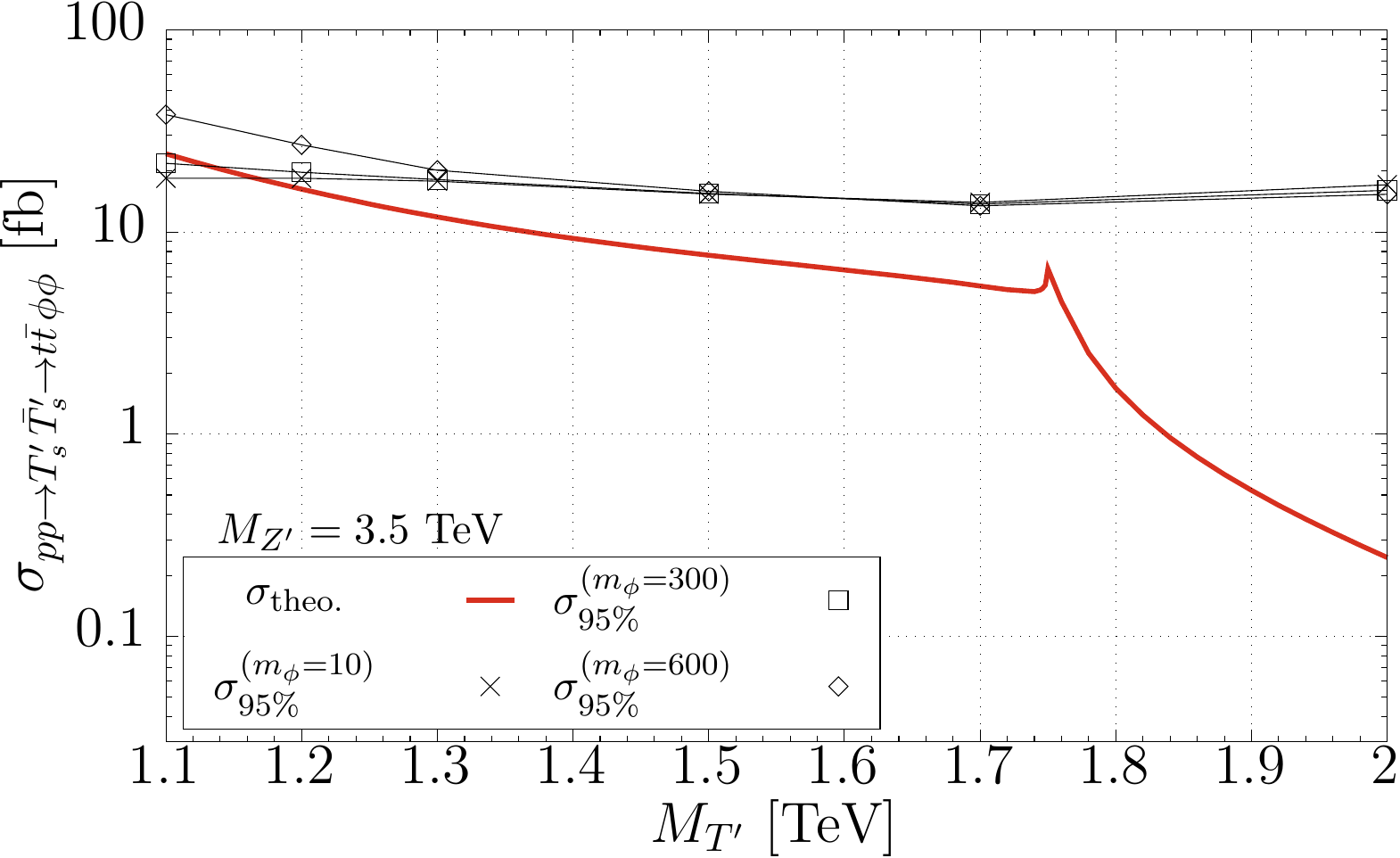}\hfill%
  \includegraphics[width=0.5\textwidth]{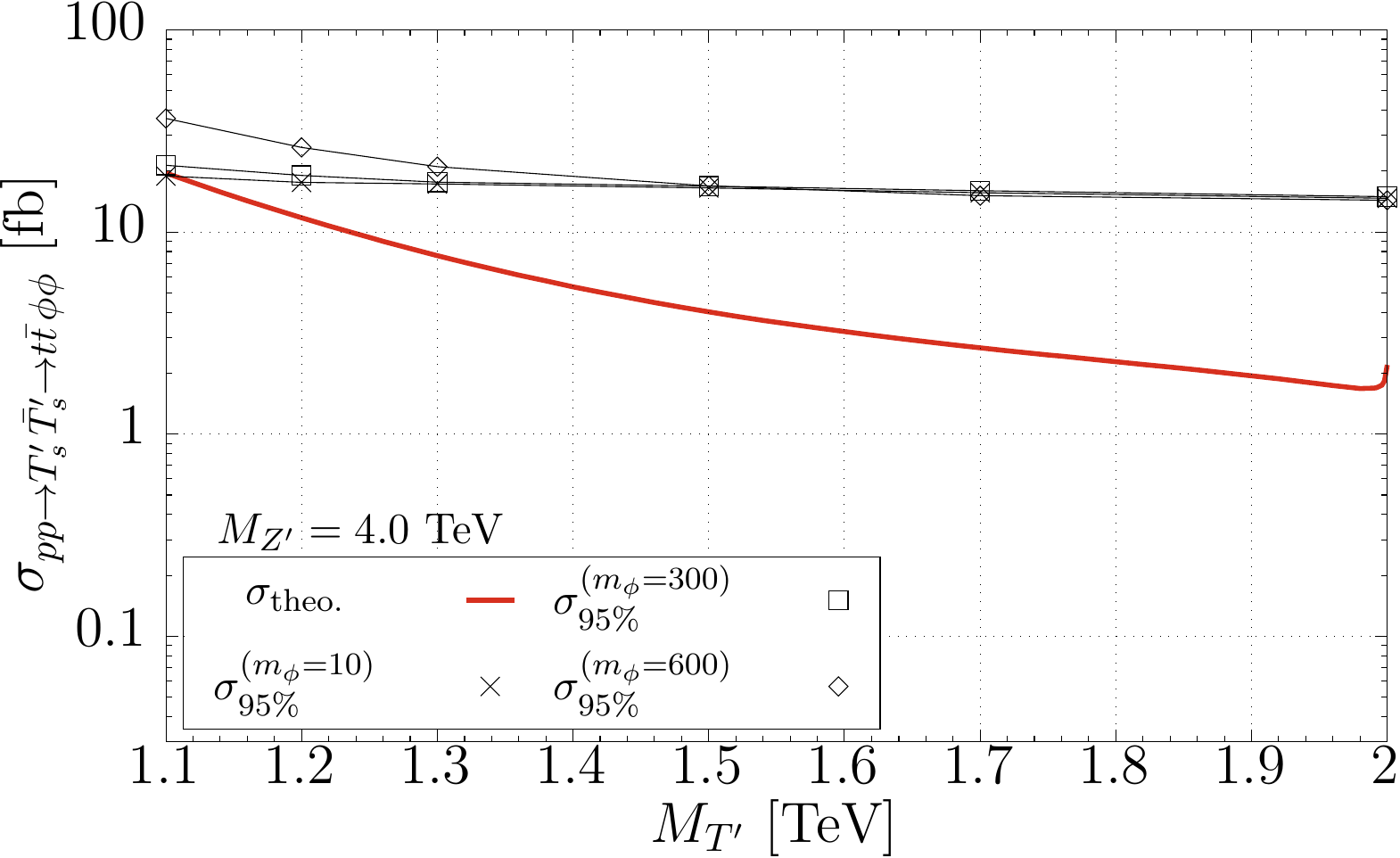}\\
  \includegraphics[width=0.5\textwidth]{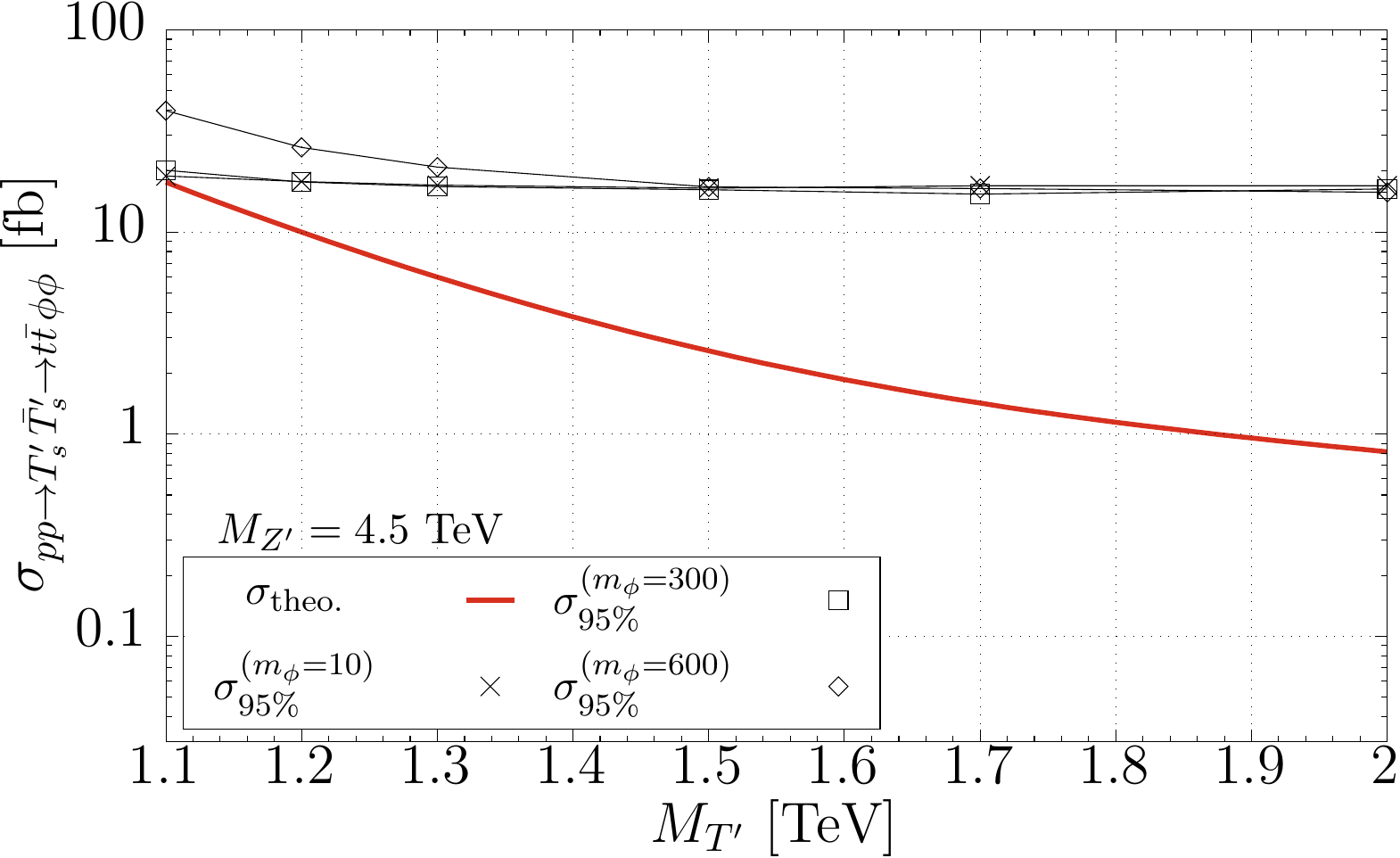}\hfill%
  \includegraphics[width=0.5\textwidth]{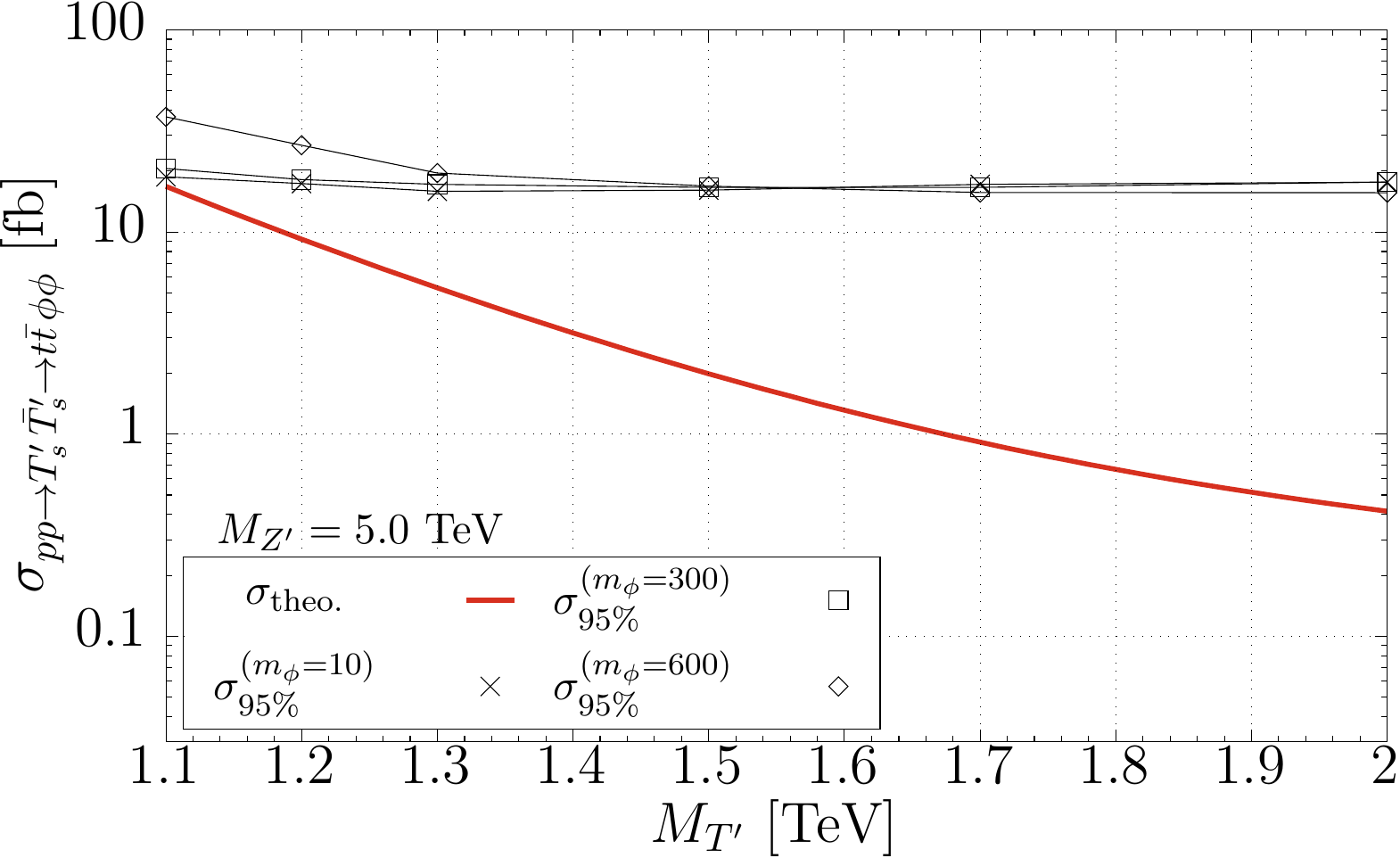}
  \vspace*{-5mm}
  \caption{Theoretical (red) and experimental (black) cross sections for $pp \to T'_s\overline{T'_s} \to t \bar{t} \phi \phi$ in fb in dependence of $M_{T'_s}$ and $M_\phi$ for our benchmark point (same data as in figure~\ref{fig:benchmark-map}). $M_{\phi}$ is given in GeV and the couplings read $\lambda_{Z'q\bar{q}} = 0.25 = \lambda_{Z'\ell^+\ell^-}$ and $\lambda_{Z'T'_s\overline{T'_s}} = 2.5$.}
  \label{fig:benchmark-1D}
\end{figure}

To get a better idea about the experimental and signal cross sections, the same data as in figure~\ref{fig:benchmark-map} has been plotted again in figure~\ref{fig:benchmark-1D}, but this time for constant $M_{Z'}$, variable $M_{T'_s}$ and $m_\phi$ up to 600 GeV.
The red curves show the signal cross section, whereas the black symbols correspond to the experimental cross section for $m_\phi = 10, 300, 600$ GeV, respectively.

In comparison to figure~\ref{fig:benchmark-map}, it is now possible to spot the signal exclusion for each $M_{Z'}$ more easily. For $M_{Z'} = 2.5$ TeV, the signal is excluded up to $M_{T'_s} \approx 1.3$~TeV, therefore excluding the entire regime for on-shell $T'_s$-pair production. For $M_{Z'} = 3$ TeV, the results are similar, although the exclusion now starts at roughly $M_{T'_s} = 1.51$ TeV, just after off-shell production starts. With $M_{Z'} = 3.5$ TeV, the $Z'$ contributions to the predicted cross sections start to weaken with most of the on-shell regime not being excluded (up to $M_{T'_s} = 1.16$ TeV). For larger $M_{Z'}$, the $Z'$ contributions are too small to yield any new exclusion limit and all signal curves are excluded for $M_{T'_s} = 1.08$ TeV, the limit from pure QCD $T'_s$-pair production as shown in figure~\ref{fig:crossx-1D-gluon}.

As mentioned above, this analysis only holds in that amount of detail for the chosen benchmark $\(\lambda_{Z'q\bar{q}}, \lambda_{Z'T'_s\overline{T'_s}}\) = \(0.25, 2.5\)$. It is possible, however, to do a qualitative analysis also for different benchmarks. For that, we note again that slight changes in $\lambda_{Z'T'\overline{T'_s}}$ will not significantly change the cross section (see blue, approximately horizontal limits in figure~\ref{fig:ttMET-pspaces}), which can be explained in the scope of the NWA by the saturation of BR$\(Z' \to T'_s\overline{T'_s}\) \to 1$ for increasing $\lambda_{Z'T'_s\overline{T'_s}}$ (see section~\ref{sec:ttMET}). The cross section will drop, however, once the NWA is not applied, and we took this drop into account by rescaling the NWA cross sections by a fitting function $\kappa\(\lambda^2_{Z'T'_s\overline{T'_s}}\)$. Still, the drop in cross section is rather small for changing $\lambda_{Z'T'_s\overline{T'_s}}$ and will therefore just slightly influence the limits obtained when performing a full benchmark study. When changing $\lambda_{Z'q\bar{q}}$, on the other hand, the situation changes significantly. Increasing $\lambda_{Z'q\bar{q}}$ will lead to rapidly growing cross sections and therefore a rapid shift of the limits as well, whereas the inverse holds for decreasing $\lambda_{Z'q\bar{q}}$.

Since we have positive interference between the $Z'$ and QCD $T'_s$-pair production channels (see section~\ref{sec:interference-effects}), we are also able to give an idea about the excluded $M_{T'_s}$ range for different benchmarks based on the above qualitative discussion. As can be seen in figure~\ref{fig:crossx-1D-gluon}, for $M_{T'_s} \gtrsim 1.2$ TeV, the QCD contributions only make up for about half of the experimental bound. Since including the $Z'$ channel with $M_{Z'} \leq 3$ TeV leads to an exclusion of the signal (see figure~\ref{fig:benchmark-1D}), we can deduct that the $Z'$ parts clearly dominate the signal and an increase in $\lambda_{Z'q\bar{q}}$ would only enhance this behaviour. For heavier $Z'$ up to 5 TeV, the $Z'$ and QCD channels are contributing nearly identical parts to the cross section. Not enough, however, to exceed the experimental bounds.

\begin{figure}[ht]
  \centering
  \includegraphics[width=\textwidth]{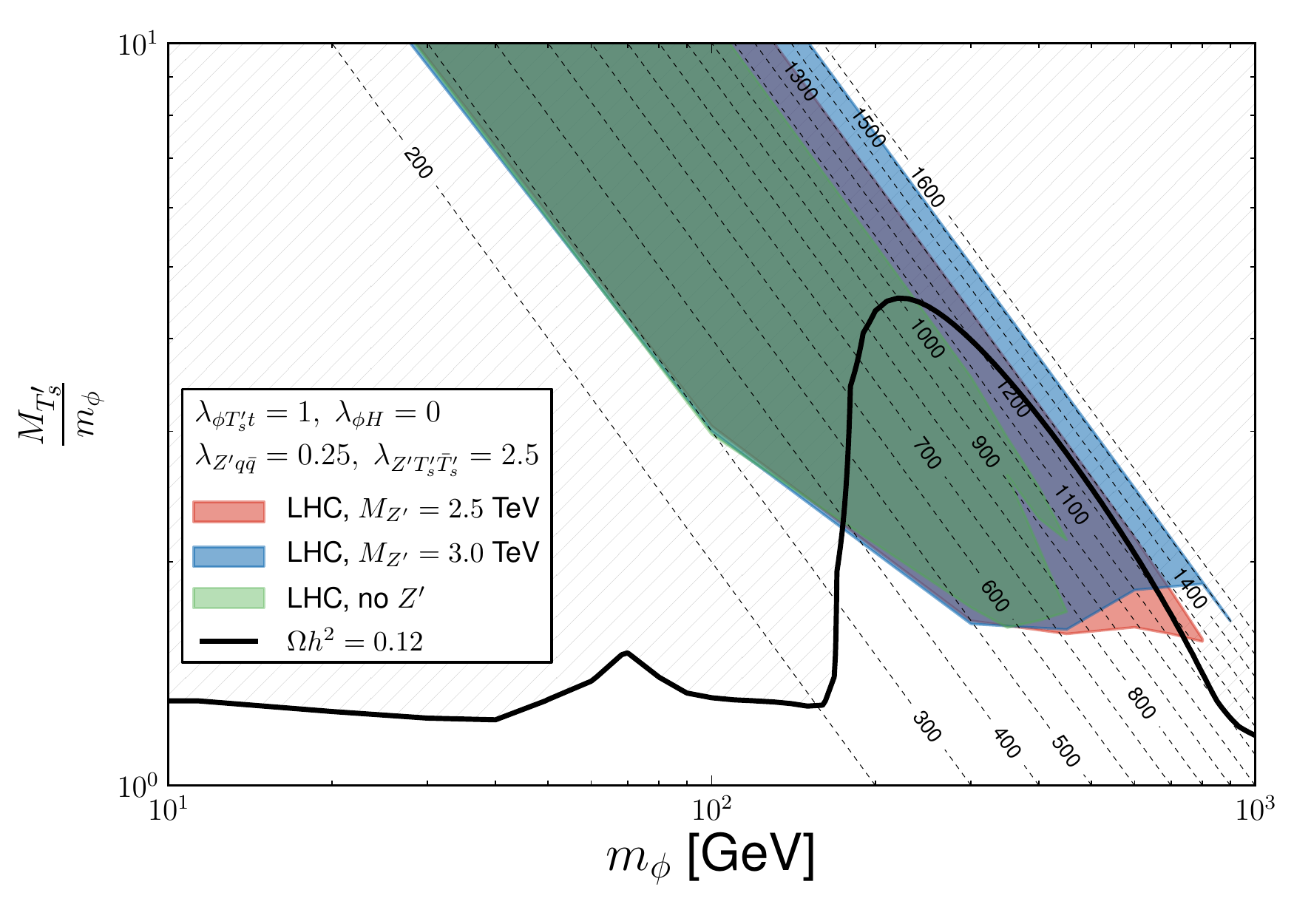}
  \vspace*{-5mm}
  \caption{LHC, DM Direct Detection and relic density constraints on the parameter space of the ZP-TP-DM  model in the $\(\frac{M_{T'_s}}{m_\phi}, m_\phi\)$ plane for $\lambda_{\phi H}=0$ and  and $\lambda_{\phi T'_s t} = 1$:
  a) the green-shaded area indicates the current LHC exclusion region for the $t\bar{t}+\MET$ signature coming from the process 
  $pp \to T'_s\overline{T'_s} \to t\bar{t}\, \phi\phi$, mediated by gluon exchange only (no $Z'$);
  b) the red- and blue-shaded areas present the extended reach of the LHC for $M_{Z'} = 2.5$ and 3 TeV, respectively, with $\lambda_{Z'q\bar{q}} = 0.25 = \lambda_{Z'\ell^+\ell^-}$ and $\lambda_{Z'T'_s\overline{T'_s}} = 2.5$;
  c) the grey-hatched parameter space above the black contour is excluded by relic density constraints. The thin dashed lines with the respective labels indicate the iso-levels of $M_{T'_s}$ in GeV.}
  \label{fig:LHC-exclusion_plus_DM}
\end{figure}

The above argument is also true for $M_{T'_s} \gtrsim m_t + m_\phi$ up to $M_{T'_s} \approx m_t + m_\phi + \mathcal{O}(50\ \text{GeV})$ (see figure~\ref{fig:crossx-1D-gluon}, left ends of coloured lines), as long as $m_\phi \lesssim 450$ GeV and $M_{Z'} \leq 3$ TeV. For the remaining $M_{T'_s}$ between $m_t + m_\phi + \mathcal{O}(50\ \text{GeV})$ and 1.2 TeV, it is hard to give any qualitative statement due to the strong signal dependence of $m_\phi$ and a dedicated benchmark analysis as the one we performed above needs to be done.

Returning to the benchmark set of couplings presented in this subsection, let us put our results into context with the dark matter bounds presented in section \ref{sec:model}. In analogy to the green LHC no-$Z'$ bounds from figures~\ref{fig:DMTP-parspace1} and \ref{fig:DMTP-parspace2}, we are now presenting these limits for our benchmark including  $Z'$ and demonstrate its role. The bounds for the no-$Z'$ case and for $M_{Z'} = 2.5$ and 3 TeV are shown in figure~\ref{fig:LHC-exclusion_plus_DM} together with the relic density for $\lambda_{\phi H} = 0$ and $\lambda_{\phi T'_s t} = 1$.
While the green limits without $Z'$ already cover a large fraction of parameter space up to $m_\phi = 450$ GeV, including the $Z'$ contributions allows us to extend this range up to $m_\phi = 800\ (900)$ GeV for $M_{Z'} = 2.5\ (3)$ TeV, therefore completely closing the gap between LHC and relic density constraints along the $m_\phi$ axis. For $M_{Z'} \geq 3.5$ TeV, the $Z'$ contributions are too small to enhance the limits visibly (see figure~\ref{fig:benchmark-1D}, middle and bottom row) and the QCD-only limits become maximal again.

To give a general idea about the kinematic properties for different $(M_{T'_s},m_\phi)$ mass pairs in the studied region, we also present four more sub-benchmarks and their QCD-only contributions in tables~\ref{tab:benchmarks} and \ref{tab:benchmarks-QCD} in appendix \ref{sec:appendix-tables}. The \texttt{CheckMATE} cutflow in the lower part of the tables shows the fraction of events surviving the listed cut (normalised to 1). The last row in the cutflow section (bold) therefore corresponds to the overall efficiency. The impact of the $Z'$ can be estimated by comparing the efficiencies between benchmarks in table~\ref{tab:benchmarks} and \ref{tab:benchmarks-QCD}. In table~\ref{tab:benchmarks}, we also present the cutflow efficicency for the BP2-RR benchmark, which is analogue to the BP2 benchmark, but with the RR $Z'$ coupling combination. One can see that the efficiencies for LL (BP2 benchmark) and RR (BP2-RR benchmark) differ only by about 2 \%. At the same time, the overall efficiencies for the ``QCD" benchmarks from table~\ref{tab:benchmarks-QCD} and benchmarks with $Z'$ from table~\ref{tab:benchmarks} differ by 10--15 \%, which is not negligible. This difference in efficiencies is related to an obvious difference in the kinematics between the QCD and $Z'$-mediated processes: a) the rapidity of a $T'_s\overline{T'_s}$-pair originating from a $Z'$ is broader than from QCD production, since the $Z'$ is produced from $q\bar{q}$ fusion; b) the $\MET$ distribution is harder when $Z'$ bosons are included, especially for heavier $Z'$ bosons; c) top partners are more boosted in case of $Z'$ mediation, which makes final state leptons more energetic but less isolated.
 
\section{Conclusions}
\label{sec:conclusions}
We have explored the phenomenology of a simplified, effective model with a vector resonance $Z'$, a fermionic vector-like coloured partner of the top quark $T'$, which carries negative DM parity, and a scalar DM candidate $\phi$ -- which we refer to as the ZP-TP-DM model.

Our main focus is the exploration of the process $pp \to Z' \to T'\overline{T'} \to t\bar{t}\,\phi\phi$ at the LHC, which to the best of our knowledge has not been studied previously. 
We have found that this process plays an important role in addition to the 
$T'\overline{T'}$ production via QCD interactions.

Because the process under study originates exclusively from a quark-anti-quark initial state
and because the $T'\overline{T'}$-pair arises from an on-shell $Z'$ decay,
its kinematical behaviour is quite different from QCD $T'\overline{T'}$-pair production.
On the one hand, the $Z'$-mediated $t\bar{t}\,\phi\phi$ signature leads to higher $p_T$ leptons
and $\MET$ (especially for heavier $Z'$) compared to the case of QCD production alone, but on the other hand, leptons for the $Z'$-mediated process have higher rapidity.
These two features affect the detector efficiency for the signature in opposite ways --
the higher $p_T$ of leptons and the higher $\MET$ increase efficiency, while the higher rapidity of the leptons decreases it. The overall effect is that the detector efficiency
for the $Z'$-mediated $t\bar{t}\,\phi\phi$ signature is about 10\% higher then for the QCD-mediated one.  We showed that the chirality of the $Z'$ couplings to SM quarks and to top partners do not play a major role in kinematical distributions, and therefore similar efficiencies apply for all chiral combinations of couplings. 
We also explored the effect of interference between the $Z'$ and QCD initiated processes and have found that the interference is positive, but very small --- at the level of about $+3\ \%$ and essentially independent of the $Z'$ width.

We have  found that  the presence of the $Z'$ can  provide an additional and even dominant contribution (about one order of magnitude larger than the QCD one) to the $t\bar{t}\,\phi\phi$ signature without conflicting with existing bounds from $Z'$ searches in di-jet and di-lepton final states. We have demonstrated that the $t\bar{t}+\MET$  signature at the LHC plays an important and complementary role to non-collider searches in setting the limits on the ZP-TP-DM parameter space. Moreover, the $Z'$,  $T'$ and DM masses can  be probed with the $t\bar{t} +\MET$ signature well beyond the reach of  QCD production alone, without being in conflict with existing $Z'$ search bounds from di-lepton and di-jet signatures, as we explicitly showed.
From figure~\ref{fig:LHC-exclusion_plus_DM}, one can see that with  $M_{Z'}=3$~TeV, the LHC is already probing DM masses up to about 900~GeV and $M_{T'}$ up to about 1.5~TeV, which is about a factor of two larger (for both particles) than for the bounds from QCD production alone. We regard this potential increase in reach quite remarkable and think that it is worth considering $Z'$-mediated top partner-pair production in future phenomenological studies and experimental searches. We provide publicly available implementations of the model at {\sc HEPMDB}~\cite{hepmdb}
under hepmdb:0717.0253~\cite{dmtp-calchep} (\texttt{CalcHEP}) and hepmdb:0717.0253~\cite{dmtp-mg} (\texttt{Madgraph}) and we would like to encourage experimental groups at the LHC to explore the potential of $Z'$-mediated top partner pair-production followed by their decays to dark matter.

In this study, we focussed on a region in parameter space with  $M_{T'} - m_\phi > m_{t}$, i.e. the case in which the decay of a top partner into a top quark and DM occurs on-shell. The case of strongly degenerate $M_{T'}$ and $m_\phi$, where the top partner only decays off-shell, is not covered by the analysis presented here, but we would like to point out that in this case, a study similar to the case of SUSY with degenerate stops and neutralinos can be performed.    
\acknowledgments
The authors would like to thank Alexander Pukhov for discussions and finding and fixing a bug in the micrOMEGAs package which was crucial in order 
to produce the correct results for the DM direct detection rates for the model under study.
The authors acknowledge the use of the IRIDIS High Performance Computing Facility, and
associated support services at the University of Southampton, in the completion of this work.
AB and PS acknowledge partial support from the InvisiblesPlus RISE from the European
Union Horizon 2020 research and innovation programme under the Marie Sklodowska-Curie grant
agreement No 690575.
AB also thanks the NExT Institute, Royal Society Leverhulme Trust Senior Research Fellowship LT140094 and Soton-FAPESP grant for partial  support.
TF's work was supported by IBS under the project code IBS-R018-D1. The work of B. J.  was partially supported by the  S\~ao Paulo
Research Foundation (FAPESP) under Grants No. 2016/01343-7  and No. 2017/05770-0

\newpage
\appendix
\section{Benchmark Cutflows}
\label{sec:appendix-tables}
The following tables hold four benchmarks for the full process shown in figure~\ref{fig:Zp-prod-Feynman-diagram} (table~\ref{tab:benchmarks}) and for the QCD-only production (table~\ref{tab:benchmarks-QCD}) together with the respective cutflows for each benchmark. The couplings of all benchmark points read $\lambda_{Z'q\bar{q}} = \lambda_{Z'\ell^+\ell^-} = 0.25$, $\lambda_{Z'T'_s\overline{T'_s}} = 2.5$, $\lambda_{\phi H} = 0$ and $\lambda_{  \phi T'_s t} = 0.1$. Additionally, we provide the benchmark "BP2-RR", which is the chiral counterpart to the LL choice we used throughout this work (see section~\ref{sec:PLA}) and has $\lambda_{Z'q\bar{q},R} = \lambda_{Z'\ell^+\ell^-,R} = 0.25$, $\lambda_{Z'T'_s\overline{T'_s},R} = 2.5$ and $\lambda_{Z'q\bar{q},L} = 0 = \lambda_{Z'\ell^+\ell^-,L}$, $\lambda_{Z'T'_s\overline{T'_s},L} = 0$.
\begin{table}[htbp]
  \centering
  \begin{tabular}{clcccccc}
    \toprule
                                                       & Parameter                            & & BP1          & BP2          & BP2-RR       & BP3          & BP4          \\
    \midrule
    \multirow{3}{*}{\rotatebox{90}{\textsc{Input}}}    & $M_{Z'}$   \ [GeV]                   & & 2500         & 3000         & 3000         & 3000         & 3500         \\
                                                       & $M_{T'_s}$ \ [GeV]                   & & 1150         & 1200         & 1200         & 1500         & 1700         \\
                                                       & $m_\phi$   \ \, [GeV]                & & 600          & 300          & 300          & 300          & 500          \\
    \midrule
                                                       & $\Gamma_{Z'}$ \ \, [GeV]             & & 241.75       & 435.61       & 435.61       & 59.61        & 227.40       \\
                                                       & $\frac{\Gamma_{Z'}}{M_{Z'}}$ \, [\%] & & 16.11        & 14.52        & 14.52        & 1.99         & 6.50         \\
                                                       & $\sigma$ \ \ \ \; [fb]               & & 66.24        & 28.29        & 28.29        & 20.46        & 5.40         \\
    \midrule
    \multirow{18}{*}{\rotatebox{90}{\textsc{Cutflow}}} & \texttt{0\_trigger\_etmiss}          & & 0.81518      & 0.90046      & 0.89952      & 0.92820      & 0.93820      \\
                                                       & \texttt{1\_lepton\_onelepton}        & & 0.17616      & 0.18460      & 0.18678      & 0.18566      & 0.18072      \\
                                                       & \texttt{2\_mt}                       & & 0.17184      & 0.17992      & 0.18190      & 0.18166      & 0.17734      \\
                                                       & \texttt{3\_jets}                     & & 0.16436      & 0.16782      & 0.17030      & 0.17042      & 0.16630      \\
                                                       & \texttt{tN\_high\_01\_tauVeto}       & & 0.14892      & 0.15640      & 0.15716      & 0.16108      & 0.15782      \\ 
                                                       & \texttt{tN\_high\_02\_nJets}         & & 0.08944      & 0.08940      & 0.09160      & 0.08970      & 0.08618      \\ 
                                                       & \texttt{tN\_high\_03\_JetsPT}        & & 0.07292      & 0.07766      & 0.07934      & 0.08020      & 0.07594      \\ 
                                                       & \texttt{tN\_high\_04\_etmiss}        & & 0.03268      & 0.05156      & 0.05208      & 0.06118      & 0.06030      \\ 
                                                       & \texttt{tN\_high\_05\_etmissVcal}    & & 0.03268      & 0.05156      & 0.05208      & 0.06118      & 0.06030      \\ 
                                                       & \texttt{tN\_high\_06\_htmiss}        & & 0.03250      & 0.05118      & 0.05162      & 0.06092      & 0.05970      \\ 
                                                       & \texttt{tN\_high\_07\_mt}            & & 0.02930      & 0.04672      & 0.04706      & 0.05612      & 0.05526      \\ 
                                                       & \texttt{tN\_high\_08\_amt2}          & & 0.02872      & 0.04578      & 0.04626      & 0.05542      & 0.05456      \\ 
                                                       & \texttt{tN\_high\_09\_no}            & & 0.02872      & 0.04578      & 0.04626      & 0.05542      & 0.05456      \\ 
                                                       & \texttt{tN\_high\_10\_no}            & & 0.02872      & 0.04578      & 0.04626      & 0.05542      & 0.05456      \\ 
                                                       & \texttt{tN\_high\_11\_dR}            & & 0.02638      & 0.04082      & 0.04046      & 0.05090      & 0.04934      \\ 
                                                       & \texttt{tN\_high\_12\_LRJET\_PT}     & & 0.02354      & 0.03846      & 0.03808      & 0.04888      & 0.04766      \\ 
                                                       & \texttt{tN\_high\_13\_LRJET\_M}      & & 0.02146      & 0.03530      & 0.03480      & 0.04580      & 0.04412      \\ 
                                                       & \texttt{tN\_high\_14\_dphi}          & & {\bf0.02030} & {\bf0.03334} & {\bf0.03250} & {\bf0.04272} & {\bf0.04104} \\ 
    \bottomrule
  \end{tabular}
  \caption{Benchmarks for the full process (see figure \ref{fig:Zp-prod-Feynman-diagram}) together with the \texttt{CheckMATE} cutflow efficiencies (fraction of events surviving a certain cut, normalised to 1). The couplings for all points read $\lambda_{Z'q\bar{q}} = \lambda_{Z'\ell^+\ell^-} = 0.25$, $\lambda_{Z'T'_s\overline{T'_s}} = 2.5$, $\lambda_{\phi H} = 0$ and $\lambda_{  \phi T'_s t} = 0.1$. The cutflow corresponds to the SR \texttt{tN\_high} from \texttt{ATLAS\_CONF\_2016\_050}, which yields the best limits.}
  \label{tab:benchmarks}
\end{table}

\begin{table}[htbp]
  \centering
  \begin{tabular}{clccccc}
    \toprule
                                                       & Parameter                            & & BP1-QCD      & BP2-QCD      & BP3-QCD      & BP4-QCD      \\
    \midrule
    \multirow{3}{*}{\rotatebox{90}{\textsc{Input}}}    & $M_{Z'}$   \ [GeV]                   & & -            & -            & -            & -            \\
                                                       & $M_{T'_s}$ \ [GeV]                   & & 1150         & 1200         & 1500         & 1700         \\
                                                       & $m_\phi$   \ \, [GeV]                & & 600          & 300          & 300          & 500          \\
    \midrule
                                                       & $\Gamma_{Z'}$ \ \, [GeV]             & & -            & -            & -            & -            \\
                                                       & $\frac{\Gamma_{Z'}}{M_{Z'}}$ \, [\%] & & -            & -            & -            & -            \\
                                                       & $\sigma$ \ \ \ \; [fb]               & & 11.62        & 8.49         & 1.49         & 0.51         \\
    \midrule
    \multirow{18}{*}{\rotatebox{90}{\textsc{Cutflow}}} & \texttt{0\_trigger\_etmiss}          & & 0.81470      & 0.88948      & 0.93012      & 0.93932      \\
                                                       & \texttt{1\_lepton\_onelepton}        & & 0.17338      & 0.18490      & 0.17670      & 0.17674      \\
                                                       & \texttt{2\_mt}                       & & 0.16814      & 0.17936      & 0.17280      & 0.17274      \\
                                                       & \texttt{3\_jets}                     & & 0.15880      & 0.16744      & 0.16024      & 0.15952      \\
                                                       & \texttt{tN\_high\_01\_tauVeto}       & & 0.14244      & 0.15376      & 0.15060      & 0.15032      \\ 
                                                       & \texttt{tN\_high\_02\_nJets}         & & 0.09092      & 0.09196      & 0.08616      & 0.08528      \\ 
                                                       & \texttt{tN\_high\_03\_JetsPT}        & & 0.07704      & 0.08022      & 0.07682      & 0.07574      \\ 
                                                       & \texttt{tN\_high\_04\_etmiss}        & & 0.03576      & 0.05076      & 0.05804      & 0.05862      \\ 
                                                       & \texttt{tN\_high\_05\_etmissVcal}    & & 0.03576      & 0.05076      & 0.05804      & 0.05862      \\ 
                                                       & \texttt{tN\_high\_06\_htmiss}        & & 0.03550      & 0.05028      & 0.05714      & 0.05772      \\ 
                                                       & \texttt{tN\_high\_07\_mt}            & & 0.03136      & 0.04468      & 0.05164      & 0.05316      \\ 
                                                       & \texttt{tN\_high\_08\_amt2}          & & 0.03036      & 0.04368      & 0.05060      & 0.05240      \\ 
                                                       & \texttt{tN\_high\_09\_no}            & & 0.03036      & 0.04368      & 0.05060      & 0.05240      \\ 
                                                       & \texttt{tN\_high\_10\_no}            & & 0.03036      & 0.04368      & 0.05060      & 0.05240      \\ 
                                                       & \texttt{tN\_high\_11\_dR}            & & 0.02664      & 0.03830      & 0.04402      & 0.04580      \\ 
                                                       & \texttt{tN\_high\_12\_LRJET\_PT}     & & 0.02330      & 0.03596      & 0.04244      & 0.04444      \\ 
                                                       & \texttt{tN\_high\_13\_LRJET\_M}      & & 0.02082      & 0.03284      & 0.03910      & 0.04080      \\ 
                                                       & \texttt{tN\_high\_14\_dphi}          & & {\bf0.01976} & {\bf0.03062} & {\bf0.03662} & {\bf0.03844} \\ 
    \bottomrule
  \end{tabular}
  \caption{QCD benchmarks (see figure \ref{fig:Zp-prod-Feynman-diagram}, centre and right) together with the \texttt{CheckMATE} cutflow efficiencies (fraction of events surviving a certain cut, normalised to 1). The couplings for all points read $\lambda_{Z'q\bar{q}} = \lambda_{Z'\ell^+\ell^-} = 0.25$, $\lambda_{Z'T'_s\overline{T'_s}} = 2.5$, $\lambda_{\phi H} = 0$ and $\lambda_{  \phi T'_s t} = 0.1$. The cutflow corresponds to the SR \texttt{tN\_high} from \texttt{ATLAS\_CONF\_2016\_050}, which yields the best limits.}
  \label{tab:benchmarks-QCD}
\end{table}

\clearpage\relax
\bibliographystyle{JHEPmod}
\bibliography{bib}

\end{document}